\documentclass[12pt]{article}
\usepackage{epsfig,cite}

\input paperdef

\evensidemargin \oddsidemargin
\def\_{\rule{.3em}{.15ex}} 

\newcommand{\scs}{\scriptscriptstyle}

\def\slash#1{\setbox0=\hbox{$#1$}#1\hskip-\wd0\dimen0=5pt\advance
       \dimen0 by-\ht0\advance\dimen0 by\dp0\lower0.5\dimen0\hbox
         to\wd0{\hss\sl/\/\hss}}
\def\simleq{\stackrel{<}{\scs \sim}}

\newcommand{ \NPB    }[3]{{\em Nucl. Phys.}      {\bf B #1} (#2) #3}
\newcommand{ \PLB    }[3]{{\em Phys. Lett.}     {\bf B #1}  (#2) #3}
\newcommand{ \PLBold }[3]{{\em Phys. Lett.}      {\bf B #1} (#2) #3}
\newcommand{ \PRD    }[3]{{\em Phys. Rev.}     {\bf D #1}  (#2) #3}

\newcommand{ \PREP   }[3]{{\em Phys. Rept.}       {\bf #1}  (#2) #3}

\newcommand{ \EPJC   }[3]{{\em Eur. Phys. Jour.}  {\bf C #1}  (#2) #3}

\begin{document}

\thispagestyle{empty}
\setcounter{page}{0}
\def\thefootnote{\fnsymbol{footnote}}

\begin{flushright}
BNL--HET--01/18, CALT--68--2315\\
CERN--TH/2000-348\\
DCPT/01/56, IPPP/01/28\\
hep-ph/0106255 \\
\end{flushright}

\vspace{1mm}

\begin{center}

{\large\sc {\bf Implications of the Higgs boson searches}}

\vspace{0.4cm}

{\large\sc {\bf on different soft SUSY-breaking scenarios}}
 
\vspace{.6cm}

{\sc 
S.~Ambrosanio$^{1,2}$%
\footnote{email: Sandro.Ambrosanio@bancaroma.it}%
, A.~Dedes$^{3}$%
\footnote{email: dedes@th.physik.uni-bonn.de}%
, S.~Heinemeyer$^{4}$%
\footnote{email: Sven.Heinemeyer@bnl.gov}%
, S.~Su$^{5}$%
\footnote{email: shufang@theory.caltech.edu}%
~and G.~Weiglein$^{1,6}$%
\footnote{email: Georg.Weiglein@cern.ch}
}

\vspace*{.6cm}

{\sl
$^1$CERN, TH Division, CH-1211 Geneva 23, Switzerland

\vspace*{0.2cm}

$^2$Banca di Roma, Direzione Generale, Linea Finanza,\\
viale U.~Tupini 180, I-00144 Roma, Italy
 (since Jan.\ 1, 2001)

\vspace*{0.2cm}

$^3$Physikalisches Institut der Universit\"at Bonn, 
 Nu\ss allee 12, 
 D-53115 Bonn, Germany

\vspace*{0.2cm}

$^4$HET, Physics Department, Brookhaven Natl.\ Lab., Upton, NY
11973, USA

\vspace*{0.2cm}

$^5$California Institute of Technology, Pasadena, CA 91125, USA

\vspace*{0.2cm}

$^6$Institute for Particle Physics Phenomenology, University of Durham,\\
Durham DH1~3LR, UK

}

\end{center}

\vspace*{0.2cm}

\begin{abstract}
We investigate the Higgs boson sector of the Minimal Supersymmetric
Standard Model (MSSM) in the framework of the three most
prominent soft SUSY-breaking scenarios, mSUGRA, mGMSB and mAMSB.
For each scenario, we determine the parameters at the electroweak scale
from the set of input variables at higher energy
scales (depending on the specific scenario) and evaluate the Higgs boson
properties. The latter are based on results
obtained within the Feynman-diagrammatic approach
by taking into account the complete one-loop and the dominant two-loop
contributions.
The maximum value of the mass of the lightest neutral $\cp$-even MSSM Higgs
boson, $\mh$, is determined in the three scenarios, and the behavior of
the Higgs couplings to fermions and gauge bosons is investigated. 
Restrictions on $\tb$ and on the set of higher-energy scale
parameters are derived from the lower limits arising from the Higgs
search at LEP2. We furthermore discuss the regions of parameter space
in the three scenarios compatible with 
interpreting the excess observed at LEP2 as a Higgs
signal, $\mh = 115^{+1.3}_{-0.9} \gev$.
The case where the events observed at LEP2 could originate from the 
production of the heavier neutral $\cp$-even
Higgs boson is also considered. The implications of a possible Higgs
signal at 115~GeV for SUSY searches at future colliders are briefly discussed 
for each of the three scenarios.

\end{abstract}

\def\thefootnote{\arabic{footnote}}
\setcounter{page}{0}
\setcounter{footnote}{0}

\newpage


\section{Introduction}

The search for the light neutral Higgs boson is a crucial test of
Supersymmetry (SUSY) that can be performed with the present and the next
generation of high-energy colliders. The prediction of a relatively light
Higgs boson is common to all supersymmetric models whose couplings
remain in the perturbative regime up to a very high energy
scale~\cite{susylighthiggs}.  Finding the Higgs boson is thus one of
the main goals of today's high-energy physics.  
The data taken during the final year of LEP running at $\sqrt{s} \gsim
206 \gev$, while establishing a 95\% C.L.\ exclusion limit for the Standard 
Model (SM) Higgs boson of $\mH > 113.5$~GeV, showed at about the
$3\sigma$ level an excess of signal-like events over the background 
expectation which is in agreement with the 
expectation for the production of a SM Higgs
boson of $\mH = 115^{+ 1.3}_{-0.9}$~GeV~\cite{LEPHiggs}.
A Higgs mass value of about 115~GeV would indicate that the SM can only
be valid up to a scale $\La \lsim 10^6 \gev$ (or the vacuum must me
meta stable), since new physics
contributions are necessary in order to prevent the effective Higgs
potential from becoming unstable~\cite{vacstabil}. 
In the Minimal Supersymmetric Standard Model (MSSM), on the other hand,
the mass of the lightest $\cp$-even Higgs boson, $\mh$, is bounded 
from above by $\mh \simleq 135 \gev$~\cite{mhiggslong} (taking into account 
radiative corrections up to \twol\
order~\cite{mhiggslong,mhiggsRG1a,mhiggsRG1b,mhiggsRG2,mhiggsEP1,mhiggsEP2,mhiggsletter,maulpaul,maulpaul2,mhiggsEP3}). 
The effective Higgs potential is stabilized by 
contributions of the Supersymmetric partners of the SM
particles~\cite{ellisross}. Within the MSSM, the LEP excess can be
interpreted as the production of the lightest $\cp$-even Higgs boson, which
over a wide parameter range has SM-like couplings, or of the heavier 
$\cp$-even Higgs boson, in a region of parameter space where the $\cp$-odd
Higgs boson $A$ is light and the ratio of the vacuum expectation values
of the two Higgs doublets, $\tan\be$, is relatively
large.

In the MSSM no specific assumptions are made about the underlying
Supersymmetry- (SUSY)-breaking mechanism, and a parameterization of all 
possible SUSY-breaking terms is used. This gives rise to the huge number of 
more than 100 new parameters in addition to the SM, 
which in principle can be chosen
independently of each other. A phenomenological analysis of this model
in full generality would clearly be very involved, and one usually
restricts to certain benchmark scenarios~\cite{benchmark}.
On the other hand, models in which all the low-energy parameters are
determined in terms of a few parameters at the Grand Unification
scale (or another high-energy scale), 
employing a specific soft SUSY-breaking scenario, are
much more predictive. The most prominent scenarios in the literature
are minimal Supergravity (mSUGRA)~\cite{Hall,mSUGRArev}, 
minimal Gauge Mediated SUSY Breaking (mGMSB)~\cite{GR-GMSB} 
and minimal Anomaly Mediated SUSY Breaking 
(mAMSB)~\cite{lr,giudice,wells}. 
Analyses of the Higgs sector in these scenarios have been
performed~\cite{GMSBmodels2,higgsmsugra2,higgsgmsb2,higgsamsb,higgsmsugra,higgsgmsb,higgsmsugra3},
mostly focusing only on the maximum value of $\mh$. 
Within the mSUGRA scenario more recently some implications of the LEP2
results on the Higgs search have been investigated in the context of 
further constraints arising from the requirement that the lightest
Supersymmetric particle (LSP) should give rise to an acceptable dark matter
relic density, and that the predictions of the model should be in
agreement with the experimental results on $b \to s \ga$ and the
anomalous magnetic moment of the muon.

In this paper we investigate in detail the predictions in the Higgs sector
arising from the three SUSY-breaking scenarios mSUGRA, mGMSB and mAMSB. 
We relate the input from these scenarios in a uniform way to the
predictions for the low-energy phenomenology in the Higgs sector,
allowing thus a direct comparison of the predictions arising from the
different scenarios. The high-energy parameters given in the three
scenarios are related to the low-energy SUSY parameters via 
renormalization group (RG) running, taking into account contributions up
to two-loop order. After transforming the parameters obtained in this
way into the corresponding
on-shell parameters~\cite{bse,FDRG1,maulpaul3}, they are used as input
for the program \fh~\cite{feynhiggs}, which contains the complete
one-loop and dominant two-loop corrections in the MSSM Higgs sector
evaluated in the Feynman-diagrammatic (FD)
approach~\cite{mhiggslong,mhiggsletter,mhiggsf1l}. Further restrictions
such as from precision observables and the non-observation
of SUSY particles are also taken into account.
Based on these predictions for the Higgs sector phenomenology, we
analyze the consequences of the results obtained from the Higgs
search at LEP on the parameter space of the three scenarios. This is
done by considering
both the LEP exclusion bound~\cite{mssmhiggs}
and the interpretation of the LEP excess as
a possible signal. For the latter case we furthermore discuss the
corresponding spectra of the SUSY particles in view of the SUSY searches
at the next generation of colliders. 

The rest of the paper is organized as follows. In Sect.~2 the three
soft SUSY-breaking scenarios as well as the evaluation of the
$\cp$-even Higgs boson sector of the MSSM 
are briefly reviewed. Details about the combination of renormalization-group
equation (RGE) and FD calculation
are given, and the parameter restrictions used are listed. The
description of our data sets and the 
numerical analyses for the three scenarios is given in Sect.~3. The
conclusions can be found in Sect.~4.


\section{The Higgs sector in soft SUSY-breaking scenarios}
\label{sec:higgs}

The fact that no SUSY partners of the SM particles have so far been
observed means that low-energy SUSY cannot be realized as an unbroken
symmetry in nature, and SUSY models thus have to incorporate
extra supersymmetry breaking interactions. 
This is achieved by adding to the Lagrangian (defined by the 
given ${\rm SU(3)}_C\times {\rm SU(2)}_L \times  {\rm U(1)}_Y$ gauge symmetry
and the superpotential $W$)
some extra interaction terms that respect the gauge symmetry but break 
supersymmetry.
This breaking however should be such that no quadratic divergences appear
and the technical ``solution'' to the hierarchy problem is not spoiled.
Such terms are generally called ``soft SUSY-breaking'' terms.
The most general supersymmetry breaking interaction Lagrangian
resulting from spontaneously broken Supergravity in the flat 
limit ($M_P \to \infty$, where $M_P$ is the Planck mass) contains
just four types of soft SUSY-breaking terms~\cite{softSUSYbreaking}, 
i.e.\ gaugino masses, $\Phi^*\Phi$-scalar masses, 
$\Phi\Phi\Phi$-scalar cubic superpotential interactions and 
$\Phi\Phi$-scalar quadratic superpotential interactions.
Assuming that $R$-parity~\cite{herbi,rpv2} 
is conserved, which we do in this paper
for all SUSY breaking scenarios, reduces the amount of 
new soft terms allowed in the Lagrangian.
Choosing a particular soft SUSY-breaking pattern allows further
reduction of the number of free parameters and the construction
of predictive models.

In this section, we first explain how we employ the principle of
radiative electroweak symmetry breaking (REWSB). Then we
introduce the three most commonly studied soft 
SUSY-breaking scenarios and describe the general method used to derive
predictions for the low-energy Higgs sector, which applies to all scenarios.


\subsection{Radiative electroweak symmetry breaking}
\label{subsec:rewsb}

The investigation of REWSB
in the MSSM~\cite{REWSB} relies on a RG analysis. The Higgs boson
``running'' mass-squared matrix, although 
positive definite
at large energy scales of the
order of the Grand Unification scale $M_{\rm GUT}$,
yields a negative eigenvalue at low energies causing the spontaneous
breakdown of the electroweak (EW) symmetry. 
The result can be interpreted as a prediction of $\MZ$ in terms of 
parameters at a large energy scale.
Alternatively, one can consider $\MZ$ as being determined 
by experiment and derive in this way the 
absolute value of the $\mu$-parameter (which defines the coupling of
the two Higgs doublets) as well as the value
of the bilinear soft-SUSY breaking parameter $B$ at a scale in the
vicinity of the EW scale,
from the minimization conditions of the effective potential,
%
\begin{eqnarray}
\label{min1} 
\mu^2(Q) \ &=& \ \frac{\bar{m}_{H_1}^2-\bar{m}_{H_2}^2 \tan^2\beta}
                      {\tan^2\beta -1}
                -\frac{1}{2} \MZ^2(Q) \;, \\
B(Q) \ &=& \ -\frac{(\bar{m}_{1}^2+\bar{m}_{2}^2)\sin2\beta}
                   {2\mu(Q)} \;,
\label{min2}
\end{eqnarray}
%
where 
$Q$ is derived from the scalar fermion sector. It is usually chosen
such that radiative corrections to the effective potential are rather
small compared to other scales.
In \refeqs{min1},(\ref{min2}) $\tb \equiv v_2/v_1$ 
is the ratio of the two vacuum 
expectation values of the Higgs fields  $H_2$ and $H_1$ responsible 
for giving masses to the up-type and down-type quarks, respectively. In 
\refeqs{min1},(\ref{min2}), $\tb$ is evaluated at the scale $Q$,
from the scale $\MZ$, where it is considered as an input parameter%
\footnote{
See
for example the discussion in the Appendix of \citere{Drees} or in
\citere{mhiggsRG1a}. 
}%
. By 
$\bar{m}_{H_i}^2 = m_{H_i}^2+\Sigma_{v_i}$ in \refeqs{min1},(\ref{min2}) 
we denote the radiatively
corrected ``running '' Higgs soft-SUSY breaking masses and 
%
\begin{eqnarray}
\bar{m}_{i}^2 = {m}_{H_i}^2 + \mu^2 + \Sigma_{v_i} 
\equiv \bar{m}_{H_i}^2 + \mu^2 \;\; (i=1,2) \;,
\label{mbar}
\end{eqnarray}
%
where $\Sigma_{v_i}$ are the one-loop corrections
based on  the 1-loop Coleman-Weinberg effective potential $\Delta V$, 
$\Sigma_{v_i}=
\frac{1}{2v_i}\frac{\partial \Delta V}{\partial v_i}$,
%
\begin{eqnarray}
\Sigma_{v_i} = \frac{1}{64\pi^2} \sum_a (-)^{2J_a} (2J_a+1) 
               C_a \Omega_a
               \frac{M_a^2}{v_i} 
               \frac{\partial M_a^2}{\partial v_i} 
               \KKL \ln \frac{M_a^2}{Q^2} - 1 \KKR \;.
\label{effpot}
\end{eqnarray}
%
Here $J_a$ is the spin of the particle $a$, $C_a$ are the color degrees of
freedom, and $\Omega_a=1(2)$ for real scalar (complex scalar), 
$\Omega_a=1(2)$ for Majorana (Dirac) fermions. $Q$ is the energy scale 
and the $M_a$ are the field dependent mass matrices. Explicit formulas
of the $\Sigma_{v_i}$ are given in the Appendices of \citeres{barger,bagger}.
In our analyses contributions from all SUSY particles at the \onel\ level
are incorporated%
\footnote{
The corresponding \twol\ corrections are not available yet. Assuming
the size of these unknown higher-order corrections to be of the same
size as for the Higgs-boson mass matrix, see \refse{subsec:mssmhiggs},
the resulting values of $\mu$ and $B$ could change by
$\sim 5-10\%$. These parameters will serve as input for our numerical
analysis in \refse{sec:numanal}. The possible changes would hardly
affect our results obtained in the 
Higgs-boson sector and only mildly affect the analysis of SUSY
particle spectra in \refse{subsec:numanalspectrum}.
}%
.
With $\MZ^2$ here we denote
the tree level ``running'' $Z$~boson mass, 
$\MZ^2(Q)=\frac{1}{2}(g_1^2+g_2^2)v^2$ ($v^2 \equiv v_1^2 + v_2^2$),
extracted at the scale $Q$ from
its physical pole mass $\MZ = 91.187 \gev$. The REWSB
is fulfilled, see \refse{subsubsec:phenorest}, 
if and only if there is 
a solution to the \refeqs{min1},(\ref{min2})%
\footnote{
Sometimes in the
literature, the requirement of the REWSB is
described by the inequality $m_1^2(Q)m_2^2(Q)-|\mu(Q) B(Q)|^2 <0$. This
relation is automatically satisfied here from \refeqs{min1},(\ref{min2})
and from the fact that the physical squared Higgs masses must
be positive.
}%
.


\subsection{mSUGRA}
\label{subsec:msugra}

A dramatic simplification of the structure of the 
SUSY breaking interactions
is provided either by Grand Unification assumptions or by Superstrings.
For example, SU(5) unification implies at tree level equality relations
between the  scalar 
soft-SUSY breaking masses $m_{\tilde{Q}}=m_{\tilde{U}^c}=m_{\tilde{E}^c}$,
and $m_{\tilde{L}}=m_{\tilde{D}^c}$, equality between the soft
breaking gaugino masses $M_1=M_2=M_3$ and for two of the
trilinear soft breaking couplings, $A_d=A_e$.
On the other hand, SO(10) unification implies  equality 
of all scalar particle masses, equality of
Higgs masses and equality of the three types of trilinear couplings.
The simplest possible
choice at tree level is to take all scalar particle and Higgs 
masses equal to a common mass
parameter $M_0$, 
all gaugino masses are chosen to be equal to the parameter $M_{1/2}$ 
and all trilinear couplings flavor blind and equal to $A_0$.
This situation is common in the effective Supergravity
theories resulting from Superstrings 
but there exist more complicated alternatives.
Interestingly, the contribution of the family--anomalous U(1)
universal D-term 
to the scalar quark masses may be intra-family non-universal, 
and may differ from the usually assumed universal boundary 
conditions~\cite{faraggi}. Another alternative are for example
Superstrings with massless 
string modes of different modular weights that lead to
different scalar particle masses at tree level~\cite{Nilles}.
Thus, it seems that in almost all the ``realistic''  models 
motivated by Grand Unified Theories (GUTs)
 or Superstrings the universality assumption
is broken and each of these models has to be addressed separately
in order to study its phenomenology at low energies. On the other
hand, one should note 
that such non-minimal alternatives like flavor-dependent
scalar particle masses are constrained by limits on Flavor Changing
Neutral Currents (FCNC) processes. In what follows,
we shall therefore consider the simplest (and most commonly used) case 
of the three  parameters 
at the GUT scale, namely $M_0$, $M_{1/2}$ and $A_0$, which is
usually called the mSUGRA scenario.

In order to solve \refeqs{min1},(\ref{min2}), i.e.\ in order to impose
the constraint of REWSB, one needs as input $\tb(\MZ)$ and ${\rm sign}(\mu)$.
The running soft SUSY-breaking Higgs mass parameters, 
$m_{H_1}$ and $m_{H_2}$,
are defined at the EW scale after their evolution
from the GUT scale where we assume that they have a common value, $M_0$.
In addition the radiative corrections $\Sigma_{v_i}$
to the minimization conditions~\refeq{effpot} 
are defined from the low-energy SUSY spectrum
and the masses of the SM particles, which in turn means knowledge 
of $M_0$, $M_{1/2}$ and $A_0$ at the GUT scale. Thus, apart from
the SM masses provided by the experimental data~\cite{pdg}, 
4~parameters and a sign are required to define the mSUGRA scenario:
%
\begin{eqnarray}    
\{\; M_0\;,\;M_{1/2}\;,\;A_0\;,\;\tb \;,\; {\rm sign}(\mu)\; \} \;.
\label{mSUGRAparams}
\end{eqnarray}

In our numerical procedure we employ a two-loop 
renormalization group analysis for
all parameters involved, i.e.\ all couplings, dimensionful parameters
and VEV's. We start with the \msbar\
values for the gauge couplings at the scale $\MZ$, where for the strong
coupling constant $\als$ a trial input value in the vicinity 
of 0.120 is used. The \msbar\ values are converted into the corresponding
\drbar\ ones~\cite{drbar}.
The \msbar\ running $b$ and $\tau$ masses are run down to 
$\mb = 4.9 \gev$, $m_\tau = 1.777 \gev$ 
with the ${\rm SU(3)}_C \times {\rm U(1)}_{em}$
RGE's~\cite{arason} to derive
the running bottom and tau masses (extracted from their pole masses).
This procedure includes all SUSY corrections at the \onel\ level and
all QCD corrections at the \twol\ level as given in \citere{bagger}.
Afterwards by making use of the \twol\ RGE's
for the running masses $\overline{m}_b$, $\overline{m}_\tau$,
we run upwards to derive their \msbar\ values at $\MZ$, which
are subsequently converted to the corresponding \drbar\ values. 
This procedure provides the
bottom and tau Yukawa couplings at the scale $M_Z$. 
The top Yukawa coupling is derived from the 
top-quark pole mass, $\mt = 175 \gev$, which is
subsequently converted to the \drbar\ value, $\overline{\mt}(\mt)$,
where the top Yukawa coupling is defined.
The evolution of all
couplings from $\MZ$ running upwards to high energies 
now determines the unification scale
$M_{\rm GUT}$ and the value of the unification coupling $\alpha_{\rm GUT}$ by
\begin{eqnarray}
\alpha_1(M_{\rm GUT})|_{\drbarm}=\alpha_2(M_{\rm GUT})|_{\drbarm}
=\alpha_{\rm GUT} \;.
\label{unification}
\end{eqnarray}
At the GUT scale we set the boundary conditions for the 
soft SUSY breaking parameters, i.e.\ the values for $M_0$, $M_{1/2}$
and $A_0$ are chosen, and also $\al_3(M_{\rm GUT})$ is set equal to
$\alpha_{\rm GUT}$.
All parameters are run down again from $M_{\rm GUT}$ to $\MZ$.
For the calculation of the soft SUSY-breaking
masses at the EW scale we use the ``step function
approximation''~\cite{sakis}. Thus, if
the equation employed is the RGE for a particular running
mass $m(Q)$, then $Q_0$ is the corresponding physical mass 
determined by the condition $m(Q_0) =Q_0$. 
After running down to $\MZ$, the trial input value for $\als$ has
changed. 
At this point the value for $\tb$ is chosen and fixed.
As described in \refse{subsec:rewsb}, the parameters $|\mu|$ and $B$
are calculated from the minimization conditions (\ref{min1}) and
(\ref{min2}), respectively. Only the sign of the $\mu$-parameter is
not automatically fixed and thus chosen now.
This procedure is iterated several times until convergence is reached.


\subsection{mGMSB}
\label{subsec:gmsb}

A very promising alternative to mSUGRA is based on the hypothesis
that the soft SUSY breaking (SSB) occurs at relatively low energy scales 
and it is mediated mainly by gauge interactions through the so-called
``messenger sector'' (GMSB)~\cite{oldGMSB,newGMSB,GR-GMSB}. 
This scheme provides a natural, automatic suppression of the SUSY 
contributions to flavor-changing neutral currents and 
$\cp$-violating processes. 
Furthermore, in the simplest versions of GMSB (denoted hereafter with mGMSB), 
the MSSM spectrum and most of the observables depend on just 
4~parameters and a sign,
\begin{equation}
\{\; M_{\rm mess}, \; N_{\rm mess}, \; \Lambda, \; 
     \tb, \; {\rm sign}(\mu) \; \} \;,
\label{eq:pars}
\end{equation}
where $M_{\rm mess}$ is the overall messenger mass scale; 
$N_{\rm mess}$ is a number called the 
messenger index, parameterizing the structure of the messenger
sector; $\Lambda$ is the universal soft SUSY-breaking mass scale felt by the
low-energy sector; $\tb$ is the ratio of the vacuum expectation 
values of the two Higgs doublets; sign($\mu$) is the ambiguity
left for the SUSY higgsino mass after imposing a correct REWSB (see
\refse{subsec:rewsb} and e.g.\ 
\citeres{GMSBmodels2,GMSBmodels1,AKM-LEP2,AB-LC,AMPPR}).

The phenomenology of GMSB (and more in general of any theory with low-energy
SSB) is characterized by the presence of a very light gravitino 
$\tilde{G}$ with mass~\cite{Fayet} given by  
$m_{3/2} = m_{\tilde{G}} = \frac{F}{\sqrt{3}M'_P} \simeq 
\left(\frac{\sqrt{F}}{100 \tev}\right)^2 2.37 \; {\rm eV}$,  
where $\sqrt{F}$ is the fundamental scale of SSB and 
$M'_P = 2.44 \times 10^{18}$~GeV is the reduced Planck mass.
Since $\sqrt{F}$ is typically of order 100 TeV, the $\tilde{G}$ is always the 
LSP in these theories. Hence, if $R$-parity is 
conserved, any MSSM particle will decay into the gravitino. 
Depending on $\sqrt{F}$, the interactions 
of the gravitino, although much weaker than gauge and Yukawa interactions,
can still be strong enough to be of relevance for collider physics. 
In most cases, the last step of any SUSY decay chain is 
the decay of the next-to-lightest SUSY particle (NLSP), which can  
occur either outside or inside a typical detector, possibly 
close to the interaction point. 
The nature of the NLSP -- or, more precisely, of the SUSY particle(s) 
having a large branching ratio for decaying into the gravitino and the 
relevant SM partner -- determines four main scenarios giving rise to 
qualitatively different phenomenology \cite{AKM-LEP2}. 

The low-energy parameter sets for this scenario have been calculated
by using the program 
{\em SUSYFIRE}~\cite{SUSYFIRE} and adopting the phenomenological 
approach of \citeres{AKM-LEP2,AB-LC,AMPPR}, see also \citere{higgsgmsb}.
The origin of $\mu$ is not specified, nor the assumption $B\mu = 0$ is
made at the messenger scale. Instead, correct REWSB is imposed to trade $\mu$
and $B\mu$ for $\MZ$ and $\tb$, leaving the sign of $\mu$
undetermined, see \refse{subsec:rewsb}. 
However, note that to build a fully coherent GMSB model, one should also 
find a more fundamental solution to the latter problem, perhaps providing a 
dynamical mechanism to generate $\mu$ and $B\mu$, possibly with values
of the same order of magnitude. This might be accomplished radiatively 
through some new interaction. In this case, the other
soft terms in the Higgs potential, namely $m^2_{H_{1,2}}$, will be also  
affected and this will in turn change the values of $|\mu|$ and $B\mu$ 
coming from REWSB conditions. 
We have checked this circumstance in detail in the case where one
has an extra term in the Higgs potential of the type $\Delta_+$
(see the parameterization of \citere{GMSBmodels1}) and also performed some
checks in the general case. In all cases we did not find any big changes
in $\mh$.

To determine the MSSM spectrum and low-energy parameters, the RGE evolution 
is solved with boundary conditions at the $M_{\rm mess}$ scale, where
\begin{eqnarray}
M_a(M_{\rm mess}) & = & N_{\rm mess}\; \Lambda\; 
g\left(\frac{\Lambda}{M_{\rm mess}}\right) \alpha_a, 
\; \; \; (a=1, 2, 3) \nonumber \\
\tilde{m}^2(M_{\rm mess}) & =  & 2 N_{\rm mess}\; \Lambda^2\; 
f\left(\frac{\Lambda}{M_{\rm mess}}\right) 
\sum_a \left(\frac{\alpha_a}{4\pi}\right)^2 C_a,
\label{eq:bound}
\end{eqnarray} 
for the gaugino and the scalar masses, respectively. 
The exact expressions for $g$ and $f$  at the one- and two-loop level
can be found, e.g., in Ref.~\cite{AKM-LEP2},
and $C_a$ are the quadratic Casimir invariants for the scalar fields.
As usual, the scalar trilinear couplings $A_f$ are assumed to vanish
at the messenger scale, as suggested by the fact that they (and not
their square) are generated via gauge interactions with the messenger 
fields at the two loop-level only. 

The interesting region of the GMSB parameter space is selected as follows.
Barring the case where a neutralino is the NLSP and decays outside
the detector (large $\sqrt{F}$), the GMSB signatures are very spectacular
and the SM background is generally negligible or easily subtractable. 
Therefore, also in accordance with negative results in the LEP2
searches~\cite{lepsusySep2000},  
only models where the NLSP mass is larger than 100~GeV are considered.
Other requirements are: $M_{\rm mess} > 1.01 \Lambda$, to prevent 
an excess of fine-tuning of the messenger masses; the mass of the 
lightest messenger scalar be at least 10 TeV; 
$M_{\rm mess} > M_{\rm GUT} \times \exp(-125/N_{\rm mess})$,
to ensure the perturbativity of gauge interactions up to the 
GUT scale; $M_{\rm mess} \simleq 10^{5} \Lambda$, for simplicity. 
As a result, 
the messenger index 
$N_{\rm mess}$, which is assumed to be an integer independent of the gauge 
group, cannot be larger than~8. To prevent the top Yukawa coupling from 
blowing up below the GUT scale, $\tb > 1.5$ is required.
Models with $\tb \gsim 55$ (with a mild dependence on 
$\Lambda$) are forbidden by the REWSB requirement, see
\refse{subsubsec:phenorest}, and typically fail 
to give $\MA^2 > 0$.

The models are generated using {\em SUSYFIRE} with the following
prescriptions for the high-energy input parameters. Logarithmic
steps have been used for $\Lambda$ (between about 45 TeV/$N_{\rm mess}$ 
and about 220 TeV/$\sqrt{N_{\rm mess}}$), $M_{\rm mess}/\Lambda$ 
(between about 
1.01 and $10^5$) and $\tb$ (between 1.5 and about 60), subject
to the constraints described above. 
{\em SUSYFIRE} starts from the values of particle masses and gauge
couplings at the weak scale and then evolves them up to the messenger
scale through RGEs. At the messenger scale, it imposes the boundary 
conditions (\ref{eq:bound}) for the soft particle masses and then 
evolves the RGEs back to the electroweak scale. The decoupling of each 
SUSY particle at the proper threshold is taken into account. 
Two-loop RGEs are used for gauge couplings, third generation Yukawa
couplings and gaugino soft masses. The other RGEs are taken at the
one-loop level%
\footnote{
Contrary to the mSUGRA scenario in \refse{subsec:msugra} 
the scalar masses are treated only at the \onel\ level in mGMSB
and the mAMSB scenario in \refse{subsec:amsb}.
However, the main effects arise from the
Yukawa couplings, which are consistently treated at the \twol\
level. The \twol\ effects on the scalar masses have been shown to be
at the 5\% level~\cite{sakis}.
}%
. At the scale 
$Q$, derived from the scalar fermion sector,
REWSB conditions are imposed by means of the one-loop effective potential 
approach. For the $\Si_{v_i}$ in \refeq{effpot} all dominant
corrections from the stop, sbottom and stau sector are included.
The program then evolves up again to $M_{\rm mess}$ and so on.
Three or four iterations are usually enough to get a good approximation
for the MSSM spectrum.


\subsection{mAMSB}
\label{subsec:amsb}

The most recently proposed Anomaly Mediated SUSY Breaking (AMSB) 
scenario \cite{lr,giudice} provides an alternative way to give mass to 
all the SUSY particles. In this model, SUSY breaking happens on a
separate brane and is  
communicated to the visible world via the super-Weyl anomaly. 
The overall scale of SUSY particle masses is set by $m_{\rm aux}$, 
which is the VEV of the auxiliary field in the supergravity multiplet.
In the AMSB scenario, the low-energy soft supersymmetry 
breaking parameters $M_i$ (gaugino masses, $i$=1$-$3),  
$m_{\rm scalar}^2$  and $A_{y}$ at the GUT scale are given by \cite{lr,wells}
\begin{eqnarray}
M_i        &=& \frac{\beta_{g_i}}{g_i}m_{\rm aux},
 \label{m1} \\
m_{\rm scalar}^2 &=& -\frac{1}{4}
 \left(\frac{\partial\gamma}{\partial{g}}\beta_{g} + 
       \frac{\partial\gamma}{\partial{y}}\beta_{y}\right)m_{\rm aux}^2
 + m_{0}^2,
 \label{m2} \\
A_{y}            &=& -\frac{\beta_{y}}{y}m_{\rm aux}.
 \label{m3}
\end{eqnarray}

Notice that the slepton squared-masses would be negative if $m_{0}$ 
were absent. 
There have been several proposals to solve this tachyonic slepton 
problem: bulk contributions \cite{lr}, non-decoupling effects 
of ultra-heavy vectorlike matter fields \cite{negative}, 
coupling of extra Higgs doublets to the leptons \cite{clm}, contributions
from the $R$-parity violating couplings 
in \refeq{m2} (with $m_0=0$) \cite{ben}, and a heavy mass 
threshold contribution at higher orders \cite{kss}. 
Here we have adopted a phenomenological approach and have introduced
an additional  
mass scale $m_{0}$ at the GUT scale in order to keep the 
slepton masses positive \cite{wells}.  
For simplification, we choose $m_{0}$ to be the 
same for all the super scalar particles.  Therefore, 
in the minimal case (mAMSB), the particle spectrum can be determined by 
3~parameters and a sign:  
\begin{equation}
\{m_{\rm aux},\ m_{0},\ \tb,\ {\rm sign}(\mu) \} .
\end{equation}

Eqs.~(\ref{m1}), (\ref{m2}) and (\ref{m3}) would hold at all 
scales if  $m_{0}$ were absent.  However, once $m_{0}$ is introduced
at the GUT scale,  the above definitions of $M_{i}$, $m_{\rm
scalar}^2$  and $A_{y}$ set the boundary conditions and
the entire SUSY spectrum can be obtained via the running of
supersymmetric RGEs down to a lower
scale. 

Once the squark threshold is crossed, the squarks decouple and one is
left with an effective field  theory with two Higgs doublets and all the
standard model particles%
\footnote{
Gluinos  are also decoupled since their
masses are close to the squark masses.  The contributions of Bino and Winos to
the Higgs sector can be neglected   since the 
${\rm U(1)}_{Y}$ and SU(2) gauge couplings are small.%
}%
.  
The two unknown parameters $|\mu|$
and $B$ are determined by the minimization of the Higgs effective
potential as explained in \refse{subsec:rewsb}. 
Therefore, the low-energy
spectrum  is fixed once the values of $m_{\rm aux}$, $m_{0}$, 
$\tb$ and the sign of $\mu$ are known.


\subsection{Evaluation of predictions in the Higgs boson sector 
of the MSSM}
\label{subsec:mssmhiggs}

The most relevant parameters for Higgs boson phenomenology in the MSSM
are the mass of the $\cp$-odd Higgs boson, $\MA$, the ratio of the two
vacuum expectation values, $\tb$, the scalar top masses and mixing
angle, $\mste, \mstz, \tst$, for large $\tb$ also the scalar
bottom masses and mixing angle, $\msbe, \msbz, \tsb$, the Higgs
mixing parameter, $\mu$, the gluino mass, $\mgl$, and the 
U(1) and SU(2) gaugino masses,
$M_1$ and $M_2$. The way in which 
these low-energy parameters are derived in each of the soft
SUSY-breaking scenarios has been described in \refses{subsec:msugra} --
\ref{subsec:amsb}. Since the RG running employed in the three scenarios
is based on the \drbar\ scheme, the corresponding low-energy parameters
are \drbar\ parameters. In order to derive predictions for observables,
i.e.\ particle masses and mixing angles, these parameters in general have 
to be converted into on-shell parameters.

For the predictions in the MSSM Higgs sector we use results obtained 
in the Feynman-diagrammatic (FD) approach (see below) within the on-shell
renormalization scheme. Since they incorporate two-loop contributions in
the $t$--$\Stop$ sector, the parameters in the scalar top sector (which
enter at one-loop order in the predictions for the Higgs boson masses)
have to be appropriately converted from \drbar\ to on-shell
parameters~\cite{bse,FDRG1,maulpaul3}. We perform this conversion using
the full $\oas$ contributions.

In the FD approach the masses of the two $\cp$-even
Higgs bosons, $\mh$ and $\mH$, are derived beyond tree level 
by determining the poles of the $h-H$-propagator
matrix whose inverse is given by
\BE
\left(\Delta_{\rm Higgs}\right)^{-1}
= - i \ML q^2 -  m_{H,{\rm tree}}^2 + \hSi_{H}(q^2) &  \hSi_{hH}(q^2) \\
     \hSi_{hH}(q^2) & q^2 -  m_{h,{\rm tree}}^2 + \hSi_{h}(q^2) \MR,
\label{higgspropagatormatrixnondiag}
\EE
where the $\hSi$ denote the renormalized Higgs boson
self-energies. Determining the
poles of the matrix $\Delta_{\rm Higgs}$ in
\refeq{higgspropagatormatrixnondiag} is equivalent to solving
the equation
\BE
\left[q^2 - m_{h,{\rm tree}}^2 + \hat\Sigma_{h}(q^2) \right]
\left[q^2 - m_{H,{\rm tree}}^2 + \hat\Sigma_{H}(q^2) \right] - 
\left[\hat\Sigma_{hH}(q^2)\right]^2 = 0 . 
\EE

We use the result for the Higgs boson self-energies consisting
of the complete \onel\ result for the
Higgs boson self-energies in the on-shell scheme~\cite{mhiggsf1l} 
combined with the dominant \twol\ contributions of
$\oaas$~\cite{mhiggslong,mhiggsletter}
and further subdominant corrections~\cite{mhiggsRG1b,mhiggsRG2}, see
\citere{mhiggslong} for details.
The matrix \refeq{higgspropagatormatrixnondiag} therefore
contains the renormalized Higgs boson self-energies
\BE
\hSi_s(q^2) = \hSie_s(q^2) + \hSiz_s(0), \quad s = h, H, hH,
\label{besttwoloopse}
\EE
where the momentum dependence is neglected only in the \twol\
contribution.

An effective mixing angle, 
\BE
 \aeff = {\rm arctan}\KKL 
  \frac{-(\MA^2 + \MZ^2) \Sb \Cb - \hSi_{\PePz}}
       {\MZ^2 \CQb + \MA^2 \SQb - \hSi_{\Pe} - \mh^2} \KKR~,~~
  -\frac{\pi}{2} < \aeff < \frac{\pi}{2}~,
\label{aeff}
\EE
can furthermore be obtained from diagonalizing the mixing matrix in the
basis of the unrotated neutral $\cp$-even fields $\Pe, \Pz$, 
neglecting the momentum dependence everywhere
(alternatively, the mass matrix in the $h, H$ basis is diagonalized by
the angle $\De\al$, where $\aeff = \al_{\rm tree} + \De\al$, see
e.g.~\citere{hff}).
Inserting $\aeff$ in the tree-level formulas for Higgs production and
decay, the dominant universal corrections in the Higgs sector are 
taken into account~\cite{hff,eehZhA}.

The results for the Higgs boson masses and $\aeff$ as well as the
conversion from \drbar\ to on-shell parameters using the full
$\oas$ contributions for both the parameters in the $t$--$\Stop$ and
$b$--$\Sbot$ sector
are implemented in the 
Fortran code \fh~\cite{feynhiggs}. 

For the investigation of the mSUGRA scenario \fh\ has
been interfaced to the program {\em SUITY}~\cite{sakis2},
used for the evaluation of the low-energy spectrum of the mSUGRA scenario.
A combined program, {\em FeynSSG}~\cite{FeynSSG}, has been created on
the basis of \fh\ and {\em SUITY}, in which the two subprograms run
automatically.

\smallskip
As a further check of our results for the Higgs boson sector, we have
(in addition to the \fh\ calculation) evaluated all results with a code
based on an independent \onel\ calculation~\cite{mhiggs1lrosiek}, but
using the \twol\ routines of \fh. The difference of the \onel\
calculation based on \citere{mhiggsf1l}, used in \fh, and the ones
given in \citere{mhiggs1lrosiek} are only due to different
renormalization prescriptions and thus of higher
order~\cite{mhiggs1lren}. The results we found in both approaches are
as expected very similar and lead to the same conclusions.


\subsection{Other constraints}
\label{subsec:constraints}

While our main focus in this paper is on the physics in the Higgs
sector, we also take into account some further (relatively mild)
constraints when determining the allowed parameter values. These
constraints are discussed in the following.

\subsubsection{Precision observables}
\label{subsubsec:precobs}

The electroweak precision observables are affected by the whole spectrum
of SUSY particles. The main SUSY contributions to the $W$~boson mass,
$\MW$, the effective leptonic weak mixing angle, $\sweff$, and other
$Z$~boson observables usually arise from $\Stop/\Sbot$ contributions.
They enter via the leading contribution to the
$\rho$-parameter~\cite{rhoparameter}. In our analysis we take into account 
the corrections arising from $\Stop/\Sbot$ loops up to two-loop
order~\cite{delrhosusy2loop}.
A value of $\De\rho$ outside the experimentally preferred region of 
$\dr^{\SU} \simleq 3\times 10^{-3}$~\cite{pdg} indicates experimentally
disfavored $\Stop$  and $\Sbot$ masses%
\footnote{
Since the $\dr^{\SU}$ evaluation involves scalar bottoms at the \twol\
level, also the parameters in the $\Sbot$~sector have to be
transformed from \drbar\ to on-shell.
}%
.
The evaluation of $\dr^{\SU}$ is implemented in \fh.

We have verified that in our analysis below the $\De\rho$ constraint
does not play a significant role, i.e.\ nearly all generated model points
give rise to an acceptable contribution to the electroweak precision
observables. As a conservative approach, we do not apply any further 
constraints from $g_{\mu} -2$ or $b \to s \ga$.


\subsubsection{Experimental bounds on SUSY particle masses}
\label{subsubsec:susymassbounds}

The search for SUSY particles has been one of the main tasks pursued
at Run I of the Tevatron and at LEP. The searches all turned out to be
negative, thus lower limits on the SUSY particle masses have been
set. In order to restrict the allowed parameter space in the three
soft SUSY-breaking scenarios we employed the following constraints on
their low-energy mass 
spectrum~\cite{pdg,lepsusySep2000,lepFestOct2000,CDF,Iashvili,deVivie}:
\BEA
m_{\tilde e} &>& 95 \gev {\rm ~(mSUGRA,~AMSB)} \non \\ 
m_{\tilde \mu} &>& 85 \gev {\rm ~(mSUGRA,~AMSB)} \non \\ 
m_{\tilde \tau} &>& 71 \gev {\rm ~(mSUGRA,~AMSB)} \non \\
m_{\tilde \nu} &>& 43 \gev {\rm ~(mSUGRA,~AMSB)} \non \\
\mst &>& 95 \gev {\rm ~(mSUGRA,~AMSB)} \non \\
\msb &>& 85 \gev {\rm ~(mSUGRA,~AMSB)} \non \\
\mgl &>& 190 \gev {\rm ~(mSUGRA,~AMSB)} \non \\
m_{\tilde{\chi}^\pm} &>& 103 \gev ~({\rm mSUGRA}, 
                                      m_{\tilde\nu} > 300 \gev) \non \\
m_{\tilde{\chi}^\pm} &\gsim& 84.6 \gev ~({\rm mSUGRA}, 
                                      m_{\tilde\nu} < 300 \gev) \non \\
m_{\tilde{\chi}^\pm} &\gsim& 45 \gev {\rm ~(AMSB)} \non \\
m_{\tilde{\chi}^0_1} &\gsim& 36 \gev {\rm ~(mSUGRA)} \non \\
m_{\tilde{\chi}^0_1} &\gsim& 45 \gev {\rm ~(AMSB)} \non \\
m_{\rm NLSP} &>& 100 \gev {\rm ~(GMSB)} .
\label{susyexp}
\EEA
Note that the NLSP condition in the GMSB scenario (applying either to
the lightest neutralino or to the lighter stau) automatically
imposes stronger bounds on the other particle masses than the purely
experimental bounds.  


\subsubsection{Other phenomenological restrictions}
\label{subsubsec:phenorest}

Besides constraints from precision observables and from unsuccessful
direct search for SUSY particles, we also take into account the
following restrictions (if not indicated otherwise):

\begin{itemize}
\item
For the top-quark mass, throughout this paper we use the value
$\mt = 175 \gev$. A variation of $\mt$ directly affects the result for
$\mh$, while its influence on the other quantities studied here is more
moderate. 
As a rule of thumb, a change in $\mt$ by $\pm1 \gev$ also results in a
change in $\mh$ of about $\pm 1 \gev$~\cite{tbexcl}.

\item 
The GUT or high-energy scale parameters are taken to be real, no SUSY
$\cp$-violating phases are assumed. 

\item
In all models under consideration the $R$-parity symmetry~\cite{herbi,rpv2}
is taken to be conserved.

\item
Parameter sets that do not fulfill the condition of 
radiative electroweak symmetry breaking (REWSB), 
i.e.\ the \onel\ minimization conditions of
\refeqs{min1},(\ref{min2}), are discarded (already at the level of
model generation.)

Within all soft SUSY-breaking scenarios considered here,
the condition of REWSB leads to restrictions
on $\tb$. For example, it is almost impossible
(or a huge fine tuning is required)
to find very large values of $\tb$, $\tb \gsim 60$ which 
pass this constraint. This is because in that 
region both the RGEs of the Higgs soft breaking masses receive large
corrections not only from the top Yukawa coupling but also
from the bottom and the tau Yukawa couplings. This drives the 
numerator (and thus $\mu^2$) of \refeq{min1} negative, thus excluding this
parameter set from our analysis. 
Another possibility of not fulfilling the REWSB condition is a very
heavy soft SUSY-breaking spectrum, so that $m_{H_2}^2$ does not reach
negative values and thus does not trigger REWSB.

\item
Parameter sets that do not fulfill the ``strong CCB'' constraints are
discarded (already at the level of model generation), i.e.\ models for
which the physical vacuum would be charge or color breaking.
In our analysis this corresponds to cases where the squared scalar
quark or charged lepton masses 
are becoming negative at the scale $Q$, where $Q$ is the energy scale
at which the low-energy parameters are decoupled.
However, we do not test the models for local or global charge or color
breaking minima in general. In most cases the tunneling time from our
charge and color conserving vacuum to the charge or color
breaking minimum is much longer than the present age of the universe,
and thus they are in practice not 
dangerous~\cite{riotto,kusenko,abel,subir}.

\item
The original motivation for the introduction of SUSY into particle
physics was the solution of the 
``hierarchy problem''. This sets a natural upper
bound on the SUSY particle masses, which of course depends on how much
fine tuning one is willing to accept. In our analysis we have imposed a
(rather mild) ``naturalness bound''. This upper bound on the SUSY
particle masses has been chosen to be equal for all three soft
SUSY-breaking scenarios. Thus the low-energy mass spectra are
directly comparable.
We have imposed
\BE
\msq \lsim 1.5 \tev , \quad 
\mgl \lsim 2 \tev .
\label{eq:natbounds}
\EE
These bounds give rise to upper bounds also on the other scalar masses
and on the electroweak gaugino masses
(depending on the specific scenario). 

The bounds imposed in \refeq{eq:natbounds} can of course not be 
considered as strict upper bounds (although constraints from cosmology,
$b \to s \ga$ and $g_{\mu} -2$ in general also favor a relatively light
particle spectrum~\cite{bsgammaexp,gminus2,higgs115msugra2,higgs115msugra3}), 
but carry a certain degree of arbitrariness. In particular, we do not
consider here the scenario of focus point supersymmetry~\cite{focusp},
in which squarks and sleptons in the multi-TeV range can occur.
It should be noted, however, that in all three soft SUSY-breaking
scenarios an upper bound of \order{5 \tev} is obtained by the
requirement of REWSB. On the other hand, saturating this upper bound
of \order{5 \tev} requires severe fine tuning to satisfy the
minimization conditions given in \refeqs{min1},~(\ref{min2}).

The imposed upper bounds on SUSY masses also naturally result in a
limit for the soft SUSY-breaking parameters at the high energy scale,
$M_0, M_{1/2}$ in mSUGRA, $\La$ in mGMSB and $m_0$, $m_{\rm aux}$ in mAMSB.
If the bounds in \refeq{eq:natbounds} were relaxed, heavier particle
spectra would be allowed. The effect on the Higgs boson sector is only
logarithmic and thus rather small. Concerning the collider
phenomenology as presented in \refse{subsec:numanalspectrum}, the
detection of SUSY particles would become more difficult.

\item
We demand that the lightest SUSY particle is uncolored and uncharged.
In the GMSB scenario the LSP is always the gravitino, so this 
condition is automatically fulfilled.
Within the mSUGRA and mAMSB scenario, the LSP
is required to be the neutralino. 
Parameter sets that result in a different LSP are excluded.

\item
We do not impose further restrictions arising from
${\rm BR}(b \to s \ga)$~\cite{bsgammaexp} and $g_\mu - 2$~\cite{gminus2},
which could lead to additional constraints on the three soft SUSY-breaking
scenarios. Restrictions of this kind depend on the experimental errors
of these observables and the uncertainties in their theoretical
prediction and could change considerably if the experimental central
values change in the future. Moreover, slight modifications of the
SUSY-breaking scenarios which would have only a minor impact on the
phenomenology of the models discussed here could have a strong influence
on constraints from ${\rm BR}(b \to s \ga)$ and $g_\mu - 2$. This could
happen, for instance, in the case of ${\rm BR}(b \to s \ga)$ via the
presence of small flavor mixing terms in the SUSY Lagrangian.

As a conservative approach, we therefore do not discard parameter sets
which do not fulfill the constraints from 
${\rm BR}(b \to s \ga)$~\cite{bsgammaexp} and $g_\mu - 2$. 
It should be noted, however,
that if these constraints are imposed and the AMSB scenario is taken at
face value, i.e.\ without any additional contributions, the parameter
space allowed by the experimental values of ${\rm BR}(b \to s \ga)$ and
$g_\mu - 2$ is rather restricted.
As for the mSUGRA scenario, the effect of ${\rm BR}(b \to s \ga)$
would disallow a region with small $M_{1/2}$ for large $\tb$, whereas
the effect of $g_\mu - 2$ would be to set an upper bound on the
combination of $M_0$ and $M_{1/2}$, see e.g.~\citere{higgs115msugra3}.

\item
In the same spirit, we also do not apply any further cosmological
constraints, i.e.\ we do not demand a relic density in the region favored
by dark matter constraints, see \citere{cdm} and references therein.
As in the case of ${\rm BR}(b \to s \ga)$ and $g_\mu - 2$, slight
modifications of the scenario which do not concern collider phenomenology
could have a strong impact on the bounds derived from cosmology. In the
case where the LSP relic abundance in the scenarios discussed here
is too small to explain the observed amount of cold dark matter (CDM),
a further mechanism could provide the additionally required amount of
CDM (this would certainly apply to the case of mGMSB). If on the other
hand the amount of CDM appears to be too large in a given scenario,
``thermal inflation''~\cite{thermalinflation} could offer a mechanism
for bringing the CDM density into agreement with the cosmological
bounds. Furthermore, the neutralino could turn out to be the NLSP and
decay (outside of collider detectors) into a very weakly interacting
LSP (e.g.~the axino~\cite{axinos}), or there could be a small amount of
$R$-parity violation present in the model.

\end{itemize}


\section{Numerical analyses}
\label{sec:numanal}

\subsection{Experimental bounds from the MSSM Higgs sector}
\label{subsec:higgsscenarios}

The results from the Higgs search at LEP have excluded a considerable 
part of the MSSM parameter space~\cite{mssmhiggs}. 
On the other hand, an excess at about 
the $3\sigma$ level has been observed which is compatible with the
production of a SM Higgs boson with a mass of about 115~GeV~\cite{LEPHiggs}.
For our numerical analysis we will focus on
three different cases implying different restrictions on the MSSM
parameter space. In case (I) we investigate the full parameter space
which is allowed in the three scenarios when taking into account the
exclusion bounds from the Higgs search and the further constraints
discussed in the previous section. In case (II) and case (III), on the
other hand, we specifically focus on the interpretation of the excess
observed in the Higgs search at LEP as production of the lightest
$\cp$-even Higgs boson of the MSSM (case (II)) and of the heavier 
$\cp$-even Higgs boson of the MSSM (case (III)).

\begin{itemize}
\item[(I)]
The results of the search for the MSSM Higgs bosons are usually
interpreted in three different benchmark scenarios~\cite{benchmark}.
The 95\% C.L.\ exclusion limit for the SM Higgs boson of $\mH >
113.5$~GeV applies also for the lightest $\cp$-even Higgs boson of the
MSSM in the parameter region of large $\MA$ and/or small $\tb$. In the
unconstrained MSSM this bound is reduced to 
$\mh > 91.0$~GeV~\cite{mssmhiggs} for 
$\MA \lsim 150$~GeV and $\tb \gsim 8$ as a consequence of a reduced
coupling of the Higgs to the Z~boson. For the $\cp$-odd Higgs boson a
lower bound of $\MA > 91.9$~GeV has been obtained~\cite{mssmhiggs}.
In order to correctly interpolate between the parameter regions where
the SM lower bound of $\mH > 113.5$~GeV and the bound $\mh > 91.0$~GeV
apply, we use the result for the Higgs-mass exclusion given with respect
to the reduced $ZZh$ coupling squared (i.e.\ $\sin^2(\be-\aeff)$, \
see \refeq{aeff})~\cite{Barate:2000zr}.
We have compared the excluded region with the theoretical prediction
obtained at the \twol\ level for $\mh$ and $\sqbaeff$ for each parameter
set (using $\mt = 175$~GeV).

Another important constraint on $\mh$ comes from the searches in 
$p\bar p$ collisions at Run~I of the Tevatron~\cite{cdfhiggs}.
No evidence of a signal of the type 
$p \bar p \to b\bar b h \to b\bar b b\bar b$ has been found. This 
leads to an improvement of the LEP limits in the region of large 
$\tb$, $\tb \gsim 50$. 
Since the bounds obtained at the Tevatron are given only in the
no-mixing and $\mhmax$ scenario~\cite{benchmark,cdfhiggs}, 
they do not necessarily
apply to all cases of our present analysis. As a conservative treatment,
we therefore do not use the Tevatron bound for excluding models in the
high $\tb$ region. It should be noted, however, that owing to the REWSB
constraint in our analysis below we do not find allowed models for $\tb
\gsim 60$.

\item[(II)]
In this scenario the LEP excess is interpreted as production of the 
lightest $\cp$-even Higgs boson of the MSSM, and we thus focus on the 
parameter regions in the three soft SUSY-breaking scenarios where
\BE
\mh = 115 \pm 2 \gev .
\label{mhiggs2}
\EE
The assumed error of $\pm 2 \gev$ is somewhat larger than the region
favored by the LEP data, in order to allow for some theoretical
uncertainties from unknown higher order corrections in the Higgs boson
mass calculation (note that the dominating theoretical uncertainty is
related to the experimental error of the top-quark mass; we focus in our
analysis on the value $\mt = 175$~GeV, the corresponding $\mh$ values
for different values of $\mt$ can be obtained from the approximate
relation $\de\mh/\de\mt = {\cal O}(1)$).
In order to allow an interpretation of the LEP excess in terms of the
lightest MSSM Higgs boson, it is furthermore necessary that the
production and decay rates of $h$ are similar to those of the SM Higgs
boson. We therefore in addition demand $\sin^2(\be - \aeff) \gsim 0.8$,
which results in a production cross section for the Higgs strahlung process,
$e^+e^- \to Zh \, \sim \, \sin^2(\be - \aeff)$ close to the SM cross
section. Furthermore, we require that the $hb\bar b$ coupling in the
MSSM is not strongly suppressed compared to the SM case.
The $hb\bar b$ coupling differs from the corresponding SM coupling
in two ways. Firstly, it has an additional factor 
$\Saeff/\Cb$ (appearing squared in the branching ratio). We demand
that $\sin^2\aeff/\cos^2\be \gsim 0.8$.
Secondly, the $hb\bar b$ vertex can be affected by gluino loop
corrections (and less importantly also gaugino loop
corrections)~\cite{deltamb1,deltamb2}. Usually they are parameterized via 
$\De m_b$, 
\BE
\De m_b \simeq {2\als \over 3\pi} \mgl
\mu\tan\beta~I(\msbe, \msbz, \mgl)
 + {Y_t \over 4\pi} A_t\mu\tan\beta~I(\mste, \mstz, \mu) ,
\label{eq:Delmb}
\EE
where $Y_t= h_t^2/(4\pi)$ and
\BE
I(a,b,c) = {a^2b^2\ln(a^2/b^2)+b^2c^2\ln(b^2/c^2)+c^2a^2\ln(c^2/a^2)
\over
(a^2-b^2)(b^2-c^2)(a^2-c^2)} .
\EE
The main correction to the $hb\bar b$ coupling is
proportional to $1/(1 + \De m_b)$. In our analysis of case (II) we 
additionally demand that $|\De m_b| < 0.5$.

\item[(III)]
In this scenario we investigate whether the LEP excess can be
interpreted as the production of the heavy $\cp$-even Higgs boson in the
MSSM. In order to allow this interpretation, $H$ has to have SM-like
couplings to the $Z$, i.e.\ $\cos^2(\be-\aeff) \gsim 0.8$.
In this case the $h$ production via Higgs
strahlung, $e^+e^- \to Zh$, is highly suppressed, whereas the
associated production $e^+e^- \to Ah$ could be beyond the kinematic
reach of LEP. We apply a bound of $\mh + \MA > 206 \gev$ in this
scenario. It should be noted that this bound is very conservative, since
values of $\mh + \MA$ as low as about 190~GeV are not excluded from the
Higgs search at LEP~\cite{mssmhiggs}.
As in case~(II) we also require that the decay of the heavy $\cp$-even
Higgs boson is SM like, i.e.\ the dominating decay channel is
$H \to b\bar b$. Therefore we demand $\cos^2\aeff/\CQb > 0.8$.

\end{itemize}


\subsection{mSUGRA}
\label{subsec:numanalmsugra}

For the numerical analysis we have scanned over about $50000$ models,
where the parameters have been randomly chosen in the intervals
\begin{eqnarray}
50~{\rm GeV} \le &M_0& \le 1~{\rm TeV} \;, \nonumber \\
50~{\rm GeV} \le  &M_{1/2}& \le 1~{\rm TeV} \;, \nonumber \\
-3~{\rm TeV} \le &A_0& \le 3~{\rm TeV} \;, \non \\
1.5 \le &\tan\beta & \le 60 \;, \non \\
 &{\rm sign}\, \mu& = \pm 1 .
\label{msugraparam}
\end{eqnarray}
Although we have scanned over about
50000 models, we show in the figures of this paper a subset of around 5000
(randomly) selected points to keep the density of the points at a
reasonable level. This has been done for all three soft SUSY-breaking
scenarios. No reduction of the data points is applied for the cases
(I)--(III) in parameter regions
with a small density of points, i.e.\ in particular for $\mh < 113$~GeV
and $\sin^2(\be - \aeff) < 0.99$.


\begin{figure}[ht!]
\begin{center}
\epsfig{figure=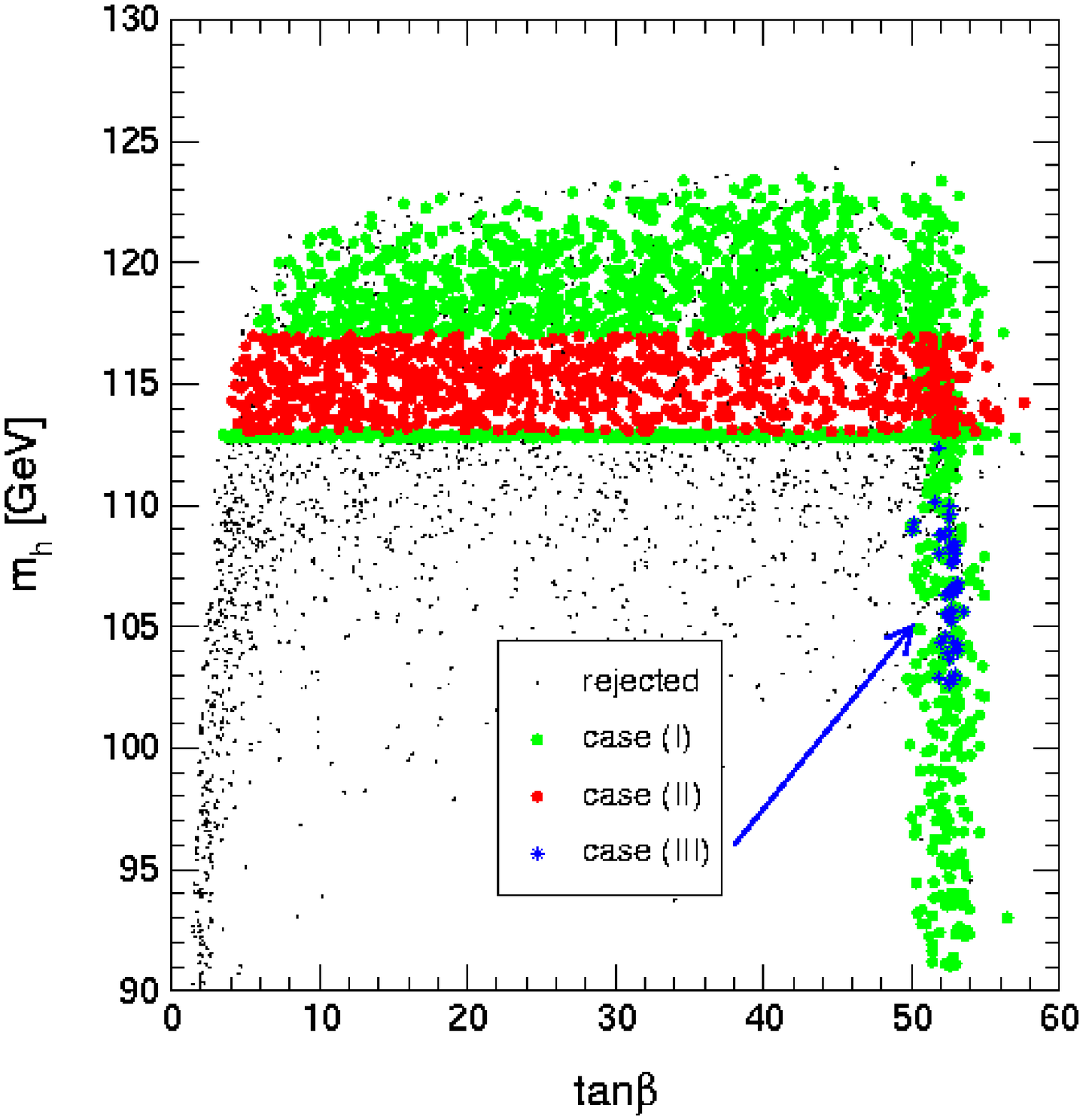,width=12cm,height=8cm}
\caption{
The light $\cp$-even Higgs boson mass $\mh$ as a function of 
$\tb$ in the mSUGRA scenario. The three cases as discussed in 
\refse{subsec:higgsscenarios} are displayed together with the rejected
models. 
Case~(I) corresponds to the models that have passed all theoretical
and experimental constraints. Case~(II) is the subset of case~(I) with
$\mh$ values in the region favored by recent LEP Higgs searches, 
$113 \gev \le \mh \le 117 \gev$, and SM like couplings of the $h$. 
In case~(III), whose parameter points
are indicated by an arrow for better readability, the heavier $\cp$-even
Higgs boson lies in the region $113 \gev \le \mH  \le 117 \gev$, while 
the lighter one has a suppressed coupling to the $Z$~boson and is too
heavy to be produced in associated production. 
}
\label{fig11mSUGRA}
\end{center}
\end{figure}
%
\begin{figure}[ht!]
\begin{center}
\epsfig{figure=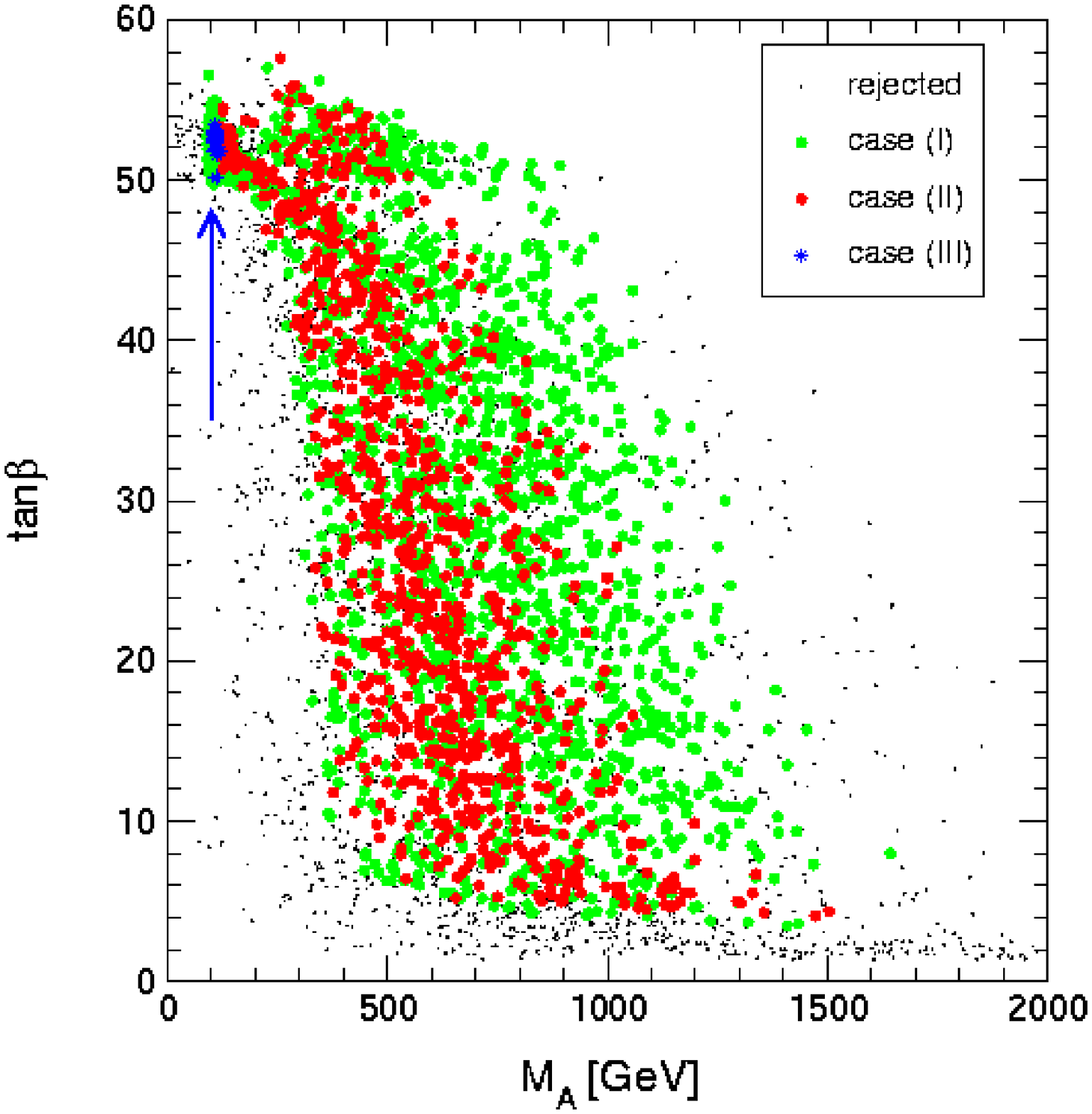,width=12cm,height=8cm}
\caption{Allowed parameter space within the mSUGRA scenario
in the $\MA-\tb$ plane for the three cases defined in
\refse{subsec:higgsscenarios}.
}
\label{fig15mSUGRA}
\end{center}
\end{figure}
%
\begin{figure}[ht!]
\begin{center}
\epsfig{figure=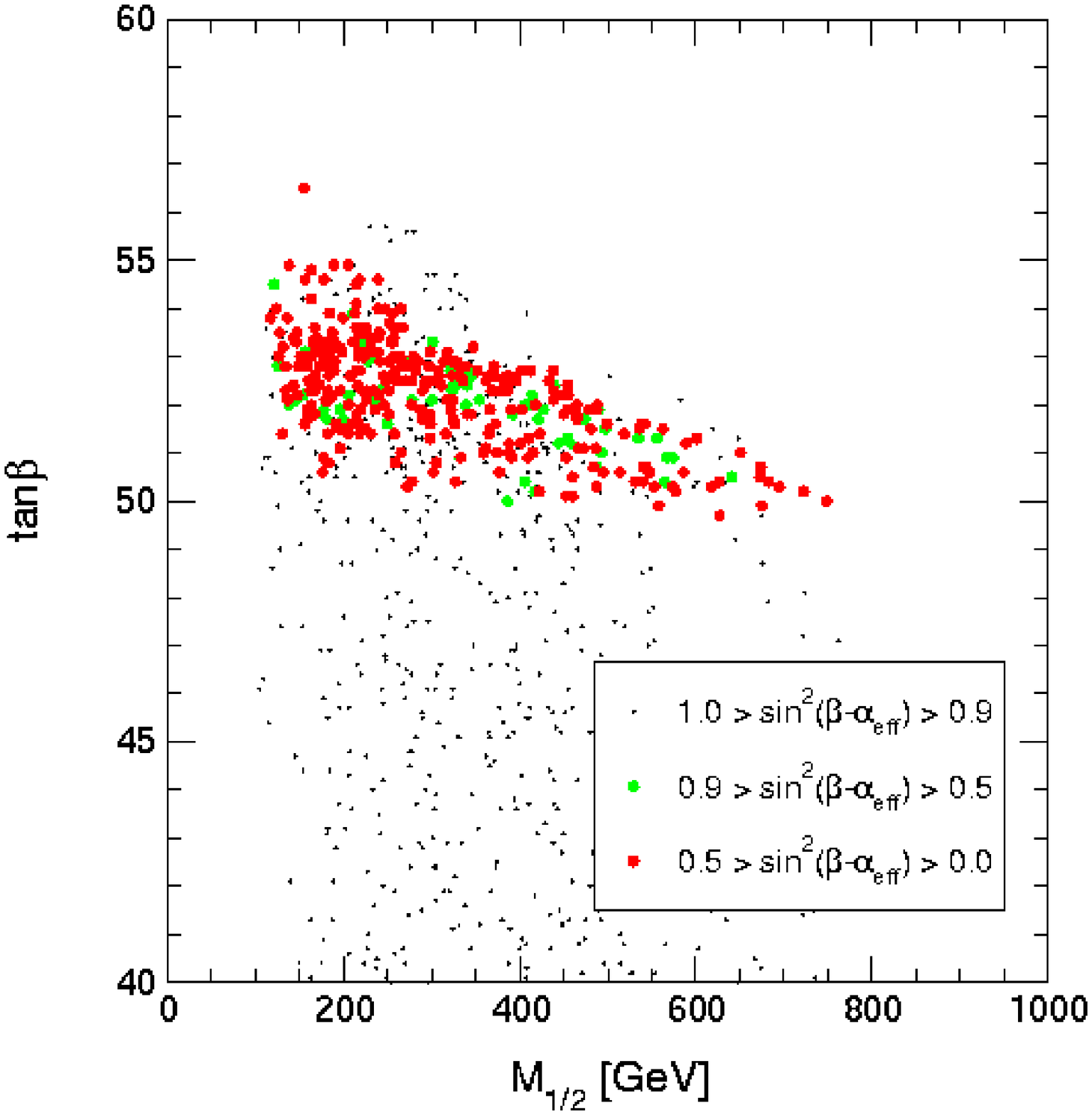,width=12cm,height=10cm}
\caption{The values for $\sin^2(\be - \aeff)$ realized in the mSUGRA
scenario are given in the $M_{1/2}-\tb$ plane.
}
\label{fig23mSUGRA}
\end{center}
\end{figure}


We first analyze the allowed parameter region in the Higgs sector 
of the mSUGRA scenario. 
In \reffi{fig11mSUGRA} we show the variation of the 
light Higgs boson mass with respect to $\tb$ 
for the three cases defined in \refse{subsec:higgsscenarios}. 
\reffi{fig15mSUGRA} shows the allowed parameter space in the $\MA-\tb$
plane.
Case (I), corresponding to the models that have passed all
experimental and theoretical constraints, is indicated in the figures 
by big green
(light shaded) points. Big red (dark shaded) points indicate case (II),
i.e.\ the subset of case~(I) in which $h$ has SM like couplings and
its mass lies within
$113 \gev \le \mh \le 117 \gev$. The points corresponding to case (III),
for which an interpretation of the LEP excess in terms of production of
the heavier $\cp$-even Higgs boson is possible, are displayed as blue
stars (indicated by arrows in the plots). 
The little black dots indicate parameter points which, while in
principle possible in the mSUGRA scenario, are rejected because of the 
experimental and theoretical constraints discussed above.

As a general feature, \reffi{fig11mSUGRA} shows that 
$\mh$ sharply increases with $\tb$ in the region
of low $\tb$, while for $\tb \gsim 10$ the $\mh$ values saturate. 
Values of $\tb \gsim 60$ are not allowed due to the REWSB constraint.
The LEP2 Higgs boson searches 
exclude the models with $\mh \lsim 113 \gev$ and $\tb \lsim 50$.
This is contrary to the general LEP2 Higgs boson searches in the
$\mhmax$ scenario~\cite{benchmark,mssmhiggs,tbexcl},
where the exclusion bound on the SM Higgs boson mass applies to $\mh$
only for $\tb \lsim 8$. For larger values of $\tb$ and 
small $\MA$ in the unconstrained MSSM a suppression of the $hZZ$ 
coupling is possible, giving rise to a reduced production rate compared to
the SM case. In the mSUGRA scenario a significant suppression of
$\sin^2(\be - \aeff)$ (i.e.\ the $hZZ$ coupling) occurs only for a small
allowed parameter region with $\tb \gsim 50$, see \reffi{fig23mSUGRA}. 
This feature can be
understood from the correlation between $\MA$ and $\tb$ shown in
\reffi{fig15mSUGRA}. Small values of $\MA$ with $\MA \lsim 150$~GeV, 
which are necessary for values of $\sin^2(\be - \aeff) \ll 1$, are only
possible for $\tb \gsim 50$. For $\tb \lsim 45$ we find that $\MA$ is
always larger than about 300~GeV, giving thus rise to a SM like behavior
of the $hZZ$ coupling.

As one can see in \reffis{fig11mSUGRA},~\ref{fig15mSUGRA} case (III)
can indeed be realized in the mSUGRA scenario in a small parameter
region where $50 \lsim \tb \lsim 55$, 
$103 \gev \lsim \mh, \MA \lsim 113 \gev$ and 
$\mH = 115 \pm 2 \gev$. It should be noted, however, that this parameter
region is close to the exclusion bounds obtained at Run~I of the
Tevatron~\cite{cdfhiggs} (which, as discussed above, we have not imposed
in the present analysis). With the upcoming results from Run~II of the
Tevatron it should be possible to fully cover the parameter space
compatible with case (III).


\begin{figure}[ht!]
\begin{center}
\epsfig{figure=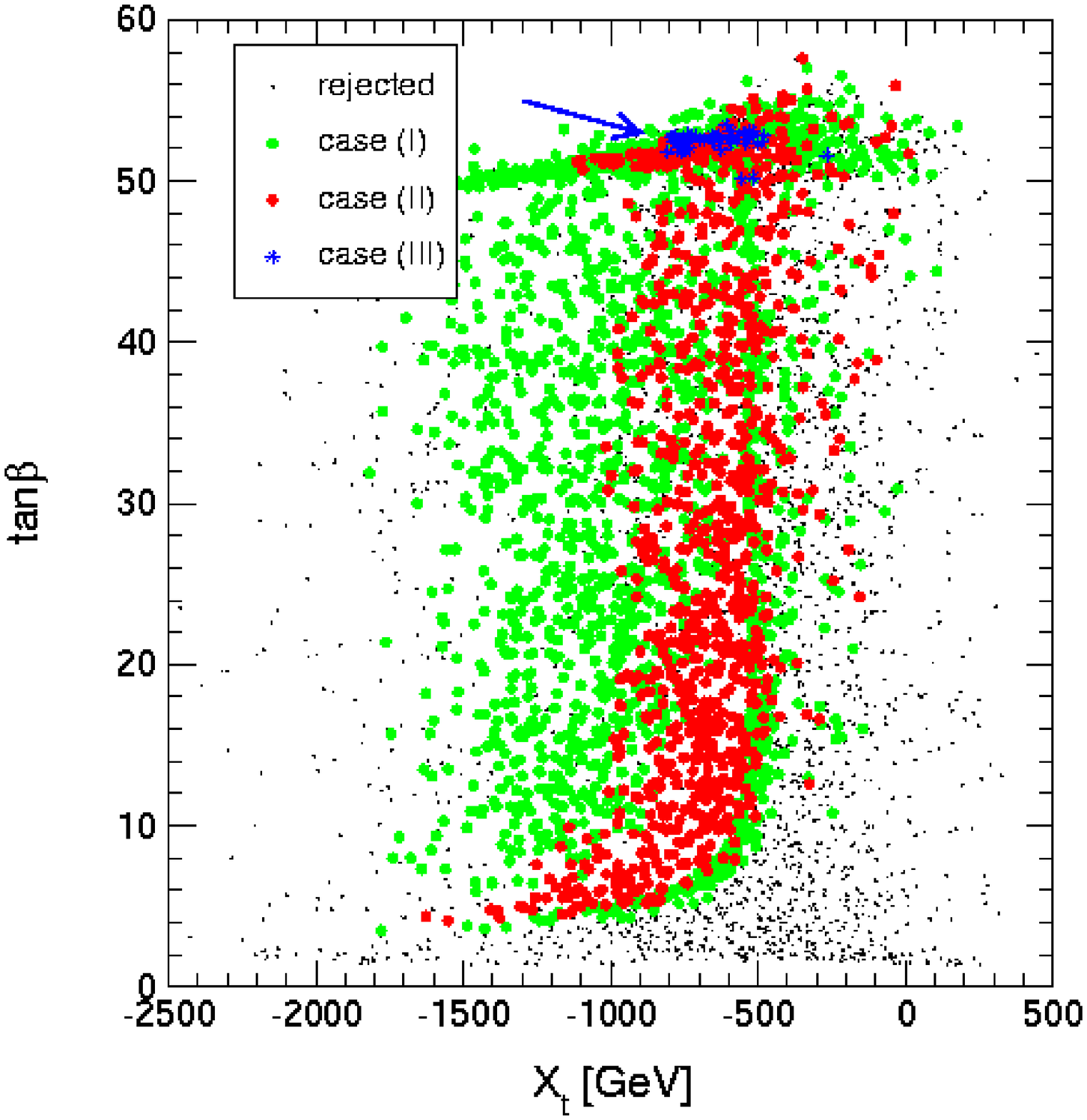,width=12cm,height=8cm}
\caption{Allowed parameter space within the mSUGRA scenario 
in the plane of the mixing parameter in
the $\Stop$~sector, $\Xt$, and $\tb$ for the three cases
defined in \refse{subsec:higgsscenarios}.
}
\label{fig15xmSUGRA}
\end{center}
\end{figure}


{}From \reffi{fig11mSUGRA} one can read off an upper bound%
\footnote{
This bound is $\sim 3 \gev$ lower than the one obtained in
\citere{higgsmsugra} due to additional constraints imposed 
in the present analysis, see \refse{subsec:constraints}.}
on the light $\cp$-even Higgs boson mass in the mSUGRA scenario of
\begin{equation}
\mhmax  \simleq 124~{\rm GeV} \;\;\; {\rm (mSUGRA)} \;.
\end{equation}
Values close to this upper limit on $m_h$ are reached in a large region
of moderate and large values of $\tb$, 
$20 \lsim \tb \lsim 50$. 

A lower bound on $\tb$ is inferred in the mSUGRA scenario,
\BE
\tb \gsim 3.3 \;\;\; {\rm (mSUGRA)} .
\EE
It should be noted that the two bounds quoted here refer to $\mt =
175$~GeV.

The upper bound on $\mh$ is about 6~GeV lower than the one in the
unconstrained MSSM~\cite{mhiggslong,mssmhiggs}, 
and the limit on $\tb$ is also more 
restrictive. This is caused by the fact that not all parameter
combinations of the unconstrained MSSM can be realized in the mSUGRA
scenario. In order to obtain the largest values for $\mh$ in particular
large values of the parameter $\Xt$,
\BE
\Xt \equiv A_t - \mu/\tb ,
\label{eq:Xt}
\EE
are necessary, which appears in the off-diagonal element of the
$\Stop$~mass matrix. Non-logarithmic genuine two-loop contributions to
$\mh$ give rise to an asymmetry with respect to the sign of $\Xt$, and
the maximum value obtained for $\mh$ is about 5~GeV higher for $\Xt > 0$
than for $\Xt < 0$~\cite{mhiggslong,mhiggslle}.

\begin{figure}[ht!]
\begin{center}
\epsfig{figure=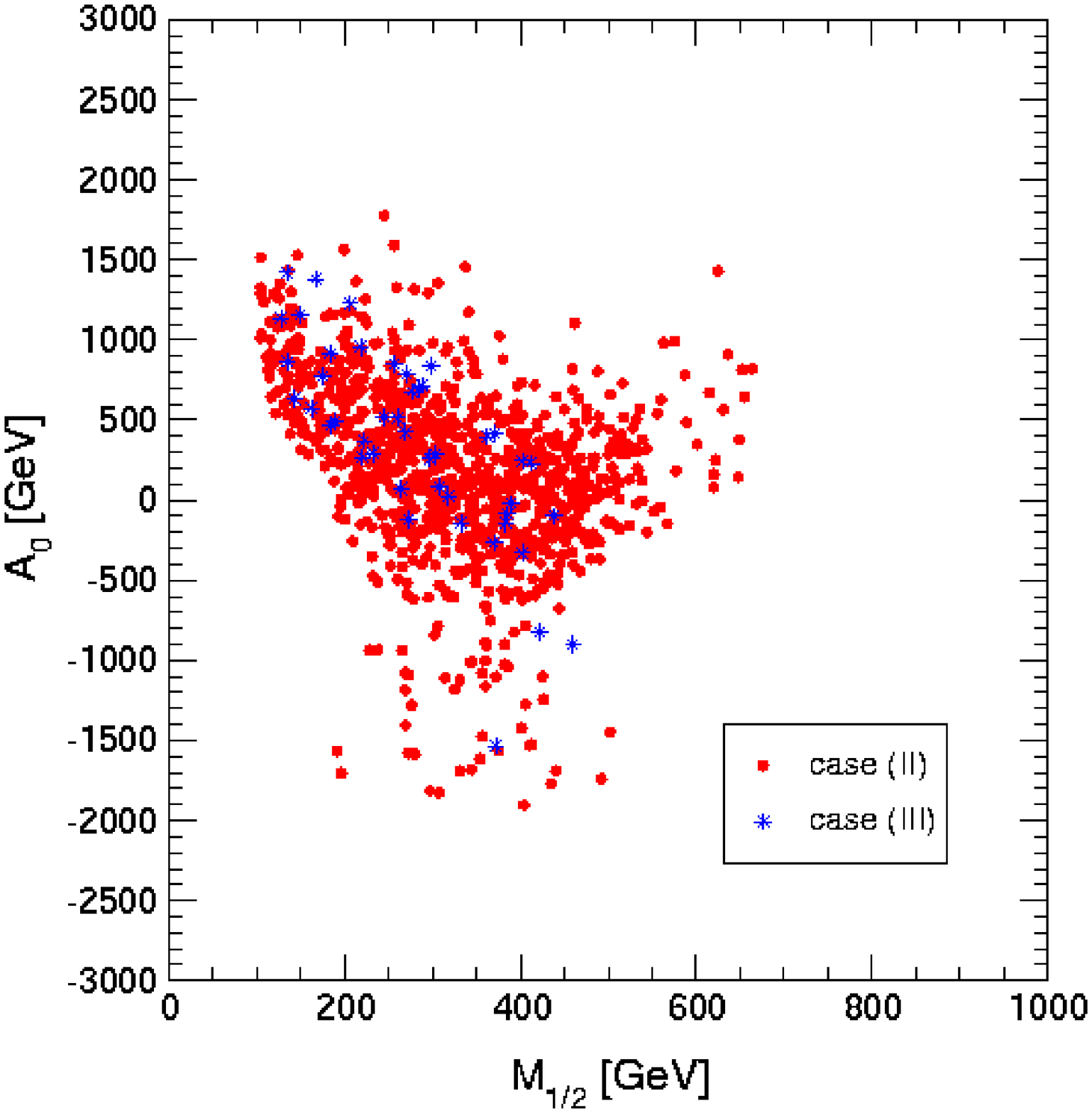,width=12cm,height=9.5cm}
\caption{
Cases (II) and (III) in the mSUGRA scenario are shown 
in the $M_{1/2}-A_0$ plane.
}
\label{fig17xmSUGRA}
\end{center}
\end{figure}
%
\begin{figure}[ht!]
\vspace{-1em}
\begin{center}
\epsfig{figure=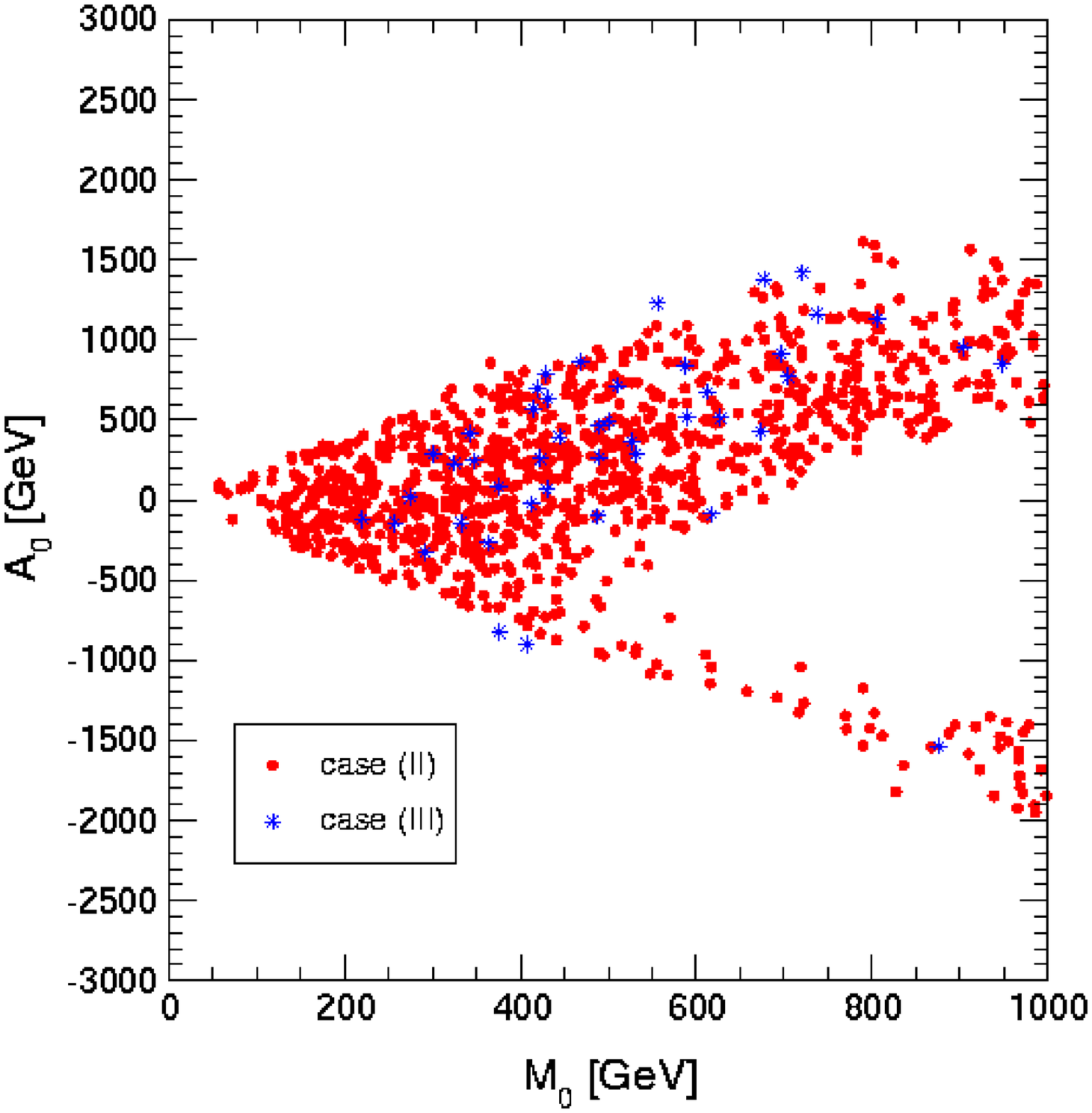,width=12cm,height=9.5cm}
\caption{
Cases (II) and (III) in the mSUGRA scenario are shown 
in the $M_0-A_0$ plane.}
\label{fig17mSUGRA}
\end{center}
\end{figure}

In \reffi{fig15xmSUGRA} the allowed parameter space in the 
$\Xt-\tb$ plane of the mSUGRA scenario is depicted for the three cases 
discussed above. The figure shows that the mSUGRA scenario strongly
favors negative values of $\Xt$. The absence of models with large positive 
$\Xt$ is the main reason for the decrease in the upper bound within the
mSUGRA compared to the unconstrained MSSM.

Concerning the lower bound on $\mh$, we find the same bound as in the
unconstrained MSSM, i.e.\
$m_h \gsim 91$~GeV.
As discussed above, $\mh$ values below 113~GeV being compatible with the
LEP exclusion bounds are only possible in a small parameter region with
$\tb \gsim 50$ in the mSUGRA scenario.

We now turn to the restrictions on the parameter space of the underlying
mSUGRA parameters, $M_0$, $M_{1/2}$ and $A_0$, which are obtained if the
LEP excess is interpreted as a signal, i.e.\ for the cases~(II)
and~(III). \reffis{fig17xmSUGRA}, \ref{fig17mSUGRA} show that the
cases~(II) and~(III) result in similar allowed regions of parameter
space. While for $M_0$ the whole
range up to $1 \tev$ is allowed, $M_{1/2}$ is restricted to
$M_{1/2} \lsim 650 \gev$, see \reffi{fig17xmSUGRA}. In
\reffi{fig17mSUGRA} the cases~(II) and~(III) are shown in the
$M_0-A_0$ plane. $|A_0|$ is restricted to $|A_0| \lsim 2 M_0$. 
For the special case $A_0 = 0$ we find that $M_0$ is bounded from above
by $M_0 \lsim 700$~GeV.

\begin{figure}[ht!]
\begin{center}
\epsfig{figure=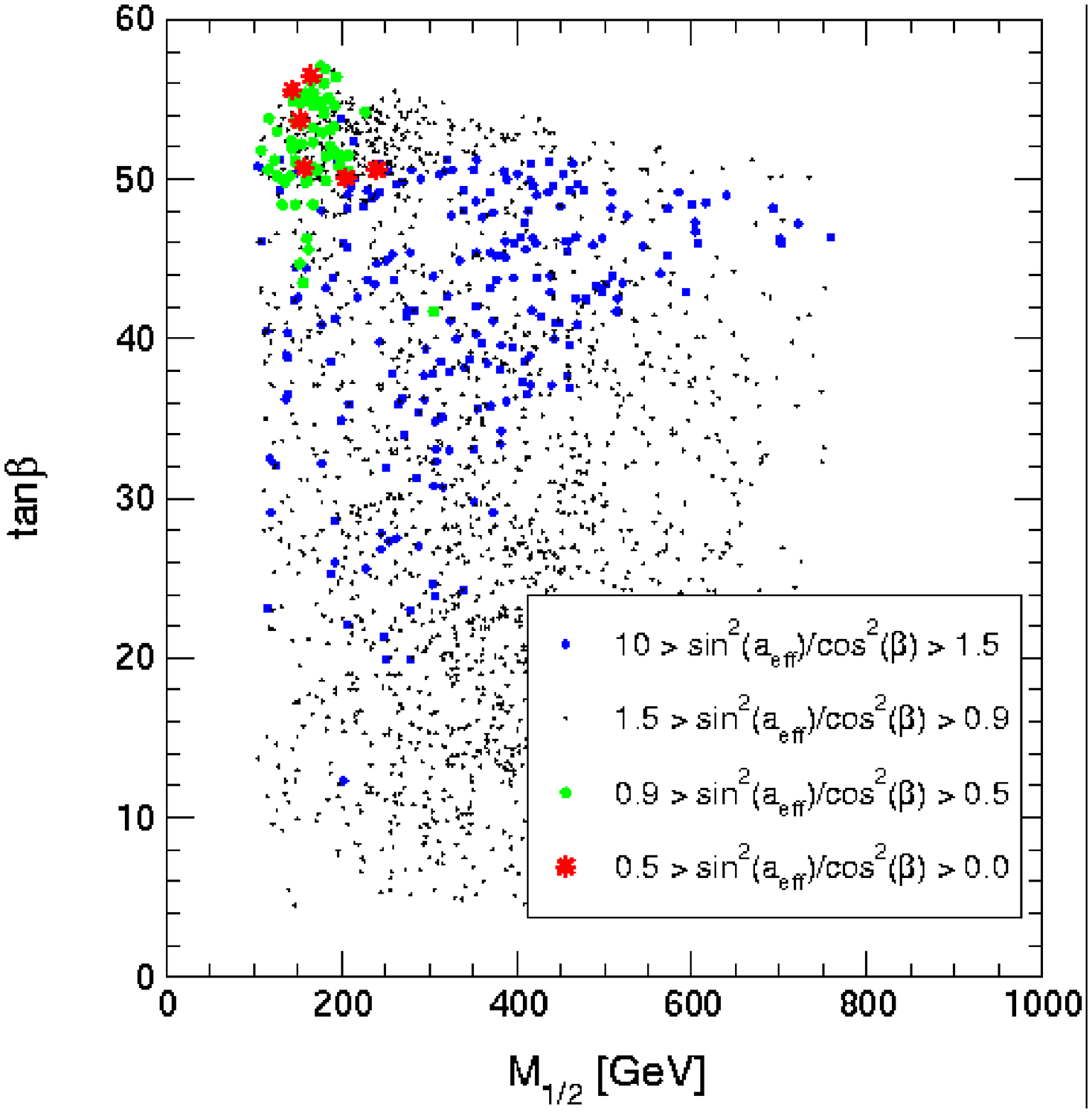,width=12cm,height=10cm}
\caption{The values for $\sin^2\aeff/\cos^2\be$ realized in the mSUGRA 
scenario are given in the $M_{1/2}-\tb$ plane.
}
\label{fig24bmSUGRA}
\end{center}
\end{figure}
%
\begin{figure}[ht!]
\begin{center}
\epsfig{figure=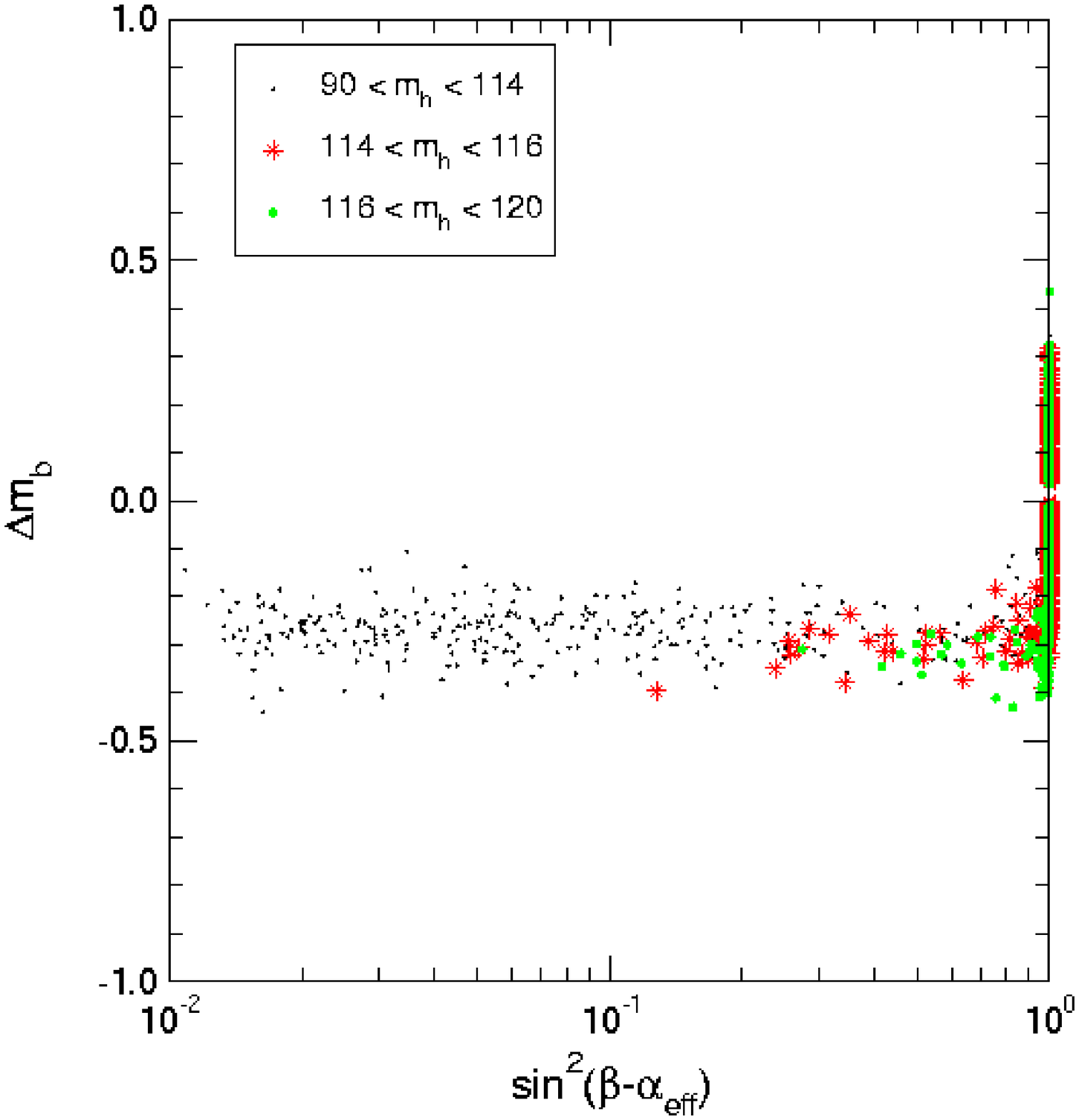,width=12cm,height=10cm}
\caption{The allowed values in the mSUGRA scenario for 
$\sin^2(\be - \aeff)$ and the quantity 
$\De m_b$ (see \refeq{eq:Delmb}). The corresponding values for $\mh$ are
also indicated.
}
\label{fig50mSUGRA}
\end{center}
\end{figure}

We find that $M_{1/2}$ values as low as 100~GeV (these values are
only obtained for not too small $M_0$, $M_0 \gsim 300$~GeV; 
the lower bound on $M_{1/2}$
for smaller values of $M_0$ is about $M_{1/2} \gsim 200$~GeV) are compatible
with the interpretation of the LEP excess as a signal of the lightest
$\cp$-even Higgs boson in the MSSM. This result is in contrast to the
analysis in \citere{higgs115msugra1}, where for $\mt = 175$~GeV and
$\mh \ge 113$~GeV a lower bound of $M_{1/2} \gsim 310$~GeV has been
found for $A_0 = 0$, while we obtain a lower bound of 
$M_{1/2} \gsim 200$~GeV, see \reffi{fig17xmSUGRA}. 
The main difference between our result and the one obtained in
\citere{higgs115msugra1} can be traced to the inclusion of genuine
non-logarithmic two-loop corrections in the result
for $\mh$ in the present paper (see also \citere{higgs115msugra3}).
The lower values found for $M_{1/2}$ in the present analysis give rise
to a different low energy spectrum for the SUSY particles.
As an exemplary case, we find that $M_0 \approx M_{1/2} \approx 200 \gev$
and $-400 \gev \lsim A_0 \lsim 400 \gev$ is compatible with 
$\mh = 115 \pm 2 \gev$ together with all the constraints listed in
\refse{subsubsec:phenorest}. In this example the heaviest 
SUSY particle is the gluino with a mass of $\sim 590 \gev$ and all the
scalar quarks have masses about $500 \gev$ or lower, 
depending on $A_0$ and on $\tb$. The SUSY spectrum in the three
soft SUSY-breaking scenarios will be analyzed in more detail in 
\refse{subsec:numanalspectrum}.

Finally, we investigate the $hb\bar b$ coupling in the mSUGRA scenario
in comparison with the SM case, where the decay into $b$~quarks is the
dominant decay channel of the Higgs boson. The $hb\bar b$ coupling is
mainly altered in two ways compared to the SM: it has an extra
factor $\sin\aeff/\cos\be$ and it receives a correction 
$\sim 1/(1 + \De m_b)$, see \refeq{eq:Delmb}. 

\reffi{fig24bmSUGRA} shows
the different values of $\sin^2\aeff/\cos^2\be$ (which enters the 
$h \to b \bar b$ decay rate) realized within the mSUGRA
scenario in the $M_{1/2}-\tb$ plane. The figure shows that a
significant enhancement of the $hb \bar b$ coupling is possible over a
wide range of the mSUGRA parameter space. In these parameter regions the
sensitivity in the Higgs search via the $h \to b \bar b$ channel is in
general slightly increased compared to the SM case. On the other hand, 
the increase in the $h \to b \bar b$ partial width leads in general to a
reduced branching ratio of $h \to \ga\ga$, making thus the search via
this channel at the LHC more difficult~\cite{higgs115msugra3}%
\footnote{
For a similar analysis for the charged Higgs bosons, see
\citere{sola}.
}%
. 

It can furthermore be seen in \reffi{fig24bmSUGRA} that
a significant suppression of the $hb\bar b$ coupling is only possible in
a small fraction of the mSUGRA parameter space. Values of 
$\sin^2\aeff/\cos^2\be < 0.7$ are only obtained in the parameter region
$\tb \gsim 50$.

In \reffi{fig50mSUGRA} the quantity $\De m_b$ is analyzed in the mSUGRA
scenario. The maximal values obtained for $\De m_b$ in the mSUGRA
scenario are about $\pm 0.4$ (where values of $|\De m_b| > 0.3$ are only
realized for $\tb \gsim 40$).
The figure shows that values of $\sin^2(\be -\aeff) \ll 1$
(corresponding to a suppressed coupling of the Higgs to vector bosons
and lower allowed values for $\mh$)
are always correlated in the mSUGRA scenario with negative values 
of $\De m_b$, giving rise to an enhancement of the $hb\bar b$ coupling.
Positive values for $\De m_b$ (corresponding to a suppression of the 
$hb\bar b$ coupling) are only possible if the Higgs boson couples with
full strength to $W$ and $Z$.


\subsection{mGMSB}
\label{subsec:numanalgmsb}

For this paper, about 40000 mGMSB models were generated under
well defined hypotheses described in \refse{subsec:gmsb}, using the program
{\em SUSYFIRE}~\cite{SUSYFIRE} and adopting the phenomenological 
approach of \citeres{AKM-LEP2,AB-LC,AMPPR}, see also \citere{higgsgmsb}.
Concerning the variation of the high-energy parameters one should keep
in mind that the lower bound on $\La$ arises from the requirement that
$m_{\rm NLSP} \ge 100 \gev$, while its upper bound as well as the
upper bound on $M_{\rm mess}$ originate mainly from the
upper bound imposed on SUSY particle masses, see
\refse{subsubsec:phenorest}. The upper bounds on $\La$ and 
$M_{\rm mess}$ automatically restrict $N_{\rm mess}$ from above, as
explained in \refse{subsec:gmsb}.
The above restrictions yield the following variations of the 
high-energy parameters:
\begin{eqnarray}
10^4 \gev \le &\La& \le 2\,\times\,10^5 \gev \;, \non \\
1.01\,\La \le &M_{\rm mess}& \le 10^5\,\La \;, \nonumber \\
1 \le  &N_{\rm mess}& \le 8 \;, \nonumber \\
1.5 \le &\tan\beta & \le 55 \;, \non \\
 &{\rm sign}\, \mu& = \pm 1 .
\label{gmsbparam}
\end{eqnarray}

\begin{figure}[ht!]
\begin{center}
\epsfig{figure=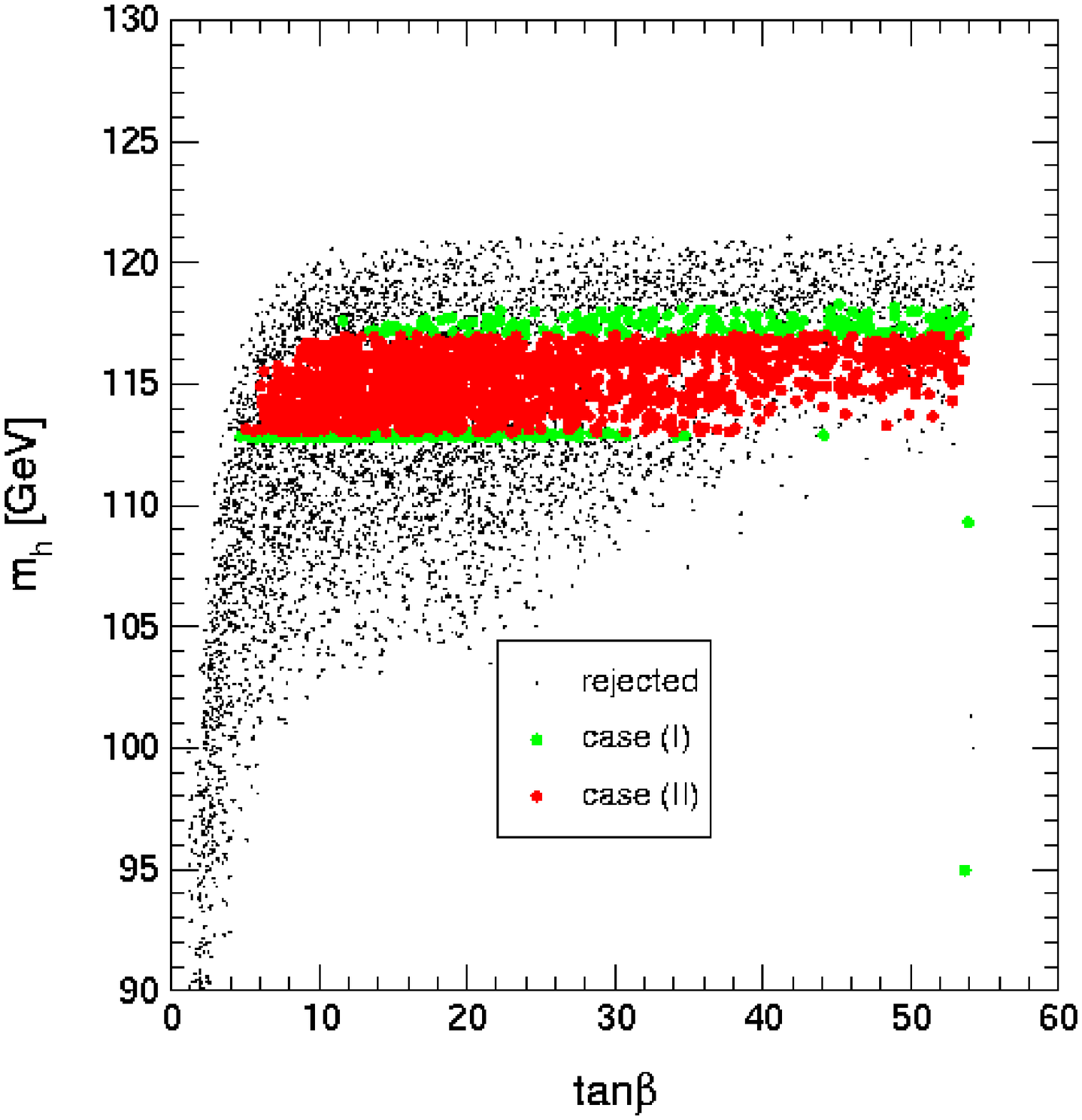,width=12cm,height=8cm}
\caption{
The light Higgs boson mass $\mh$ as a function of 
$\tb$ in the mGMSB scenario. Case~(I) and case~(II) as discussed in
\refse{subsec:higgsscenarios} are displayed together with the rejected
models. 
Case~(I) corresponds to the models that have passed all theoretical
and experimental constraints. Case~(II) is the subset of case~(I) with
$\mh$ values in the region favored by recent LEP Higgs searches, 
$113 \gev \le \mh \le 117 \gev$, and SM like couplings of the $h$. 
Case~(III), where the LEP excess is
interpreted as a signal of the heavier $\cp$-even
Higgs boson in the MSSM, is not realized in the mGMSB scenario.
}
\label{fig11GMSB}
\end{center}
\end{figure}
%
\begin{figure}[ht!]
\begin{center}
\epsfig{figure=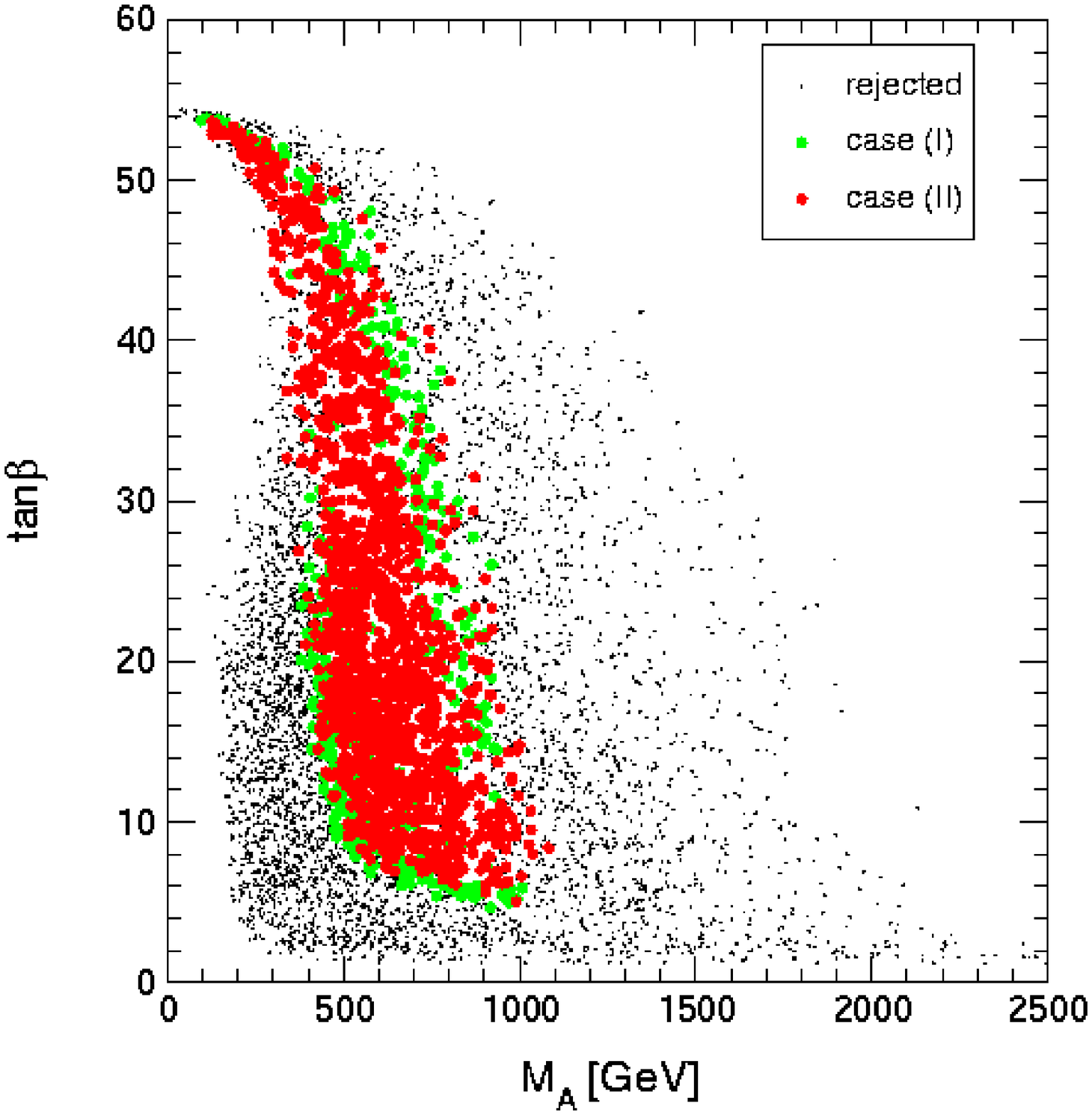,width=12cm,height=8cm}
\caption{
Allowed parameter space within the mGMSB scenario
in the $\MA-\tb$ plane for case~(I) and case~(II) defined in
\refse{subsec:higgsscenarios}.
}
\label{fig15GMSB}
\end{center}
\end{figure}
%
\begin{figure}[ht!]
\begin{center}
\epsfig{figure=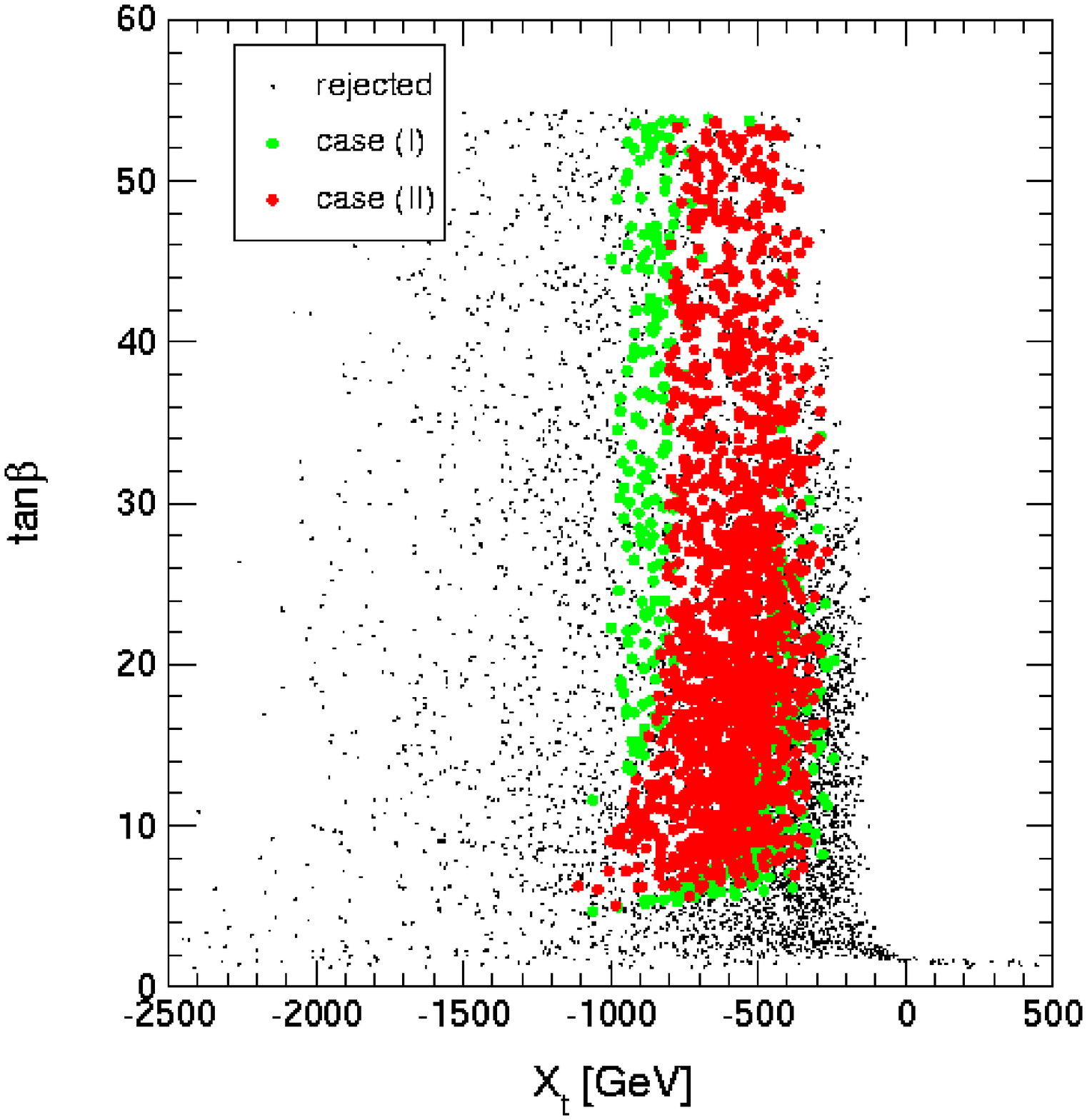,width=12cm,height=8cm}
\caption{Allowed parameter space within the mGMSB scenario
in the plane of the mixing parameter in
the $\Stop$~sector, $\Xt$, and $\tb$ for case~(I) and case~(II) defined
in \refse{subsec:higgsscenarios}.
}
\label{fig15wGMSB}
\end{center}
\end{figure}

As above, we first analyze the allowed parameter region in the Higgs
sector in this scenario. 
In \reffi{fig11GMSB} we show
the variation of the light Higgs boson mass with respect to 
$\tb$ for case~(I) and case~(II) defined in
\refse{subsec:higgsscenarios}. Case~(III), where the LEP excess is
interpreted as a signal of the heavier $\cp$-even
Higgs boson in the MSSM, is not realized in the mGMSB scenario.
\reffi{fig15GMSB} shows the allowed parameter space in the $\MA-\tb$
plane.

It can be seen from the two figures that the experimental and
theoretical constraints discussed in \refse{subsec:constraints} 
have a bigger effect on the parameter space than in the case of the
mSUGRA scenario.
In particular, they significantly influence the upper bound on $\mh$,
which is reduced in this way by about 3~GeV. 
Like in the mSUGRA case, the LEP2 Higgs boson searches exclude the
models with $\mh \lsim 113 \gev$ and $\tb \lsim 50$. A significant
suppression of $\sin^2(\be - \aeff)$ (i.e.\ the $hZZ$ coupling) occurs
only for a small allowed parameter region with $\tb \gsim 50$ (the lower
density of points with $\tb \gsim 50$ and $\mh < 113$~GeV as compared to
\reffi{fig11mSUGRA} has no direct physical meaning; it is mainly due to the
fact that $\tb$ has been varied on a logarithmic scale in the mGMSB
scenario, while a linear scale has been chosen in the mSUGRA scenario).
This can be understood from \reffi{fig15GMSB}, which shows that
values of $\MA \lsim 300 \gev$ are only realized for $\tb \gsim 50$.
Values of $\tb \gsim 55$ are not allowed due to the REWSB constraint.

For the upper bound on the light $\cp$-even Higgs boson mass in the 
mGMSB scenario we obtain%
\footnote{
This bound is $\sim 4 \gev$ lower than the one obtained in
\citere{higgsgmsb}, mainly due to additionally imposed constraints like the
``naturalness'' bound on the scalar quark
masses, see \refse{subsubsec:phenorest}.
}
%
\begin{equation}
\mhmax  \simleq 119~{\rm GeV} \;\;\; {\rm (mGMSB)} \;.
\end{equation}
Values close to this upper limit on $m_h$ are reached in a large region
of moderate and large values of $\tb$,
$20 \lsim \tb \lsim 50$.

A lower bound on $\tb$ is inferred in the mGMSB scenario,
\BE
\tb \gsim 4.6 \;\;\; {\rm (mGMSB)} .
\EE
As above, the two bounds quoted here refer to $\mt = 175$~GeV.

\begin{figure}[ht!]
\begin{center}
\epsfig{figure=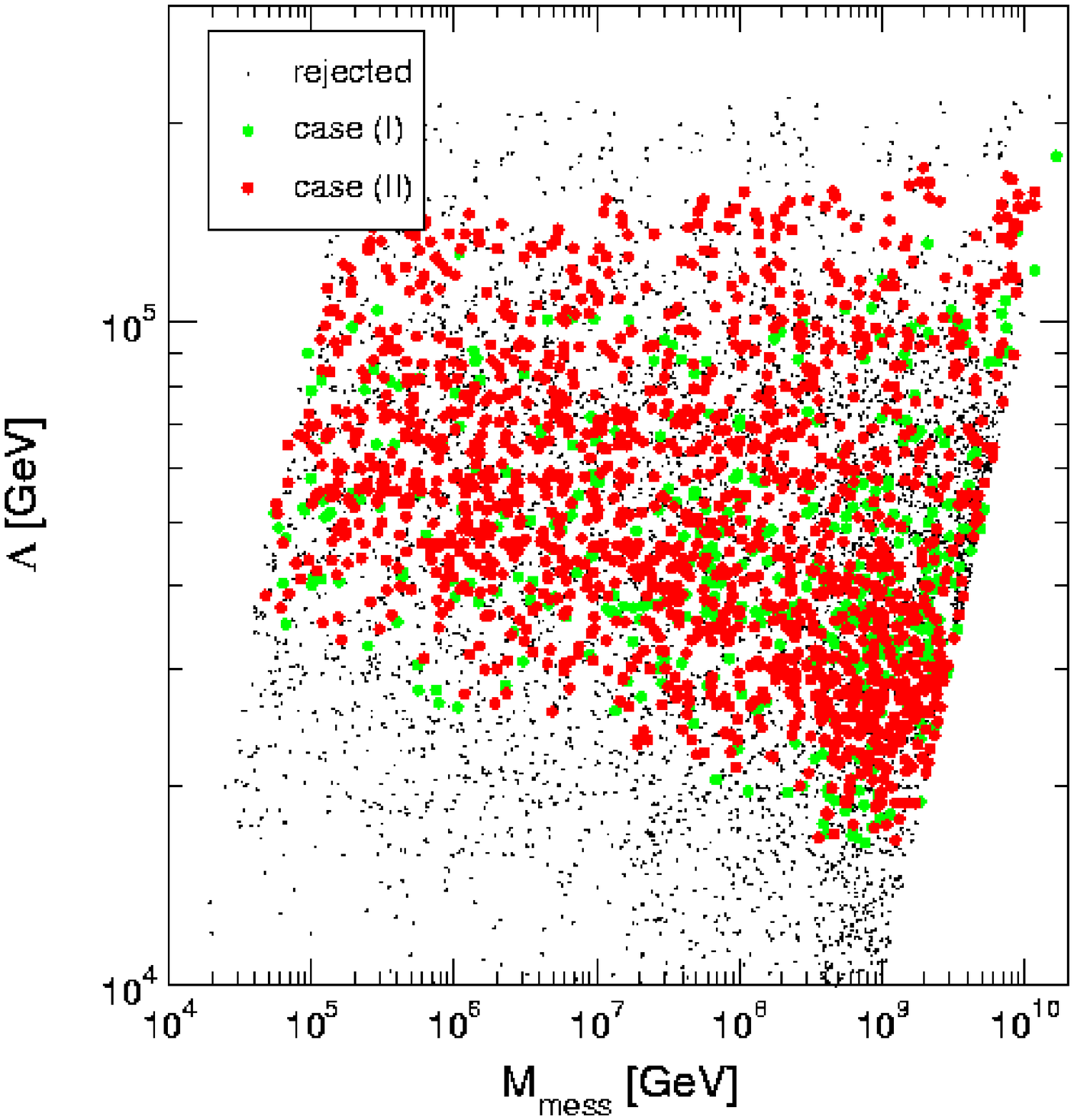,width=12cm,height=11cm}
\caption{
Cases (I) and (II) in the mGMSB scenario are shown 
in the $M_{\rm mess}-\La$ plane.
}
\label{fig16GMSB}
\end{center}
\end{figure}
%
\begin{figure}[hb!]
\vspace{-1cm}
\begin{center}
\epsfig{figure=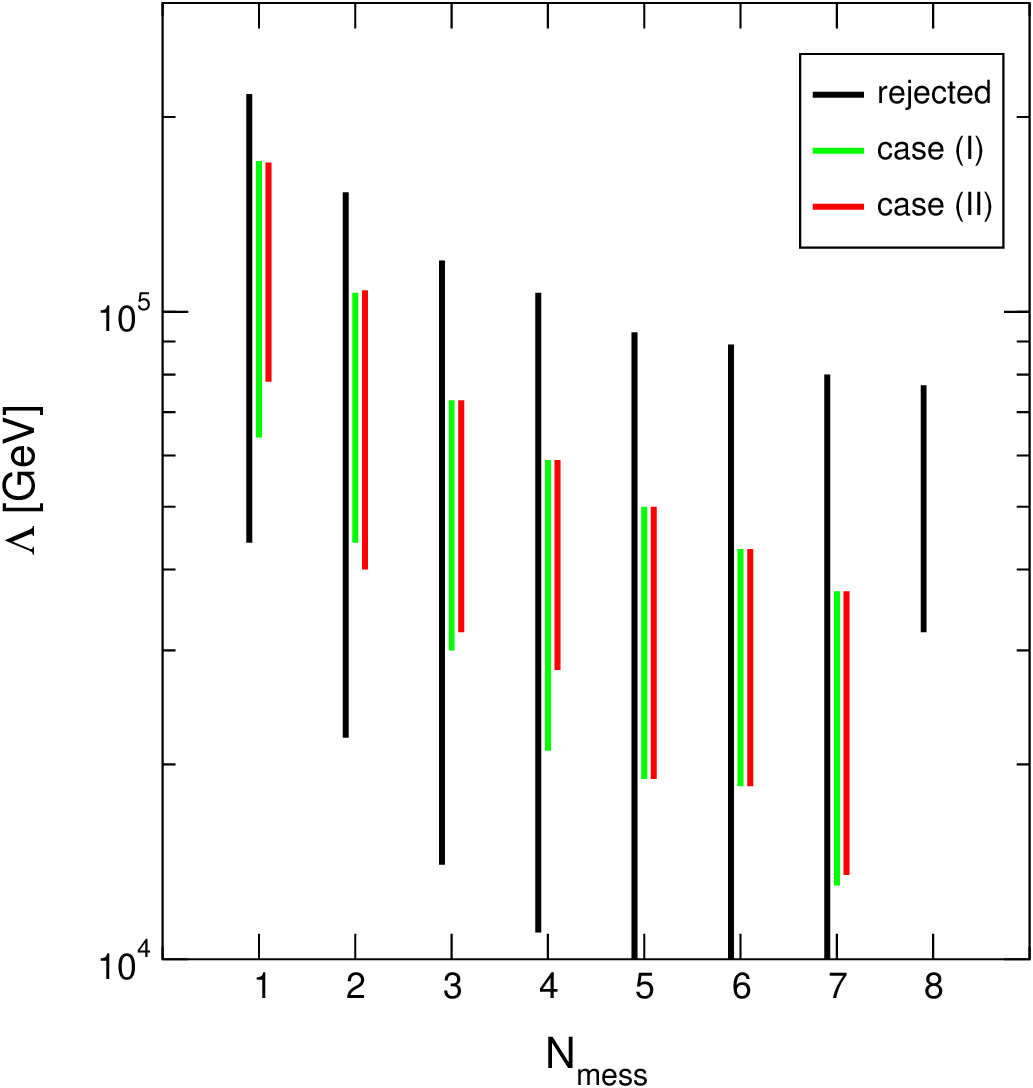,width=7.2cm,height=8cm}
\hspace{1cm}
\epsfig{figure=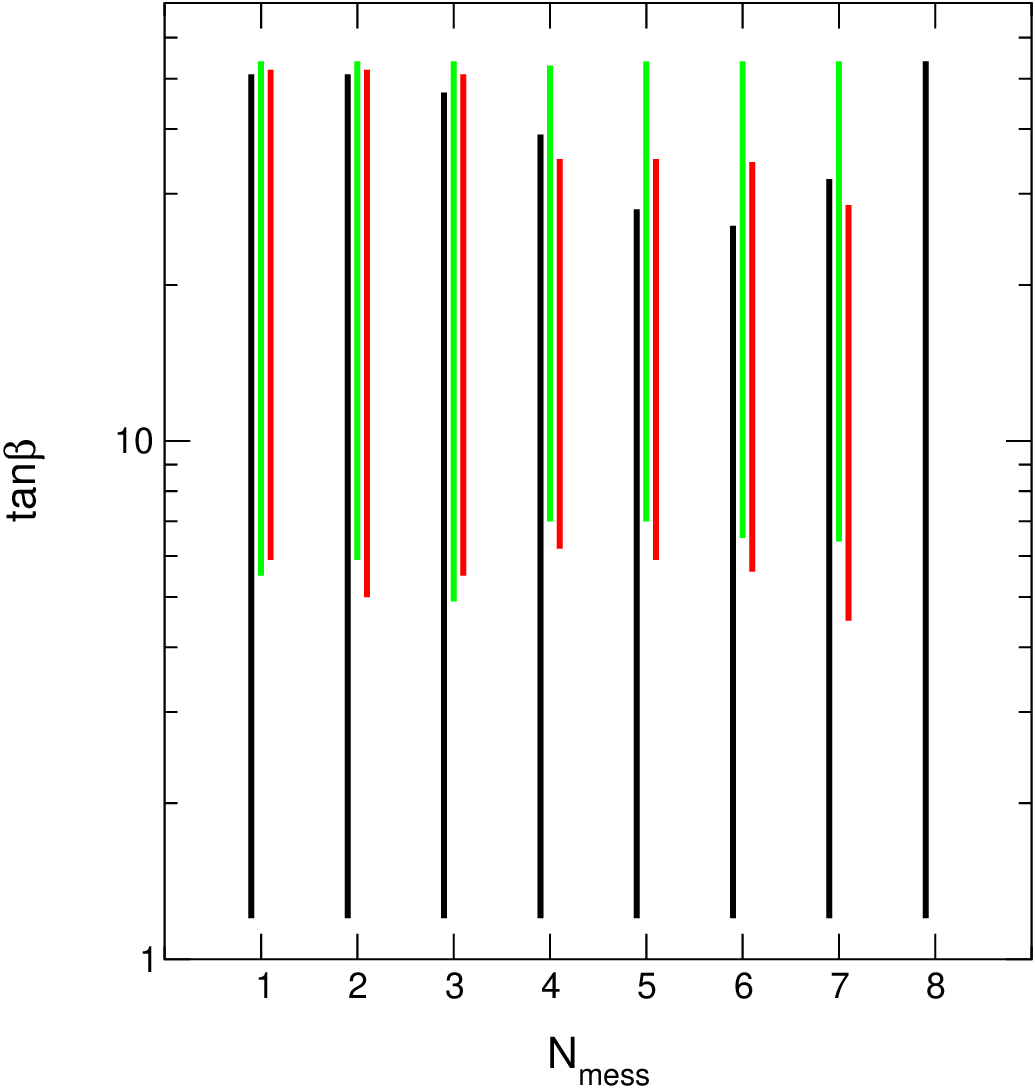,width=7.2cm,height=8cm}
\caption{
Cases (I) and (II) in the mGMSB scenario are shown 
in the $N_{\rm mess}-\La\,(\tb)$ plane in the left (right) plot. The
legend shown in the left plot applies to both plots.
}
\label{fig17GMSB}
\end{center}
\end{figure}

As in the mSUGRA scenario, the restrictions on the mixing parameter 
$\Xt$ (see \refeq{eq:Xt}) arising from the parameter correlations in the
mGMSB scenario (and the upper bound imposed in \refeq{eq:natbounds})
are the main effect causing the decrease in the upper
bound on $\mh$ as compared to the unconstrained MSSM. 
\reffi{fig15wGMSB} shows that only negative values for $\Xt$ are allowed
in the mGMSB scenario. Compared to the mSUGRA case (see
\reffi{fig15xmSUGRA}) the allowed parameter region for $\Xt$ is smaller
and is furthermore shifted towards smaller values of $|\Xt|$.

Concerning the underlying GMSB parameters, 
$M_{\rm mess}, N_{\rm mess}$ and $\La$, no severe restrictions can be
deduced for the cases~(I) and~(II), 
see \reffis{fig16GMSB}, \ref{fig17GMSB}.
In \reffi{fig16GMSB} we show the allowed regions in the 
$M_{\rm mess}-\La$ plane. The experimental and theoretical constraints
imposed in our analysis affect in particular the region of low $M_{\rm mess}$
and low $\La$. In the left plot
of \reffi{fig17GMSB} the allowed regions in the $N_{\rm mess}-\La$
plane are presented. Lower values of $N_{\rm mess}$ correspond to higher
values of $\La$. This is a consequence of the boundary values imposed on 
the physical masses in \refeq{eq:bound}.
The Higgs boson mass constraints cut away a significant part
of the $\La$ range for each value of $N_{\rm mess}$. We only find
allowed parameter regions for $N_{\rm mess} \leq 7$ (although higher
values of $N_{\rm mess}$ might be allowed if the upper bound of
$M_{\rm mess}/\La$ is relaxed.) The right plot of
\reffi{fig17GMSB} shows the $N_{\rm mess}-\tb$ plane. 
Case~(II) corresponds to about the same allowed region as case~(I),
apart from the values $N_{\rm mess} \geq 4$, where the highest values of
$\tb$ are not allowed in case~(II).

\begin{figure}[htb!]
\begin{center}
\epsfig{figure=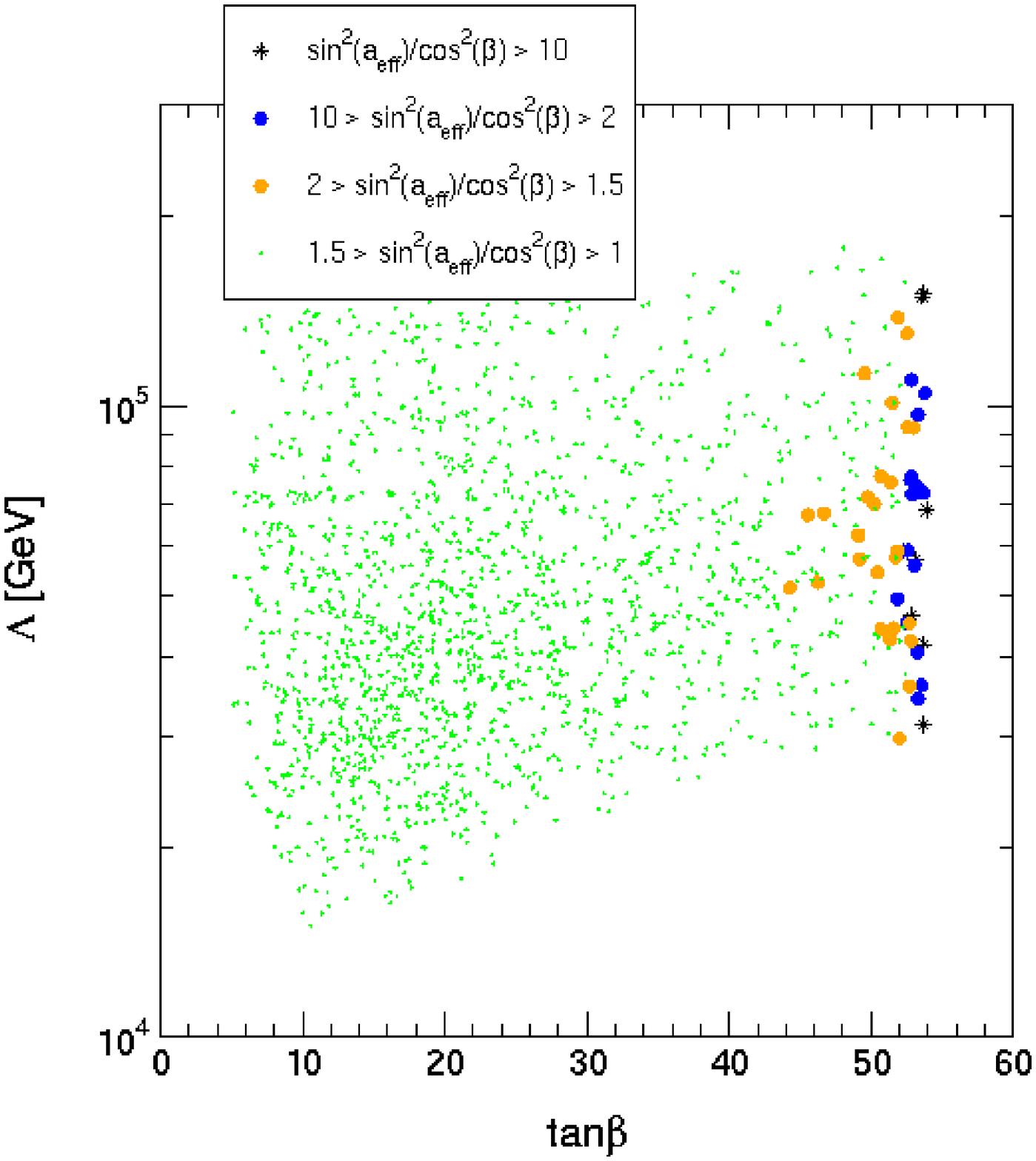,width=12cm,height=11cm}
\caption{
The values for $\sin^2\aeff/\cos^2\be$ realized in the mGMSB
scenario are given in the $\tb-\La$ plane.
}
\label{fig26GMSB}
\end{center}
\end{figure}

We also investigate the $hb\bar b$ coupling within the mGMSB scenario.
In \reffi{fig26GMSB} the different values of $\sin^2\aeff/\cos^2\be$ 
realized within the mGMSB scenario are shown in the $\tb-\La$ plane. In
contrast to the mSUGRA case, no values of $\sin^2\aeff/\cos^2\be < 1$
exist, i.e.\ no suppression of the $hb\bar b$ coupling occurs in this
way. As above, a significant enhancement of the $hb\bar b$ coupling is
possible. This applies in particular to the region of the highest values
of $\tb$.

We have also analyzed the quantity $\De m_b$ (see \refeq{eq:Delmb})
within the mGMSB scenario. The absolute value of $\De m_b$ is smaller in
the mGMSB scenario than in the mSUGRA case and does not exceed 
$|\De m_b| = 0.2$. Values of $|\De m_b| > 0.1$ are only realized for
$\tb \gsim 35$.


\subsection{mAMSB}
\label{subsec:numanalamsb}

According to the description presented in \refse{subsec:amsb} 
about 50000 models have been created. The
GUT scale parameters have been varied in the ranges
\BEA
20 \tev \le &m_{\rm aux}& \le 100 \tev , \non \\
0 \le &m_0& \le 2 \tev , \non \\
1.5 \le &\tb& \le 60 , \non \\
 &{\rm sign}\, \mu& = \pm 1 . 
\label{gmsbparams}
\EEA

\begin{figure}[ht!]
\begin{center}
\epsfig{figure=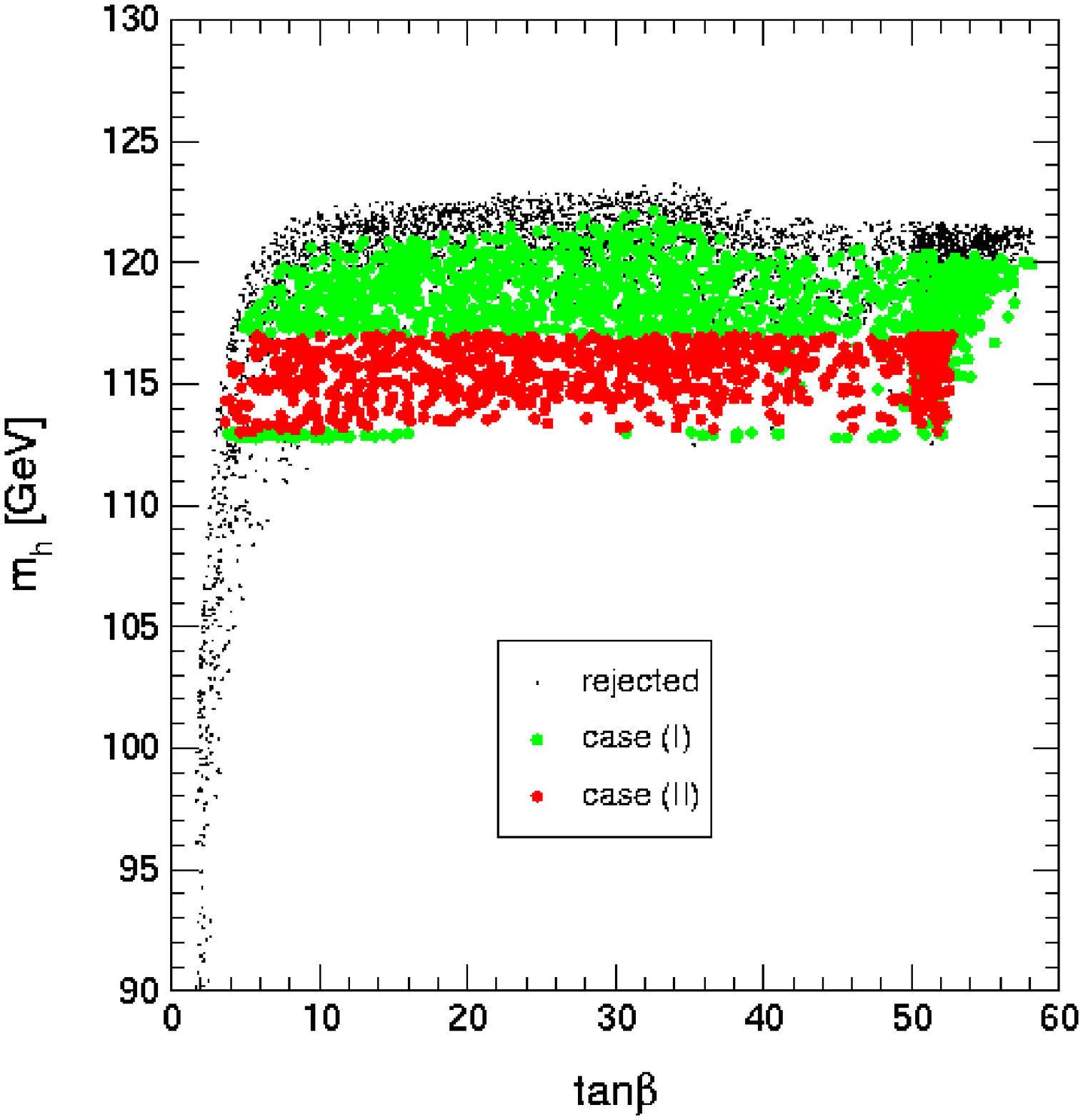,width=12cm,height=8cm}
\caption{
The light $\cp$-even Higgs boson mass $\mh$ as a function of 
$\tb$ in the mAMSB scenario. Case~(I) and case~(II) as discussed in
\refse{subsec:higgsscenarios} are displayed together with the rejected
models.
Case~(I) corresponds to the models that have passed all theoretical
and experimental constraints. Case~(II) is the subset of case~(I) with
$\mh$ values in the region favored by recent LEP Higgs searches,
$113 \gev \le \mh \le 117 \gev$, and SM like couplings of the $h$. 
Case~(III), where the LEP excess is
interpreted as a signal of the heavier $\cp$-even
Higgs boson in the MSSM, is not realized in the mAMSB scenario.
}
\label{fig11AMSB}
\end{center}
\end{figure}
%
\begin{figure}[ht!]
\begin{center}
\epsfig{figure=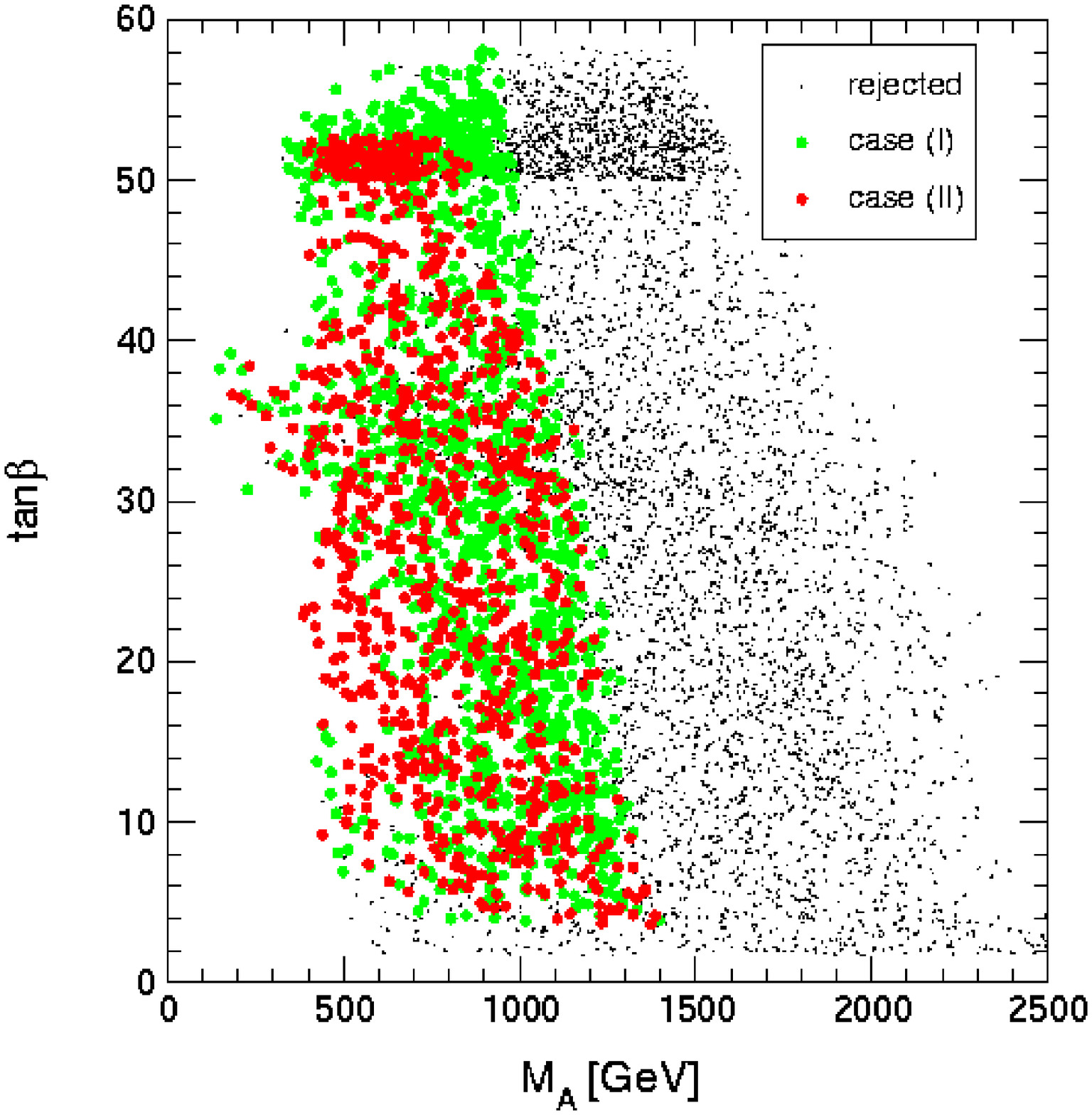,width=12cm,height=8cm}
\caption{
Allowed parameter space within the mAMSB scenario scenario
in the $\MA-\tb$ plane for case~(I) and case~(II) defined in
\refse{subsec:higgsscenarios}.
}
\label{fig15AMSB}
\end{center}
\end{figure}
%
\begin{figure}[ht!]
\begin{center}
\epsfig{figure=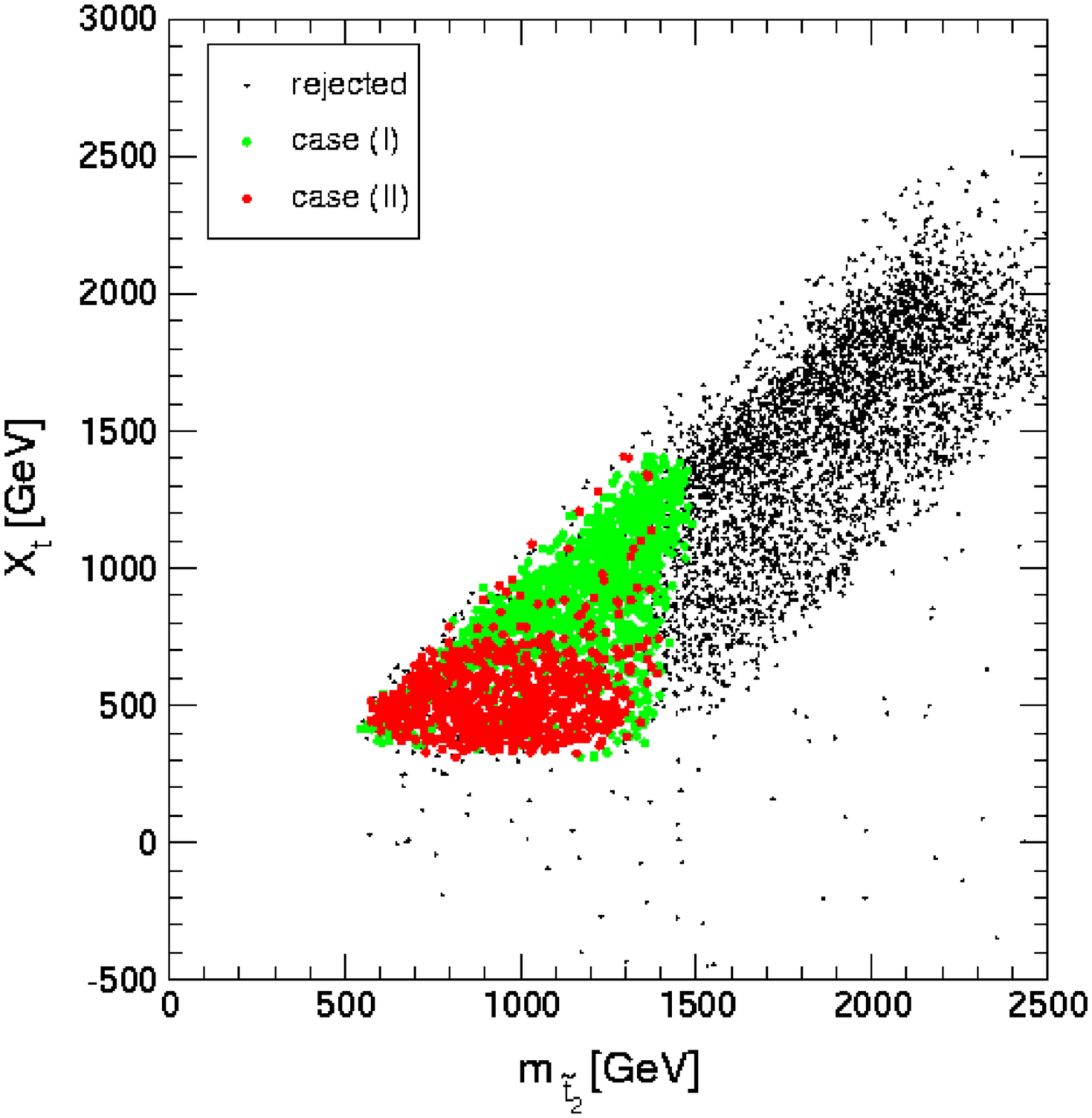,width=12cm,height=8cm}
\caption{Allowed parameter space within the mAMSB scenario
in the plane of the heavier scalar top mass, $\mstz$, and the
mixing parameter in the $\Stop$~sector, $\Xt$, 
for case~(I) and case~(II) defined
in \refse{subsec:higgsscenarios}.
}
\label{fig15xAMSB}
\end{center}
\end{figure}

The general behavior of the Higgs boson sector has already been
described in \citere{higgsamsb}.
Both $\mh$ and $\MA$ increase with $m_{\rm aux}$, which determines the 
SUSY mass scale.  $\mh$ depends only very weakly on $m_0$ and $\tb$ except 
for large $m_0$ and small $\tb$.  
$\MA$ gets larger for larger $m_0$ and $\tb$, although
the dependence on $\tb$ is rather weak.   

In \reffi{fig11AMSB} we show
the variation of the light Higgs boson mass with respect to 
$\tb$ for case~(I) and case~(II) defined in
\refse{subsec:higgsscenarios}. 
Case~(III), where the LEP excess is interpreted as a signal of the
heavier $\cp$-even
Higgs boson in the MSSM, is not realized in the mAMSB scenario (as will
be explained below).
\reffi{fig15AMSB} shows the allowed parameter space in the $\MA-\tb$
plane.

We find the highest values for $\mh$ at $\tb$ values of about 35. The
experimental and theoretical constraints discussed in
\refse{subsec:constraints} reduce the upper bound on $\mh$ by 1--2~GeV.
In the mAMSB scenario we do not find a significant suppression of 
$\sin^2(\be - \aeff)$ (i.e.\ the $hZZ$ coupling). As can be seen in
\reffi{fig15AMSB}, values of $\MA$ below 300~GeV are only realized in
the interval $30 \le \tb \le 40$, and $\MA$ stays always above about 
150~GeV,
giving thus rise to a SM like $hZZ$ coupling. 
As a consequence, the LEP2 Higgs boson searches
exclude all models with $\mh \lsim 113 \gev$. This affects mainly the
region $\tb \lsim 10$, while for larger values of $\tb$ hardly any mAMSB
model results in a Higgs boson mass lower than $113 \gev$, see
\reffi{fig11AMSB}.
Values larger than $\tb > 60$ are not allowed due to the REWSB constraint.

\reffi{fig15AMSB} furthermore shows that the experimental and
theoretical constraints discussed in \refse{subsec:constraints} 
exclude a significant fraction of the parameter space in the $\MA-\tb$ 
plane. In general larger values of $\MA$ correspond
to smaller $\tb$ (giving rise to a smaller
bottom Yukawa coupling $y_b$) because $m_{H_d}^2$, 
which contributes mainly to $\MA^2$, is larger at low energies due to the 
RGE running.  
For larger $\tb$, smaller $\MA$ cannot be realized. Otherwise the
corresponding $m_0$ would be too small to avoid negative slepton masses.
The relatively large values required for $\MA$ in the mAMSB scenario, on
the other hand, exclude the possibility of case (III).

The upper bound on the light $\cp$-even Higgs boson mass in the mAMSB 
scenario is (for $\mt = 175$~GeV)
\begin{equation}
\mhmax  \simleq 122~{\rm GeV} \;\;\; {\rm (mAMSB)} \;.
\end{equation}
Values close to this upper limit on $m_h$ are reached in a large region
of moderate and large values of $\tb$, $\tb \gsim 10$.

A lower bound on $\tb$ is inferred in the mAMSB scenario (for $\mt =
175$~GeV),
\BE
\tb \gsim 3.2 \;\;\; {\rm (mAMSB)} .
\EE
 
As above, we have analyzed the allowed values of the mixing
parameter in the scalar top sector, $\Xt$. \reffi{fig15xAMSB} shows the
allowed parameter space in the plane of the heavier $\Stop$~mass,
$\mstz$, and $\Xt$. In contrast to the mSUGRA and the mGMSB scenarios
positive values for $\Xt$ are preferred. 
The experimental and theoretical constraints discussed in
\refse{subsec:constraints} are seen to have a significant effect,
limiting the allowed values of $\Xt$ to $\Xt \lsim 1.5$~TeV.
\reffi{fig15xAMSB} shows that in the mAMSB scenario $\Xt$ is bounded
from above, $\Xt \lsim \mstz$. This is the main reason for the decrease
in the upper bound on $\mh$ compared to the unconstrained MSSM, since
the highest values for $\mh$ are reached for values of $\Xt$
significantly larger than the heavier $\Stop$~mass (see e.g.\
\citeres{mhiggslong,mhiggslle}).

\begin{figure}[ht!]
\begin{center}
\epsfig{figure=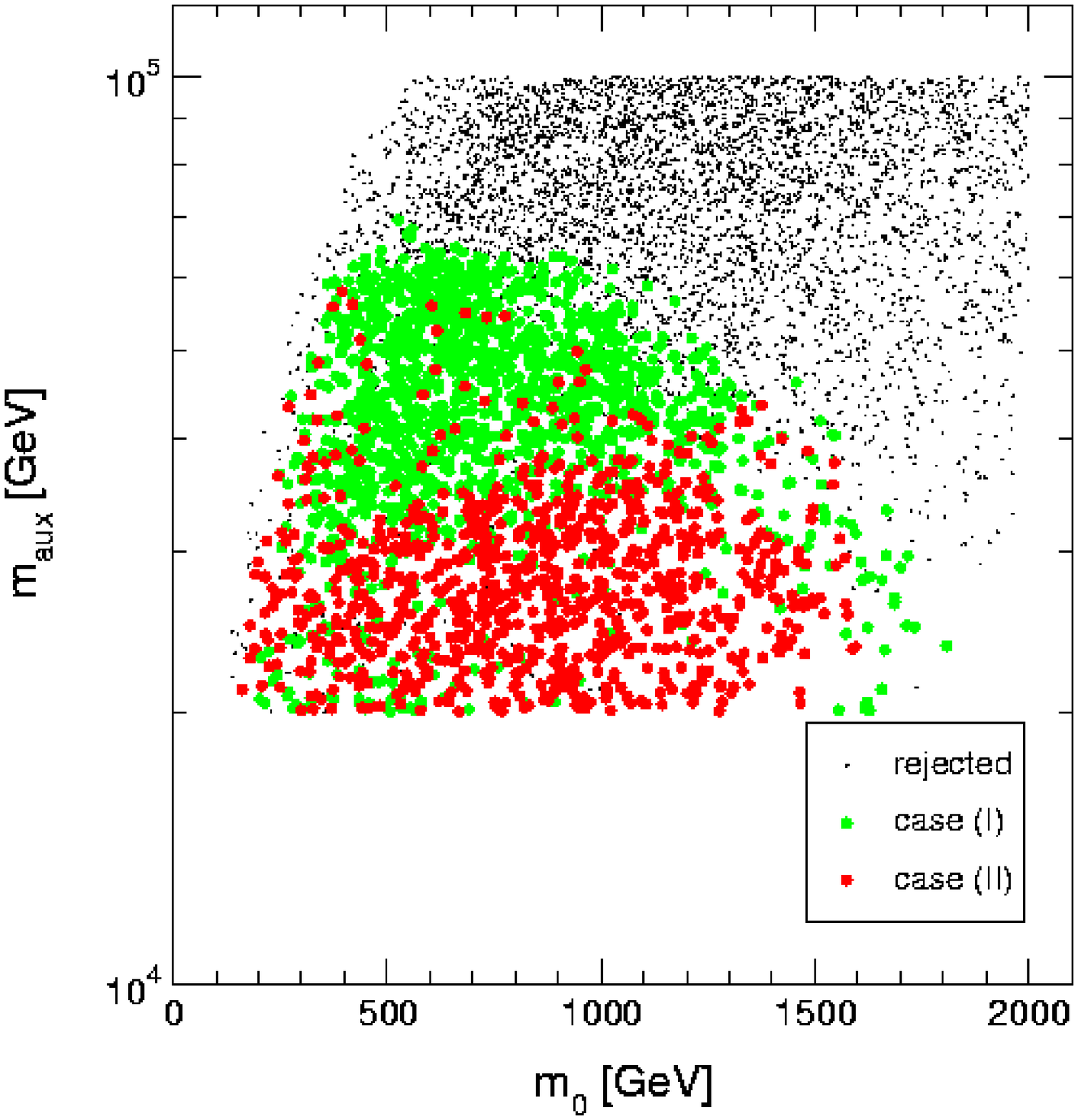,width=12cm,height=9.5cm}
\caption{
Cases (I) and (II) in the mAMSB scenario are shown 
in the $m_0-m_{\rm aux}$ plane.
}
\label{fig16AMSB}
\end{center}
\end{figure}
%
\begin{figure}[ht!]
\begin{center}
\epsfig{figure=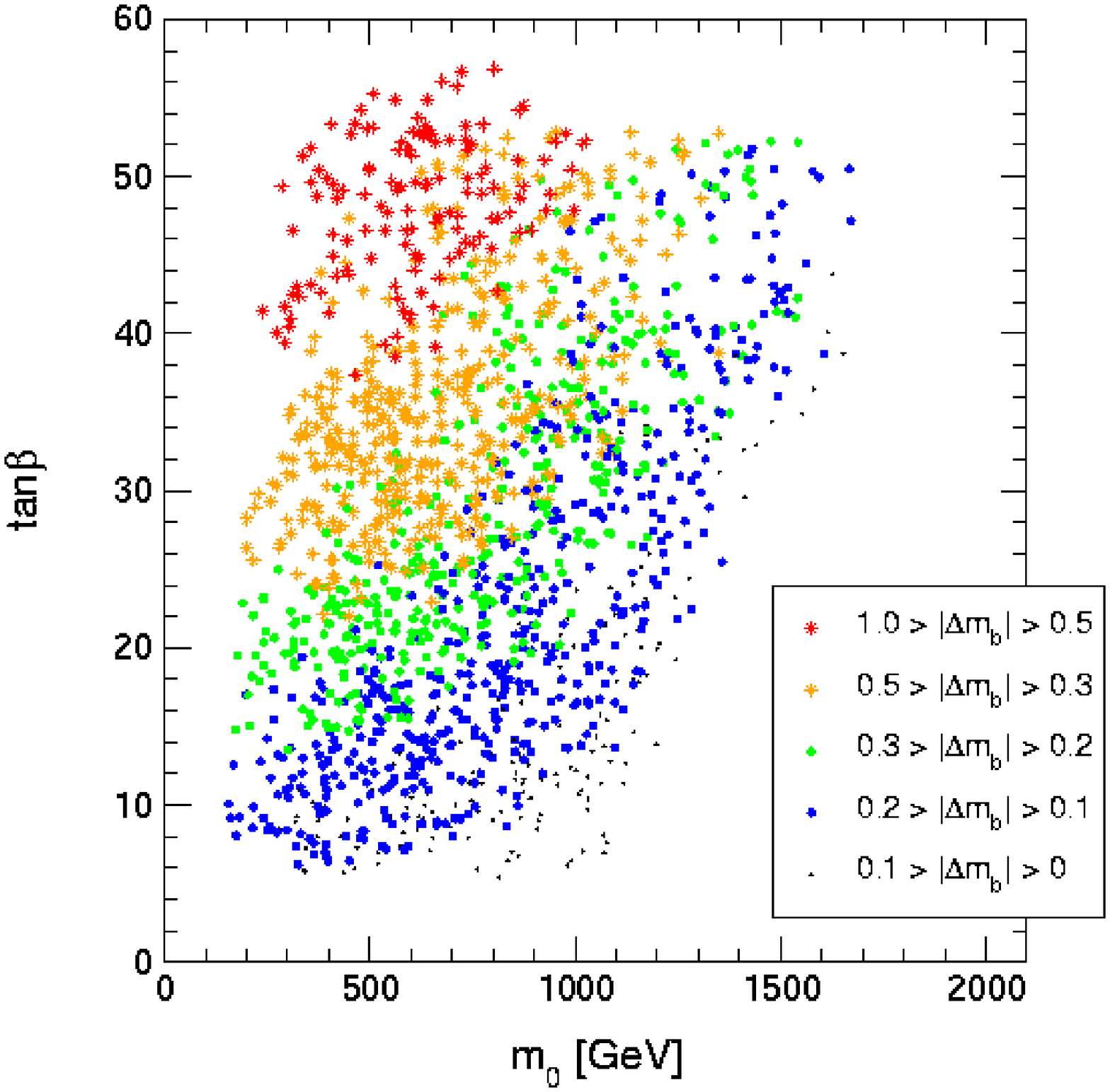,width=12cm,height=9.5cm}
\caption{The allowed values for the quantity $|\De m_b|$ (see
\refeq{eq:Delmb}) for different values of $m_0$ and $\tb$ in the
mAMSB scenario.
}
\label{fig56AMSB}
\end{center}
\end{figure}

In \reffi{fig16AMSB} we present the allowed regions in the plane of
the high energy input parameters $m_0$ and $m_{\rm aux}$.
The experimental and theoretical constraints
imposed in our analysis affect in particular the region of large $m_0$
and $m_{\rm aux}$. We find no allowed models with $m_{\rm aux} \gsim 70$~TeV.
On the other hand, even the smallest values of $m_0$ giving rise to
acceptable slepton masses lead to allowed parameter points in case~(I)
and case~(II).

Concerning the $hb\bar b$ coupling within the mAMSB scenario, we have
analyzed the possible values for $\sin^2\aeff/\cos^2\be$ and $\De m_b$.
We do not find any models with $\sin^2\aeff/\cos^2\be$ $\lsim 0.9$ (values
of $\sin^2\aeff/\cos^2\be < 1$ only occur for $\tb \gsim 40$), i.e.\ the
SUSY contributions entering via $\aeff$ do not give rise to a
significant reduction of the $h \to b \bar b$ decay rate in the mAMSB
scenario. Values of $\sin^2\aeff/\cos^2\be > 10$ are possible for
large $\tb$.

The quantity $\De m_b$ (see \refeq{eq:Delmb}) receives large contributions 
in the mAMSB scenario, in particular for large $\tb$ and relatively
small $m_0$. This is shown in
\reffi{fig56AMSB}, where the different values for $|\De m_b|$ are
indicated in the $m_0-\tb$ plane. We find that positive values of $\De m_b$, 
leading to a suppression of the $hb\bar b$ coupling, are bounded from
above by $\De m_b \lsim 0.5$. On the other hand, we obtain negative
contributions as large as $\De m_b \approx - 0.8$, giving rise to a
strongly enhanced $hb\bar b$ Yukawa coupling.


\subsection{The SUSY mass spectra in the three SUSY-breaking scenarios
compatible with a possible Higgs signal at LEP}
\label{subsec:numanalspectrum}

\begin{figure}[htb!]
\begin{center}
\mbox{}\vspace{2cm}

\epsfig{figure=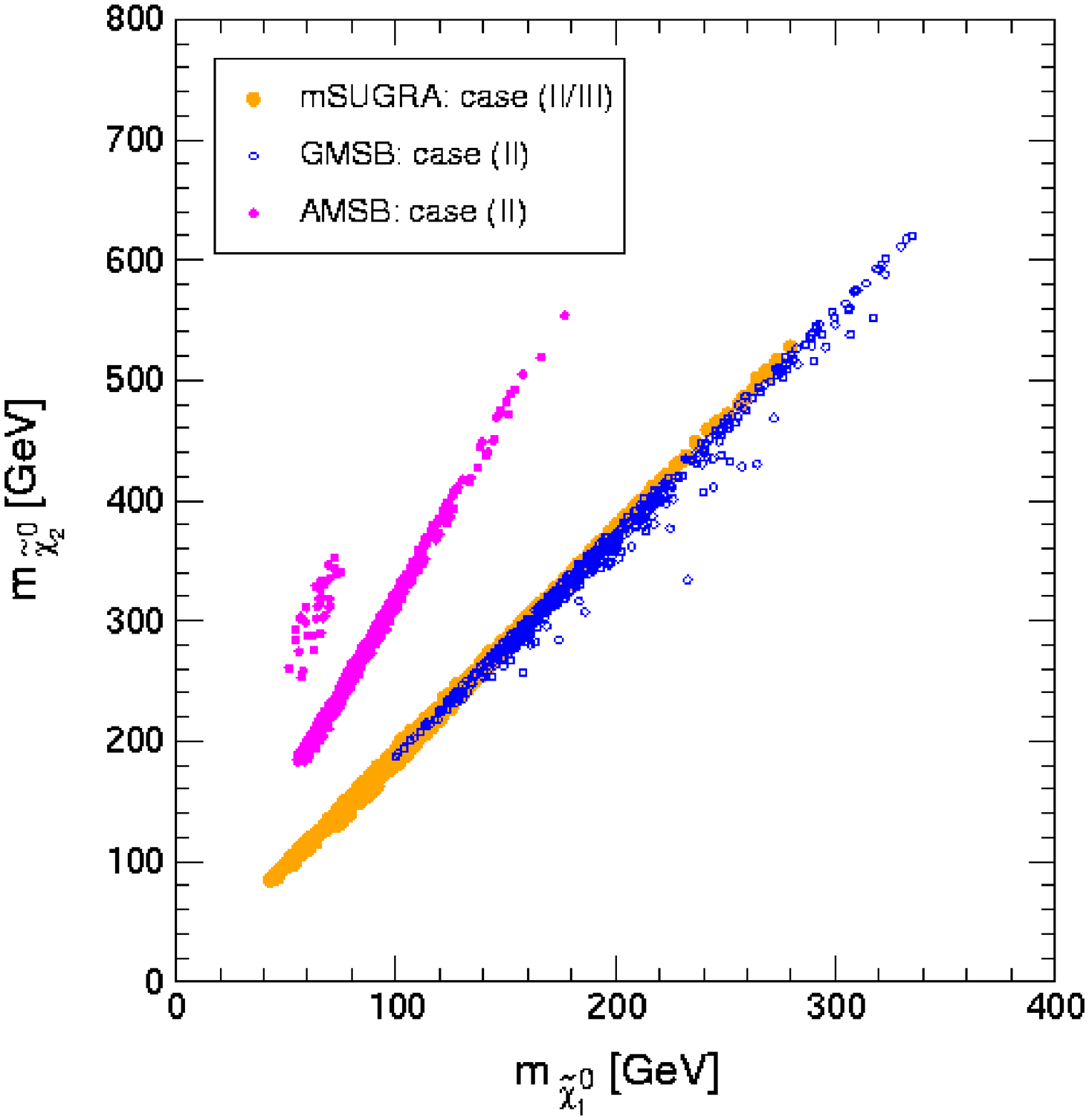,width=7.2cm,height=7.6cm}
\hspace{1cm}
\epsfig{figure=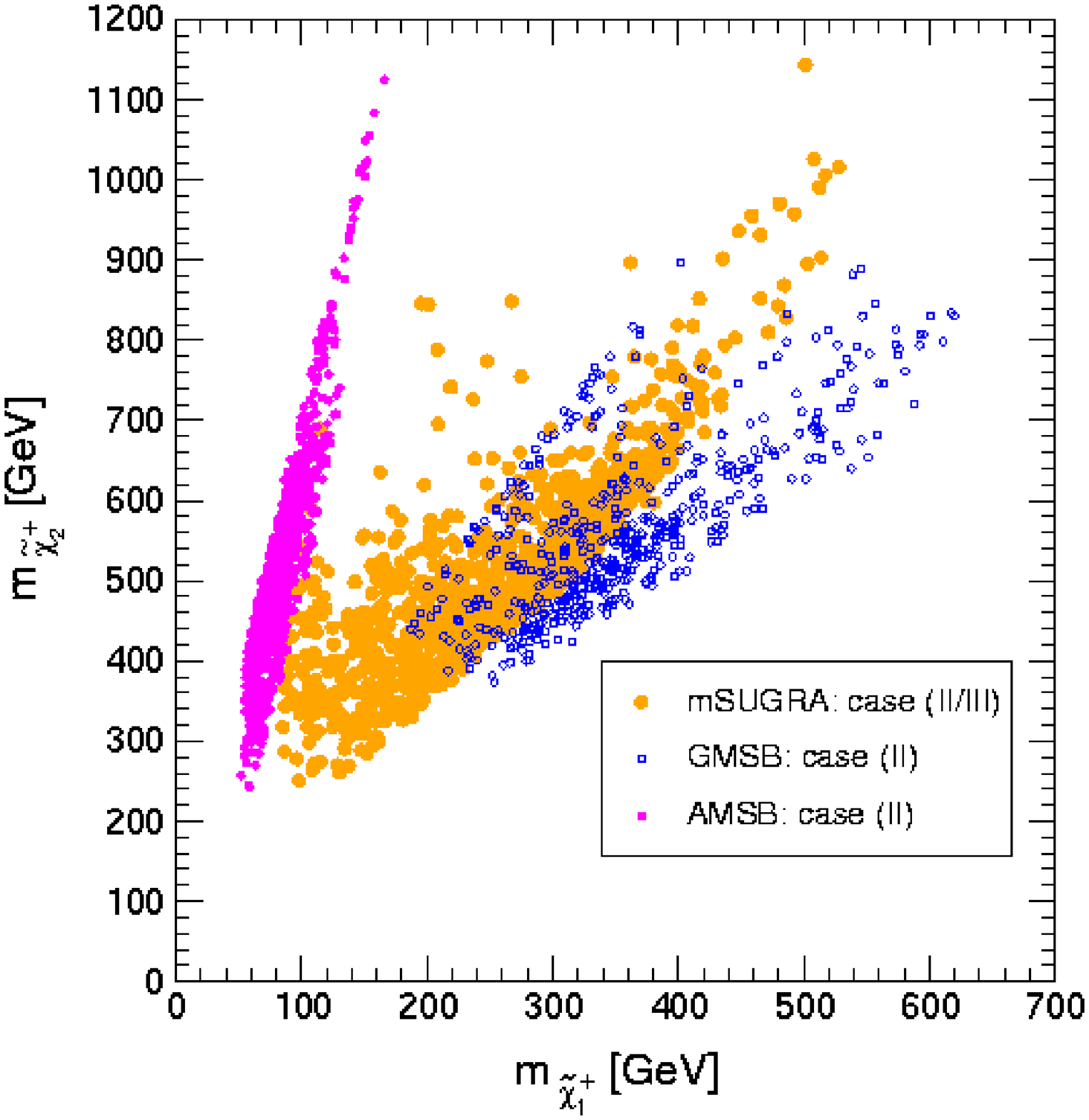,width=7.2cm,height=7.6cm}
\vspace{1.5cm}

\epsfig{figure=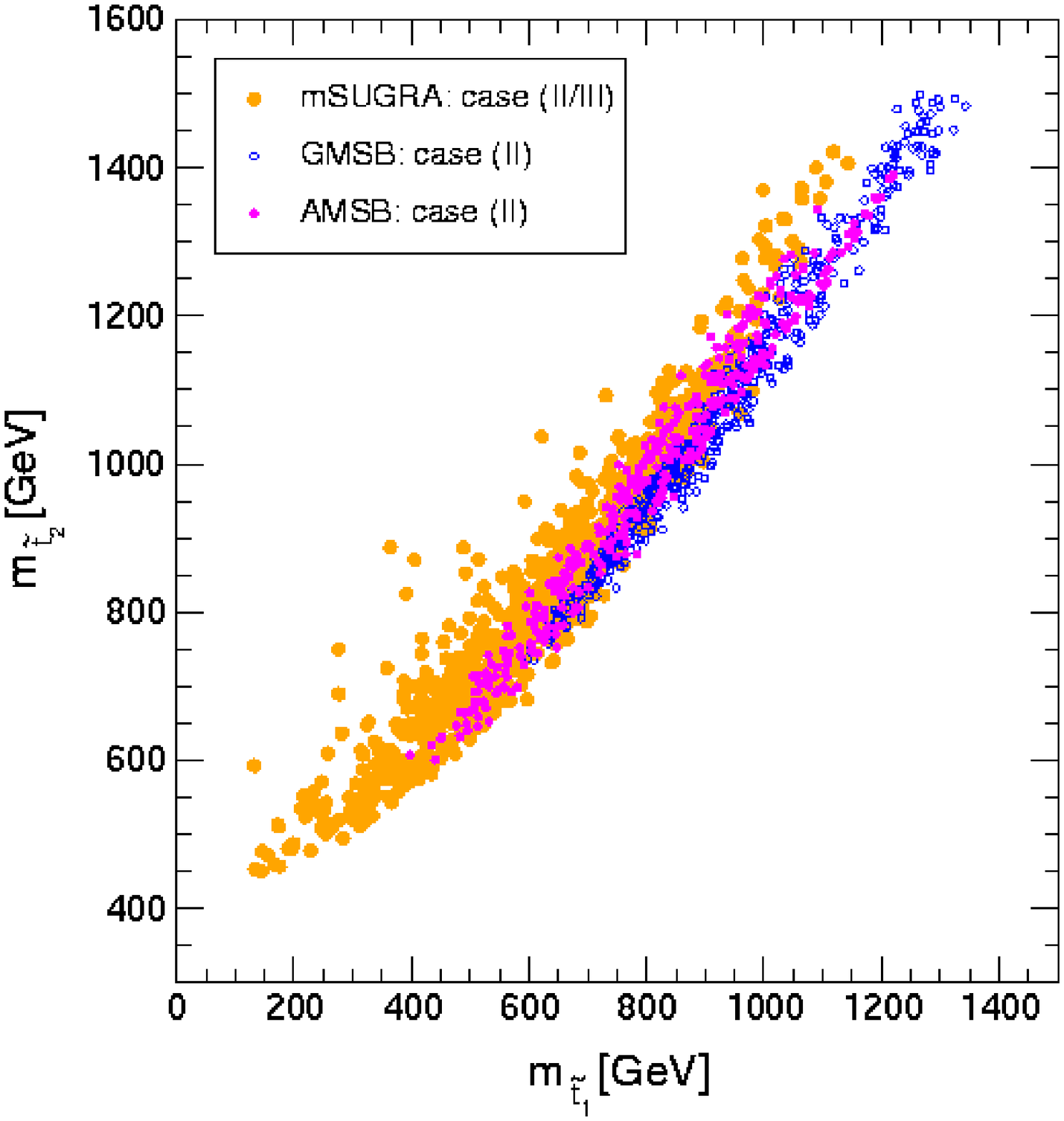,width=7.2cm,height=7.6cm}
\hspace{1cm}
\epsfig{figure=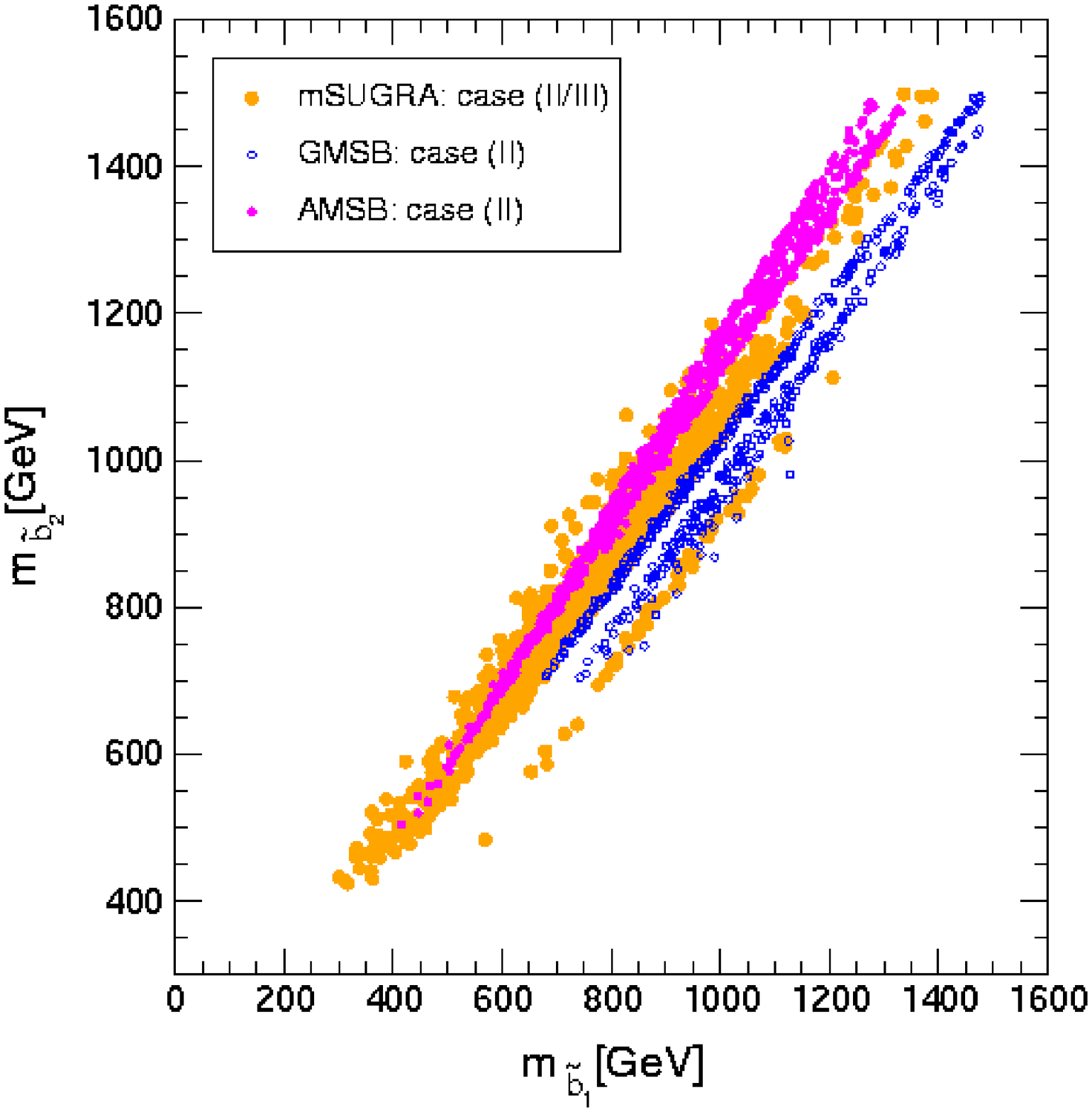,width=7.2cm,height=7.6cm}
\vspace{0.5cm}
\caption{
The allowed mass ranges for the lightest neutralinos (upper left plot), 
the charginos (upper right), the scalar top quarks (lower left) and the
scalar bottom quarks (lower right) are shown for the cases (II) and
(III) of the mSUGRA, mGMSB and mAMSB scenarios.
}
\label{fig6:massspectraA}
\end{center}
\end{figure}

\begin{figure}[ht!]
\begin{center}
\mbox{}\vspace{2cm}

\epsfig{figure=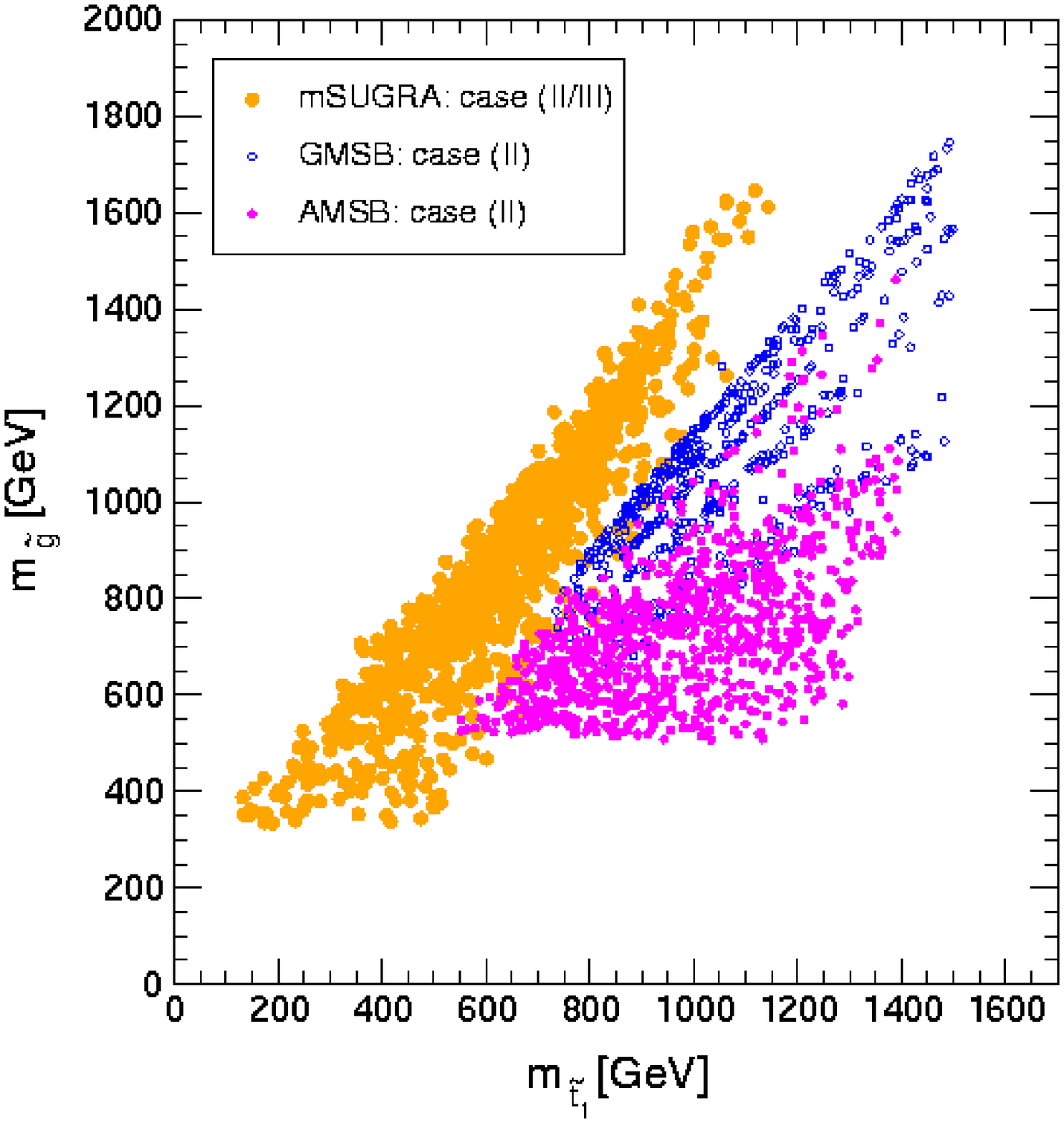,width=7.2cm,height=7.7cm}
\hspace{1cm}
\epsfig{figure=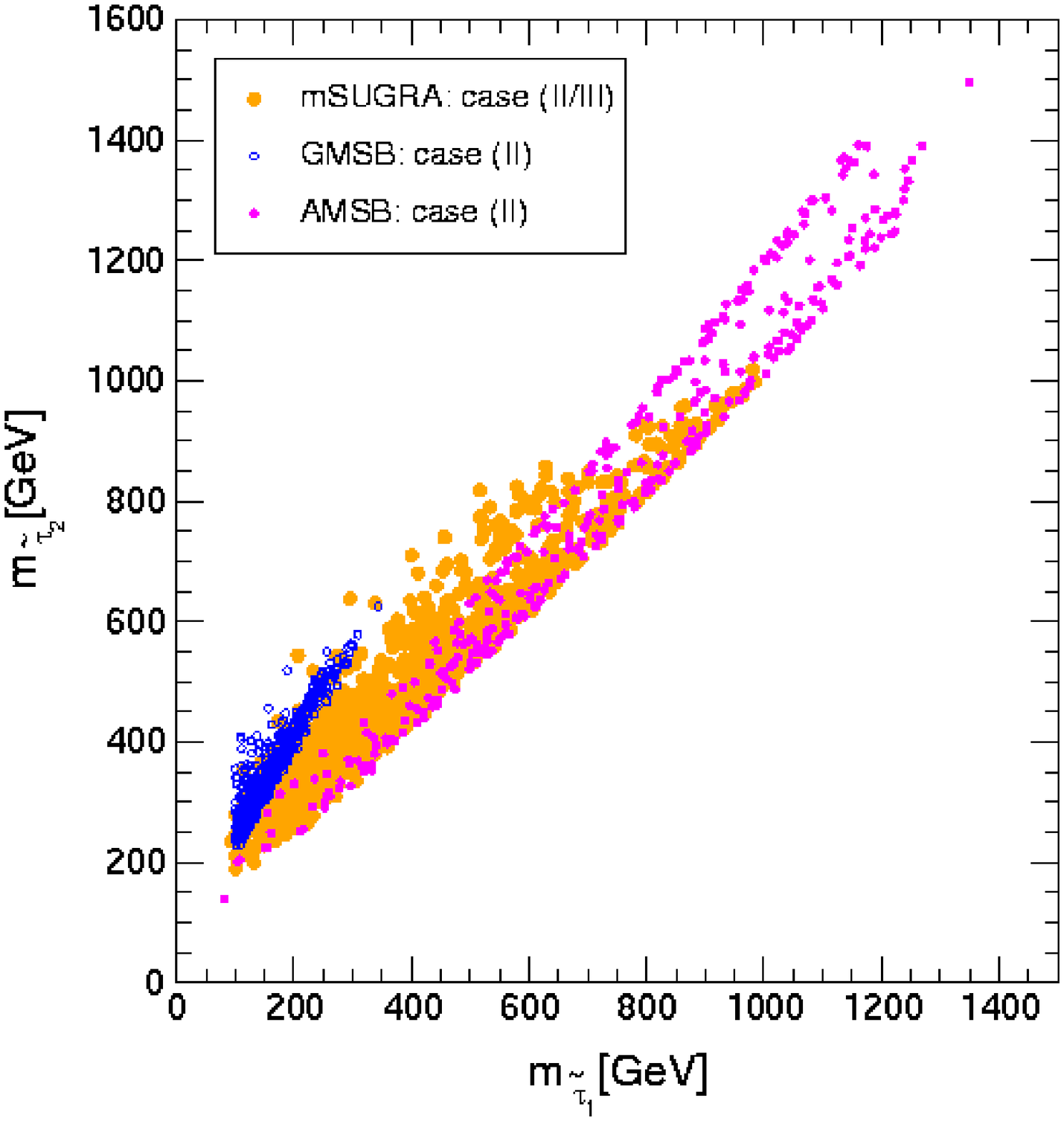,width=7.2cm,height=7.7cm}
\vspace{0.5cm}
\caption{
The allowed mass ranges for the gluino (left plot, shown in the
$\mste-\mgl$ plane) and for the scalar $\tau$ leptons (right plot)
are shown for the cases (II) and (III) of the
mSUGRA, mGMSB and mAMSB scenarios.
}
\label{fig6:massspectraB}
\end{center}
\end{figure}

We finally compare the mass spectra in the three soft SUSY-breaking
scenarios assuming that the LEP excess is due to the production of the
$h$ or $H$ boson in the MSSM (cases (II) and (III))
and briefly discuss possible implications for SUSY searches at
the next generation of colliders%
\footnote{
Phenomenological differences as well as characteristic signatures at
future experiments between the three models have
also recently discussed in \cite{hinchrich}.
}%
 . In 
\reffis{fig6:massspectraA},~\ref{fig6:massspectraB} we show the spectra
of the lightest neutralinos, the charginos, the scalar top and bottom
quarks, the scalar $\tau$ leptons and of the gluino in the mSUGRA, mGMSB
and mAMSB scenarios. The points shown for the mGMSB and mAMSB scenarios
correspond to case (II), while for the mSUGRA models we do not
distinguish in these models between cases (II) and (III) (in general
case~(III) results in about the same mass ranges as case~(II)). 

For the lightest neutralino, in the mSUGRA and mAMSB scenarios values as
low as $m_{\tilde\chi_1} \approx 50$~GeV are compatible with case~(II), while
in the mGMSB scenario the lower bound of 
$m_{\tilde\chi_1} \gsim 100 \gev$ holds in accordance with 
\refeq{susyexp}. The upper bounds on $m_{\tilde\chi_1}$ in 
\reffi{fig6:massspectraA}
are about 200, 300 and 350~GeV in the mAMSB, mSUGRA and mGMSB scenario,
respectively. For $m_{\tilde\chi_2}$, values as low as about 100~GeV are
possible in the mSUGRA scenario, while we find upper bounds between about
550 and 650~GeV in the three scenarios.

The lightest chargino is bounded from above by about 200~GeV in case~(II)
for the mAMSB scenario, by about 550~GeV in the mSUGRA scenario, and by about
650~GeV in the mGMSB scenario. For $m_{\tilde\chi^+_2}$ we find a lower
bound of about 250~GeV in the mSUGRA and mAMSB scenarios, while the
lower bound in the mGMSB scenario is about 350~GeV.

Since within the GMSB scenario the LSP is always the gravitino,
detection of the lightest neutralino via $\neone\neone$ production is
possible in this scenario if $\sqrt{F}$ is not too large, while the search 
in the mSUGRA and mAMSB scenarios has to focus on $\neone\netwo$ associated 
production.

Within the mSUGRA scenario, the neutralino and chargino searches at
Run~II of the Tevatron and the LHC will be sensitive to a significant
part of the parameter space of the models shown in
\reffi{fig6:massspectraA}~\cite{tev:susysearch,atlastdr,cms}. 
A future \epem\ linear collider (LC) with
a center of mass (CMS) energy of $\sqrt{s} \lsim 1$~TeV will have a very
good chance to observe both the associated production of
$\neone\,\netwo$ and the production of the lightest chargino,
$\chone\,\chonem$~\cite{teslatdr,orangebook}.

In the mGMSB scenario the discovery potential at the next generation of
colliders for gauginos (and to some extent also for squarks) strongly 
depends on the lifetime and other properties of the NLSP. 
The charginos are in general heavier than the two lightest neutralinos,
following the mass relation originating from the condition of a unified
gaugino mass at the high energy scale,
$m_{\chone} \approx 2\,m_{\neone} \approx m_{\netwo}$.
The Tevatron and the LHC will cover at least part of the parameter space in
\reffi{fig6:massspectraA}, and the prospects at a LC for neutralino and
chargino production in the GMSB scenario are very promising.

A peculiar feature of the mAMSB scenario is that the Wino is always lighter 
than the Bino~\cite{wells,feng}:
\BE
M_1\, :\, M_2\, :\, M_3\, \approx\, 2.8\, :\, 1\, :\, -8.3 .
\EE
In most of the parameter space, the neutral Wino is the LSP. The NLSP,
the charged Wino, is generically extremely 
mass degenerate with the LSP and decays, after its production at a 
collider, after centimeters into an LSP plus 
a very soft lepton or pion.  The detection of such a charged Wino poses
novel experimental challenges since such events escape
the conventional triggers.  The search for SUSY
in the Wino LSP scenario has been studied by several 
groups~\cite{winotev,wells,winolsp}.
It was pointed out in \citere{winotev} that hundreds of Wino pairs 
can be produced at Run~II of the Tevatron with $\sqrt{s}=2$~TeV and
$\cL = 2\, {\rm fb}^{-1}$. Tens of Wino pairs can be produced in
association 
with a jet.  The accompanying high energy jet works as a trigger and 
the detection of 5 events is possible for Wino masses up to $180 \gev$ at the
Tevatron. At an LC with a CMS energy of $\sqrt{s} \lsim 1$~TeV both
the associated production of $\neone \netwo$ and of $\chone\,\chonem$
should be observable.

We now turn to the mass spectra of the third generation squarks
compatible with cases~(II) and (III). Within the mSUGRA scenario the
lowest mass values for the third generation squarks are possible, 
around 150~GeV and 450~GeV for $\mste$ and $\mstz$, respectively, and
around 300~GeV and 450~GeV for $\msbe$ and $\msbz$, respectively, see
\reffi{fig6:massspectraA}.
Within the GMSB scenario, the low-energy masses generated for colored
particles are $\sim \als$, but only $\sim g_2$ for uncolored particles.
Thus, scalar quarks are in general heavier than sleptons or electroweak
gauginos in this scenario.
\reffi{fig6:massspectraA} shows that
in the mGMSB and mAMSB scenarios no third generation squark below about 
400~GeV is possible in case~(II). Similarly, a gluino as light as about 300~GeV
is possible in the mSUGRA scenario, while the lower bound on $\mgl$ is
about 200~GeV higher in the mGMSB and mAMSB scenarios, see
\reffi{fig6:massspectraB}. As a consequence, the prospects for the
production of third generation squarks and the gluino in these scenarios
are not very high both at Run~II of the Tevatron and a LC with a CMS
energy of $\sqrt{s} \lsim 1$~TeV. At the LHC, on the other hand,
the production of the third generation squarks and the gluino is
guaranteed (it should be noted, however, that this conclusion relies on
the ``naturalness bound'' imposed in our analysis, see 
\refse{subsubsec:phenorest}).

Finally we analyze the mass spectrum of the scalar $\tau$ leptons in the
three scenarios, see \reffi{fig6:massspectraB}. In the mGMSB scenario,
stringent upper limits on the scalar $\tau$ masses of 
$m_{\tilde\tau_1} \lsim 400$~GeV and $m_{\tilde\tau_2} \lsim 600$~GeV
apply, giving rise to good prospects for production of scalar $\tau$
at the LHC and a future LC (at Run~II of the Tevatron a discovery reach
of only up to $\sim 150 \gev$ is expected~\cite{tev:susysearch}). 
In the mSUGRA and the mAMSB
scenarios, on the other hand, much larger masses of the scalar $\tau$
leptons are possible, and their discovery in the scenario considered
here is not guaranteed at the next generation of colliders.


\section{Conclusions}

We have analyzed the three most prominent soft SUSY-breaking
scenarios, mSUGRA, mGMSB and mAMSB, regarding their phenomenology in the
Higgs sector. We have discussed the constraints arising from the
exclusion limits in the Higgs sector recently obtained at LEP and
the possible implications in the situation where the excess of events
observed at LEP is interpreted as a signal of the light or heavy neutral
$\cp$-even MSSM Higgs boson with mass of about 115~GeV.

In order to obtain the predictions for the Higgs sector in the
three scenarios, we have combined the \twol\ RG calculations, employed to
derive the low-energy mass spectrum from the high-energy parameters
in mSUGRA, mGMSB and mAMSB, with Feynman-diagrammatic results up to
two-loop order for the Higgs boson spectrum and the effective mixing
angle in the Higgs sector.
The possible values for the low-energy mass spectrum have been obtained
by
scanning over the fundamental (high-energy) parameters in the scenarios.
In addition to the constraints on the MSSM Higgs sector from the Higgs
search at LEP, which is the main issue in this paper, we have taken into
account some further phenomenological constraints on the low-energy mass
spectrum. The lower bounds on SUSY particle masses
obtained from the searches at LEP2 and the Tevatron have been
incorporated as well as the constraints from electroweak
precision observables. We have assumed $\cp$-invariance and conservation
of $R$-parity and have discarded models giving rise to
charge and color breaking minima in the scalar potential or violating
the condition for radiative electroweak symmetry breaking. We have
furthermore imposed a mild ``naturalness'' upper bound on the masses of
the scalar quarks and the gluino of 1.5--2~TeV.

As upper bound on the mass of the lightest $\cp$-even Higgs boson (for
$\mt = 175$~GeV) we have found $\mh \lsim 124, 119$ and $122 \gev$
in the mSUGRA, mGMSB and mAMSB scenario, respectively. In these
scenarios the $\tb$ values are excluded up to $\tb \gsim 3.3, 4.6$ and
3.2, respectively. The upper bound on $\mh$ in the three scenarios is
significantly reduced compared to the unconstrained MSSM. This decrease
in the upper bound on $\mh$ is in particular related to the restrictions
imposed on the mixing in the scalar top sector by the underlying
structure of the three scenarios. We have furthermore investigated the
Higgs couplings to vector bosons and fermions (in particular the $b$
quark). Using these results, we have discussed in which parameter
regions a non SM-like behavior of the Higgs production and decay
processes is possible in the three scenarios.

The set of models that passed all constraints (called ``case~(I)'' in
our terminology),
was then further analyzed in view of whether they permit the
interpretation of the excess of events observed at LEP2
as a signal of the light $\cp$-even Higgs boson (``case~(II)'') or the
heavy
$\cp$-even Higgs boson (``case~(III)'') with a mass of $115 \pm 2 \gev$
and SM-like couplings.

While the interpretation of the LEP excess as production of the light
$\cp$-even Higgs boson is possible in all three scenarios, the
interpretation as a signal of the heavy $\cp$-even Higgs boson is only
possible in the mSUGRA scenario in a small parameter region with
$50 \lsim \tb \lsim 55$, which is constrained from the Higgs
search results of Run~I of the Tevatron and is also close to the region
where no radiative electroweak symmetry breaking occurs.

Assuming the interpretation of the LEP excess as a Higgs signal in the
MSSM (according to cases (II) and (III)), we have analyzed the
restrictions on the parameter space of the fundamental parameters in the
three scenarios. We have furthermore investigated the corresponding
spectra of the SUSY particles in these scenarios in view of the SUSY
searches at Run~II of the Tevatron, the LHC and an \epem\ LC with center 
of mass energy of up to 1~TeV. 
While for the scenario studied here the Tevatron has only a limited chance 
to observe SUSY particles,
the LHC can always cover the scalar tops and bottoms (the latter is
related to the naturalness condition imposed on the models in our
analysis).
A LC offers very good prospects for gaugino and slepton production.
We find that at least a part of the gaugino and slepton spectrum
should be accessible at at LC with center of mass energy of up to 1~TeV
in all three scenarios.


\section*{Acknowledgements}
We thank K.~Olive for helpful communications concerning the comparison 
of our results. We furthermore thank L.~Covi for helpful discussion on
astro-physical constraints. 
S.S.\ has been supported by the DOE grant DE-FG03-92-ER-40701. 
A.D.\ would like to acknowledge financial support from the
Network RTN European Program HPRN-CT-2000-0014
``Physics Across the Present Energy Frontier: Probing the Origin of
Mass''.



\begin{thebibliography}{99} 


\bibitem{susylighthiggs} G.~Kane, C.~Kolda and J.~Wells, 
                         {\em Phys. Rev. Lett.} {\bf 70} (1993) 2686, 
                         hep-ph/9210242;\\ 
                         J.~Espinosa and M.~Quir\'os, 
                         {\em Phys. Lett.} {\bf B 302} (1993) 51,
                         hep-ph/9212305. 

\bibitem{LEPHiggs} ALEPH collaboration, R.~Barate {\it et al.}, 
                   {\em Phys.\ Lett.} {\bf B 495} (2000) 1, 
                   hep-ex/0011045;\\
                   L3 collaboration, M.~Acciarri {\it et al.}, 
                   {\em Phys.\ Lett.} {\bf B 495} (2000) 18, 
                   hep-ex/0011043;\\
                   DELPHI collaboration, P. Abreu {\it et al.},
                   {\em Phys.\ Lett.} {\bf B 499} (2001) 23, 
                   hep-ex/0102036;\\
                   OPAL collaboration, G. Abbiendi {\it et al.}, 
                   {\em Phys.\ Lett.} {\bf B 499} (2001) 38, 
                   hep-ex/0102036.\\
For a preliminary compilation of the LEP data presented on Nov. 3rd,
2000, see:\\
P. Igo-Kemenes, for the LEP Higgs working group,\\
{\tt lephiggs.web.cern.ch/LEPHIGGS/talks/index.html}.

\bibitem{vacstabil} G.~Altarelli and G.~Isidori, 
                    {\em Phys. Lett.} {\bf B 337} (1994) 141;\\ 
                    J.A.~Casas, J.R.~Espinosa and M.~Quir\'os,
                    {\em Phys. Lett.} {\bf B 342} (1995) 171,
                    hep-ph/9409458;\\
                    T.~Hambye and K.~Riesselmann, hep-ph/9708416;\\
                    G.~Isidori, G.~Ridolfi and A.~Strumia,
                    hep-ph/0104016.

\bibitem{mhiggslong} S.~Heinemeyer, W.~Hollik and G.~Weiglein, 
                     {\em Eur. Phys. Jour.} {\bf C 9} (1999) 343, 
                     hep-ph/9812472.

\bibitem{mhiggsletter} S.~Heinemeyer, W.~Hollik and G.~Weiglein, 
                       {\em Phys. Rev.} {\bf D 58} (1998) 091701, 
                       hep-ph/9803277; 
                       {\em Phys. Lett.} {\bf B 440} (1998) 296, 
                       hep-ph/9807423.

\bibitem{mhiggsRG1a} M.~Carena, J.~Espinosa, M.~Quir\'os and C.~Wagner, 
                     {\em Phys. Lett.} {\bf B 355} (1995) 209, 
                     hep-ph/9504316.
 
\bibitem{mhiggsRG1b} M.~Carena, M.~Quir\'os and C.~Wagner, 
                     {\em Nucl. Phys.} {\bf B 461} (1996) 407, 
                     hep-ph/9508343.

\bibitem{mhiggsRG2} H.~Haber, R.~Hempfling and A.~Hoang, 
                    {\em Z. Phys.} {\bf C 75} (1997) 539, 
                    hep-ph/9609331.

\bibitem{mhiggsEP1} R.~Hempfling and A.~Hoang, 
                    {\em Phys. Lett.} {\bf B 331} (1994) 99, 
                    hep-ph/9401219.

\bibitem{mhiggsEP2} R.-J.~Zhang, 
                    {\em Phys. Lett.} {\bf B 447} (1999) 89, 
                    hep-ph/9808299.

\bibitem{maulpaul} J.~Espinosa and R.~Zhang,
                   {\em Nucl. Phys.} {\bf B 586} (2000) 3,
                   hep-ph/0003246. 

\bibitem{maulpaul2} J.~Espinosa and I.~Navarro, hep-ph/0104047.

\bibitem{mhiggsEP3} G.~Degrassi, P.~Slavich and F.~Zwirner,
                    hep-ph/0105096.

\bibitem{ellisross} J.~Ellis and D.~Ross, 
                    {\em Phys. Lett.} {\bf B 506} (2001) 331, 
                    hep-ph/0012067.

\bibitem{benchmark} M.~Carena, S.~Heinemeyer, C.~Wagner and 
                    G.~Weiglein, hep-ph/9912223.

\bibitem{Hall} H.P.~Nilles,
               {\em Phys. Lett.} {\bf B 115} (1982) 193;
               {\em Nucl. Phys.} {\bf B 217} (1983) 366.\\
               A.H.~Chamseddine, R.~Arnowitt and P.~Nath,
               {\em Phys. Rev. Lett.}  {\bf 49} (1982) 970;\\
               R.~Barbieri, S.~Ferrara and C.A.~Savoy,
               {\em Phys. Lett.} {\bf B 119} (1982) 343;\\
               H.P.~Nilles, M.~Srednicki and D.~Wyler,
               {\em Phys. Lett.} {\bf B 120} (1983) 346.\\
               E.~Cremmer, P.~Fayet and L.~Girardello,
               {\em Phys. Lett.} {\bf B 122} (1983) 41.\\
               S.~Ferrara, L.~Girardello and H.P.~Nilles,
               {\em Phys. Lett.} {\bf B 125} (1983) 457.
               L.~Hall, J.~Lykken and S.~Weinberg,
               {\em Phys. Rev.} {\bf D 27} (1983) 2359;\\
               S.K.~Soni and H.A.~Weldon,
               {\em Phys. Lett.} {\bf B 126} (1983) 215.\\
               R.~Arnowitt, A.H.~Chamseddine and P.~Nath,
               {\em Nucl. Phys.} {\bf B 227} (1983) 121.


For more details see, S.~Weinberg,
``The quantum theory of fields.  Vol. 3: Supersymmetry,''
{\it  Cambridge University Press (2000)}. 

\bibitem{mSUGRArev} For reviews  see also, \\
                    H.-P.~Nilles,
                    {\em Phys. Rept.}  {\bf 110} (1984) 1; \\
                    H.E.~Haber and G.L.~Kane,
                    {\em Phys. Rept.}  {\bf 117} (1985) 75; \\
                    A.B.~Lahanas and D.V.~Nanopoulos,
                    {\em Phys. Rept.}  {\bf 145} (1987) 1; \\
                    S.P.~Martin, 
                    in ``Perspectives on supersymmetry'', ed. G.~Kane,
                    hep-ph/9709356 ({\tt zippy.physics.niu.edu/primer.shtml}).

\bibitem{GR-GMSB}
  For a review, see G.F.~Giudice and R.~Rattazzi, 
                    \PREP{322}{1999}{419}, hep-ph/9801271 .


\bibitem{lr} L.~Randall and R.~Sundrum, 
             {\em Nucl. Phys.} {\bf B 557} (1999) 79, 
             hep-th/9810155 .

\bibitem{giudice} G.F.~Giudice, M.A.~Luty, H.~Murayama and
                  R.~Rattazzi, 
                  {\em JHEP} {\bf 9812} (1998) 027, 
                  hep-ph/9810442 .

\bibitem{wells} T.~Gherghetta, G.F.~Giudice, J.D.~Wells,
                {\em Nucl. Phys.} {\bf B 559} (1999) 27, 
                hep-ph/9904378 .

\bibitem{GMSBmodels2}
    J.A.~Bagger, K.~Matchev, D.M.~Pierce, R.~Zhang, 
    \PRD{55}{1997}{3188}, 
    hep-ph/9609444 .

\bibitem{higgsmsugra2} K.T.~Matchev and D.M.~Pierce,
                       {\em Phys. Lett.} {\bf B 445} (1999) 331,
                       hep-ph/9805275;\\
                       W.~de~Boer,
                       hep-ph/9808448.

\bibitem{higgsgmsb2} T.A.~Kaeding and S.~Nandi,
                     hep-ph/9906342.

\bibitem{higgsamsb} S.~Su, 
                    {\em Nucl. Phys.} {\bf B 573} (2000) 87,
                    hep-ph/9910481.

\bibitem{higgsmsugra} A.~Dedes, S.~Heinemeyer, P.~Teixeira-Dias 
                      and G.~Weiglein, 
                      {\em Jour. Phys.}  {\bf G 26} (2000) 582,
                      hep-ph/9912249.

\bibitem{higgsgmsb} S.~Ambrosanio, S.~Heinemeyer and G.~Weiglein,
                    in hep-ph/0002191 
                    and hep-ph/0005142.

\bibitem{higgsmsugra3} J.~Ellis, G.~Ganis, D.V.~Nanopoulos and K.A.~Olive,
                       {\em Phys. Lett.} {\bf B 502} (2001) 171,
                       hep-ph/0009355.

\bibitem{bse} M.~Carena, H.~Haber, S.~Heinemeyer, W.~Hollik, C.~Wagner
              and G.~Weiglein, 
              {\em Nucl. Phys.} {\bf B 580} (2000) 29, 
              hep-ph/0001002.

\bibitem{FDRG1} S.~Heinemeyer, W.~Hollik and G.~Weiglein, 
                hep-ph/9910283.

\bibitem{maulpaul3} J.~Espinosa and R.~Zhang,
                   {\em JHEP} {\bf 0003} (2000) 026, 
                   hep-ph/9912236.

\bibitem{feynhiggs} S.~Heinemeyer, W.~Hollik and G.~Weiglein, {\em
                    Comp. Phys. Comm.} {\bf 124} 2000 76,
                    hep-ph/9812320; 
                    hep-ph/0002213.\\
                    The codes are accessible via
                    {\tt www.feynhiggs.de} .

\bibitem{mhiggsf1l} A.~Dabelstein, 
                    {\em Nucl. Phys.} {\bf B 456} (1995) 25, 
                    hep-ph/9503443;
                    {\em Z. Phys.} {\bf C 67} (1995) 495, 
                    hep-ph/9409375.

\bibitem{mssmhiggs}
The LEP working group for Higgs boson searches, LHWG Note 2001-2,\\
{\tt lephiggs.web.cern.ch/LEPHIGGS}.

\bibitem{softSUSYbreaking} L.~Girardello and M.T.~Grisaru,
                           {\em Nucl.\ Phys.} {\bf B 194} (1982) 65.
 
\bibitem{herbi} For a review, see H.~Dreiner, 
                hep-ph/9707435.

\bibitem{rpv2} G.~Bhattacharyya,
               {\em Nucl. Phys. Proc. Suppl.}  {\bf 52 A} (1997) 83,
               hep-ph/9608415.

\bibitem{REWSB} L.E.~Iba\~nez  and G.G.~Ross, 
                {\em Phys. Lett.} {\bf 110} (1982) 215; \\
                K.~Inoue, A.~Kakuto, H.~Komatsu and S.~Takeshita,
                {\em Progr. Theor. Phys.} {\bf 68} (1982) 927;
                {\em Progr. Theor. Phys.} {\bf 71} (1984) 413;\\
                J.~Ellis, D.V.~Nanopoulos and K.~Tamvakis,
                {\em Phys. Lett.} {\bf B 121} (1983) 123;\\
                L.E.~Iba\~nez , 
                {\em Nucl. Phys.} {\bf B 218} (1983) 514;\\
                L.~Alvarez-Gaum\'e, J.~Polchinski and M.~Wise, 
                {\em Nucl. Phys.} {\bf B 221} (1983) 495;\\
                J.~Ellis, J.S.~Hagelin, D.V.~Nanopoulos and K.~Tamvakis, 
                {\em Phys. Lett.} {\bf B 125} (1983) 275;\\
                L.~Alvarez-Gaum\'e, M.~Claudson and M.~Wise, 
                {\em Nucl. Phys.} {\bf B 207} (1982) 96.

\bibitem{Drees} M.~Drees and M.M.~Nojiri,
                {\em Phys.\ Rev.}  {\bf D 45} (1992) 2482.

\bibitem{barger} V.~Barger, M.S.~Berger and P.~Ohmann,
                 {\em Phys.\ Rev.} {\bf D 49}, (1994) 4908,
                 hep-ph/9311269.

\bibitem{bagger} D.M.~Pierce, J.A.~Bagger, K.~Matchev and R.~Zhang,
                 {\em Nucl.\ Phys.}  {\bf B 491} (1997) 3,
                 hep-ph/9606211.

\bibitem{faraggi}
The scalar particle spectrum and phenomenology of this model 
has been studied in
                  A.~Dedes and A.E.~Faraggi,
                  {\em Phys.\ Rev.}  {\bf D 62} (2000) 016010,
                  hep-ph/9907331.

\bibitem{Nilles} D.~Matalliotakis and H.-P.~Nilles, 
                 {\em Nucl. Phys.} {\bf B 435} (1995) 115, 
                 hep-ph/9407251  ; \\
                 A.~Lleyda and C.~Munoz, 
                 {\em Phys. Lett.} {\bf B 317} (1993) 82, 
                 hep-ph/9308208 ; \\
                 N.~Polonsky and A.~Pomarol, 
                 {\em Phys. Rev.} {\bf D 51} (1995) 6532, 
                 hep-ph/9410231 .

\bibitem{pdg} Part. Data Group,
              {\em Eur. Phys. Jour. }{\bf C 15} (2000) 1.

\bibitem{drbar} S.~Martin and M.~Vaughn,
                {\em Phys. Lett.} {\bf B 318} (1993) 331,
                hep-ph/9308222.

\bibitem{arason} H.~Arason, D.J.~Castano, B.Keszthelyi, S.Mikaelian, 
                 E.J.~Piard, P.~Ramond and B.D.~Wright,
                 {\em Phys. Rev.} {\bf D 46} (1992) 3945.

\bibitem{sakis} A.~Dedes, A.B.~Lahanas and K.~Tamvakis,
                {\em Phys.\ Rev.} {\bf D 53}, 3793 (1996),
                hep-ph/9504239.

\bibitem{oldGMSB}
    M.~Dine, W.~Fischler, M.~Srednicki, \NPB{189}{1981}{575};\\
    S.~Dimopoulos, S.~Raby, \NPB{192}{1981}{353};\\
    M.~Dine, W.~Fischler, \PLBold{110}{1982}{227};\\
    M.~Dine,  M.~Srednicki, \NPB{202}{1982}{238};\\
    M.~Dine, W.~Fischler, \NPB{204}{1982}{346};\\
    L.~Alvarez-Gaum\'e, M.~Claudson, M.B.~Wise, \NPB{207}{1982}{96};\\
    C.R.~Nappi, B.A.~Ovrut, \PLBold{113}{1982}{175};\\
    S.~Dimopoulos, S.~Raby, \NPB{219}{1983}{479}.

\bibitem{newGMSB}
    M.~Dine, A.E.~Nelson, \PRD{48}{1993}{1277},
                           hep-ph/9303230;\\
    M.~Dine, A.E.~Nelson, Y.~Shirman, \PRD{51}{1995}{1362},
                           hep-ph/9408384;\\
    M.~Dine, A.E.~Nelson, Y.~Nir, Y.~Shirman, \PRD{53}{1996}{2658},
                           hep-ph/9507378.

\bibitem{GMSBmodels1}
    S.~Dimopoulos, S.~Thomas, J.D.~Wells, 
    \PRD{54}{1996}{3283},
    hep-ph/9604452;\\
    \NPB{488}{1997}{39}, 
    hep-ph/9609434 .

\bibitem{AKM-LEP2} 
    S.~Ambrosanio, G.D.~Kribs, S.P.~Martin, 
    \PRD{56}{1997}{1761}, 
    hep-ph/9703211 . 

\bibitem{AB-LC} S.~Ambrosanio, G.A.~Blair, 
                \EPJC{12}{2000}{287}, 
                hep-ph/9905403 .

\bibitem{AMPPR} S.~Ambrosanio {\it et al.}, 
                {\em JHEP} {\bf 0101} (2001) 014, 
                hep-ph/0010081.

\bibitem{Fayet} P.~Fayet, 
                \PLBold{70}{1977}{461}; 
                \PLBold{86}{1979}{272};
                \PLB{175}{1986}{471} 
    and in ``Unification of the fundamental 
    particle interactions", eds.~S.~Ferrara, J.~Ellis,   
    P.~van Nieuwenhuizen (Plenum, New York, 1980) p.~587.

\bibitem{SUSYFIRE}
  An updated, generalized and {\tt Fortran}-linked version of the 
  program used in Ref.~\cite{AKM-LEP2}. It generates minimal and 
  non-minimal GMSB and SUGRA models.
  For inquiries about this software package, please send e-mail to  
  {\tt Sandro.Ambrosanio@bancaroma.it}.

\bibitem{lepsusySep2000} K.~Jacobs,
                         talk given at the 
                         {\em XXXVI Rencontres de Moriond}, 
                         11th of March 2001, Moriond, France.

\bibitem{negative} A.~Pomarol and R.~Rattazzi,
                   {\em JHEP} {\bf 9905} (1999) 013, 
                   hep-ph/9903448.

\bibitem{clm} Z.~Chacko, M.A.~Luty and E.~Ponton, 
              {\em JHEP} {\bf 0004} (2000) 001, 
              hep-ph/9905390. 

\bibitem{ben} B.C.~Allanach and A.~Dedes,
              {\em JHEP} {\bf 0006} (2000) 017,
              hep-ph/0003222.

\bibitem{kss} E.~Katz, Y.~Shadmi and Y.~Shirman,
              {\em JHEP} {\bf 9908} (1999) 015, 
              hep-ph/9906296  .

\bibitem{hff} S.~Heinemeyer, W.~Hollik and G.~Weiglein, 
              {\em Eur. Phys. Jour.} {\bf C 16} (2000) 139, 
              hep-ph/0003022.

\bibitem{eehZhA} S.~Heinemeyer, W.~Hollik, J.~Rosiek and G.~Weiglein,
                 {\em Eur. Phys. Jour.} {\bf C 19} (2001) 535, 
                 hep-ph/0102081.

\bibitem{sakis2} A.~Dedes, unpublished.

\bibitem{FeynSSG} A.~Dedes, S.~Heinemeyer and G.~Weiglein,
``{\em FeynSSG:} A program for the minimal supergravity
and Higgs spectrum", {\em in preparation}. The code will be accessible
via {\tt www.feynhiggs.de} .

\bibitem{mhiggs1lrosiek} P.~Chankowski, S.~Pokorski and J.~Rosiek,
                         {\em Phys. Lett.} {\bf B 286} (1992) 307;
                         {\em Nucl. Phys.} {\bf B 423} (1994) 423,
                         hep-ph/9303309.

\bibitem{mhiggs1lren} M.~Frank, S.~Heinemeyer, W.~Hollik and G.~Weiglein,
                      {\em in preparation}.

\bibitem{rhoparameter} M.~Veltman, 
                       {\em Nucl. Phys.} {\bf B 123} (1977) 89. 

\bibitem{delrhosusy2loop} A.~Djouadi, P.~Gambino, S.~Heinemeyer, W.~Hollik,
                          C.~J\"unger and G.~Weiglein,
                          {\em Phys. Rev. Lett.} {\bf 78} (1997) 3626,
                          hep-ph/9612363;
                          {\em Phys. Rev.} {\bf D 57} (1998) 4179,
                          hep-ph/9710438;\\
                          S.~Heinemeyer and G.~Weiglein, hep-ph/0102317.
                      
\bibitem{lepFestOct2000} LEP Fest, Oct.9-11, 2000, Talks given by  
     F.~Cerutti (ALEPH), 
     M.~Kienzle (L3), 
     R.~Hemingway (OPAL), 
     V.~Hedberg (DELPHI).

\bibitem{CDF} 
    C. Pagliarone, SUSY Searches at the Tevatron Collider,
     Proceedings of 13th Topical Conference on Hadron Collider Physics, 
     Mumbai, India, 14-20 Jan 1999, hep-ex/0011016.

\bibitem{Iashvili} I.~Iashvili,
                   hep-ex/0007001.

\bibitem{deVivie} J.B.~de Vivie,
                  hep-ex/9911032.

\bibitem{tbexcl} S.~Heinemeyer, W.~Hollik and G.~Weiglein, 
                 {\em JHEP} {\bf 0006} (2000) 009, 
                 hep-ph/9909540.

\bibitem{riotto} A.~Riotto and E.~Roulet,
                 {\em Phys. Lett.} {\bf B 377} (1996) 60,
                 hep-ph/9512401.

\bibitem{kusenko} A.~Kusenko, P.~Langacker and G.~Segre,
                  {\em Phys.\ Rev.}  {\bf D 54} (1996) 5824,
                  hep-ph/9602414.

\bibitem{abel} S.~Abel, C.A.~Savoy, 
               {\em Phys. Lett} {\bf B 444} (1998) 119,
               hep-ph/9809498; \\
               S.~Abel, T.~Falk, 
               {\em Phys. Lett.} {\bf B 444} (1998) 427,
               hep-ph/9810297 ; \\
               S.~Abel and B.C.~Allanach,
               {\em JHEP} {\bf 0007} (2000) 037,
               hep-ph/9909448.

\bibitem{subir}
For an overview on bounds see for example\\
S. Sarkar, {\em Rept. Prog. Phys.} {\bf 59} (1996) 1493, 
              hep-ph/9602260.

\bibitem{bsgammaexp} CLEO Collaboration, M.S. Alam {\it et al.}, 
                     {\em Phys. Rev. Lett.} {\bf 74} (1995) 2885 
                     as updated in
                     S.~Ahmed {\it et al.}, {CLEO CONF 99-10};\\
                     K.~Abe {\it et al.}, Belle Collaboration,
                     hep-ex/0103042.

\bibitem{gminus2} H.N.~Brown {\it et al.}, Muon $g_\mu - 2$ Collaboration,
                  {\em Phys. Rev. Lett.}  {\bf 86} (2001) 2227,
                  hep-ex/0102017;\\
                  A.~Czarnecki and W.~J.~Marciano,
                  {\em Phys. Rev.} {\bf D 64} (2001) 013014,
                  hep-ph/0102122.

\bibitem{higgs115msugra2} G.~Kane, S.~King and L.-T.~Wang, 
                          hep-ph/0010312;\\
                          A.~Bottino, N.~Fornengo and S.~Scopel, 
                          hep-ph/0012377;\\
                          J.~Ellis, D.V.~Nanopoulos and K.A.~Olive, 
                          hep-ph/0102331.

\bibitem{higgs115msugra3} J.~Ellis, S.~Heinemeyer, K.A.~Olive and G.~Weiglein, 
                          to appear in {\em Phys. Lett.} {\bf B},
                          hep-ph/0105061.

\bibitem{focusp} J.L.~Feng and T.~Moroi, 
                 {\em Phys. Rev.} {\bf D 61} (2000) 095004,
                 hep-ph/9907319;\\
                 J.L.~Feng, K.T.~Matchev and T.~Moroi, 
                 {\em Phys. Rev. Lett.} {\bf 84} (2000) 2322, 
                 hep-ph/9908309; 
                 {\em Phys. Rev.} {\bf D 61} (2000) 075005,
                 hep-ph/9909334.

\bibitem{bsgAterm} F.~Gabbiani, E.~Gabrielli, A.~Masiero and L.~Silvestrini,
                   {\em Nucl. Phys.} {\bf B 477} (1996) 321, 
                   hep-ph/9604387;\\
                   F.~Borzumati, C.~Greub, T.~Hurth and D.~Wyler,
                   {\em Phys. Rev.} {\bf D 62} (2000) 075005,
                   hep-ph/9911245.

\bibitem{cdm} J.~Ellis, J.~Hagelin, D.~Nanopoulos, K.~Olive and M. Srednicki, 
              {\em  Nucl. Phys.} {\bf B238} (1984) 453; \\
              H. Goldberg, 
              {\em Phys. Rev. Lett.} {\bf 50} (1983) 1419;\\
              J.~Ellis, T.~Falk, G.~Ganis, K.~Olive and M.~Srednicki,
              {\em Phys. Lett.} {\bf B 510} (2001) 236,
              hep-ph/0102098.

\bibitem{thermalinflation} D.~Lyth and E.~Stewart,
                          {\em Phys. Rev.} {\bf D 53} (1996) 1784,
                          hep-ph/9510204.

\bibitem{axinos} L.~Covi, J.~Kim and L.~Roszkowski,
                 {\em Phys. Rev. Lett.} {\bf 82} (1999) 4180,
                 hep-ph/9905212.

\bibitem{Barate:2000zr} R.~Barate {\it et al.}  [ALEPH Collaboration],
                        {\em Phys. Lett.} {\bf B 499} (2001) 53,
                        hep-ex/0010062.

\bibitem{cdfhiggs} T.~Affolder {\it et al.}, [CDF Collaboration]
                   {\em Phys. Rev. Lett.} {\bf D 86} (2001) 4472,
                   hep-ex/0010052.

\bibitem{deltamb1} L.J.~Hall, R.~Rattazzi and U.~Sarid,
                   {\em Phys. Rev.} {\bf D 50} (1994) 7048,
                   hep-ph/9306309.

\bibitem{deltamb2} M.~Carena, S.~Mrenna and C.~Wagner,
                   {\em Phys. Rev.} {\bf D 60} (1999) 075010,
                   hep-ph/9808312;
                   {\em Phys. Rev.} {\bf D 62} (2000) 055008,
                   hep-ph/9907422;\\
                   H.~Eberl, K.~Hidaka, S.~Kraml, W.~Majerotto and
                   Y.~Yamada,
                   {\em Phys. Rev.} {\bf D 62} (2000) 055006,
                   hep-ph/9912463;\\
                   M.~Carena, D.~Garcia, U.~Nierste and C.~Wagner,
                   {\em Nucl. Phys.} {\bf B 577} (2000) 88,
                   hep-ph/9912516.

\bibitem{mhiggslle} S.~Heinemeyer, W.~Hollik and G.~Weiglein, 
                   {\em Phys. Lett.} {\bf B 455} (1999) 179,
                   hep-ph/9903404.

\bibitem{higgs115msugra1} J.~Ellis, G.~Ganis, D.V.~Nanopoulos 
                          and K.A.~Olive,
                          hep-ph/0009355;\\
                          J.~Ellis,
                          hep-ex/0011086.

\bibitem{hinchrich} I.~Hinchliffe and P. Richardson, 
                    hep-ph/0106212.

\bibitem{sola} A.~Belyaev, D.~Garcia, J.~Guasch and J.~Sola,
               hep-ph/0105053.

\bibitem{tev:susysearch} S.~Abel {\it et al.} 
                         [SUGRA Working Group Collaboration],
{\it Report of the SUGRA working group for Run~II of the Tevatron},
hep-ph/0003154 and references therein;\\
                    R.~Culbertson {\it et al.},
                    hep-ph/0008070, and references therein.

\bibitem{atlastdr} ATLAS Collaboration, 
  {\em Detector and Physics Performance Technical Design Report},
  CERN/LHCC/99-15 (1999), see:\\
 {\tt atlasinfo.cern.ch/Atlas/GROUPS/PHYSICS/TDR/access.html} .

\bibitem{cms} CMS Collaboration, see:\\
 {\tt cmsinfo.cern.ch/Welcome.html/CMSdocuments/CMSplots/} .

\bibitem{teslatdr} TESLA TDR Part~3: ``Physics at an $e^+e^-$ 
                   Linear Collider'', 
                   eds. R.D.~Heuer, D.~Miller, F.~Richard and P.M.~Zerwas, 
                   hep-ph/0106315,
                   see: {\tt tesla.desy.de} .

\bibitem{orangebook} T.~Abe {\it et al.}  
                     [American Linear Collider Working Group Collaboration],
                     {\it Resource book for Snowmass 2001}, 
                     hep-ex/0106055, 
                     hep-ex/0106056,
                     hep-ex/0106057, 
                     hep-ex/0106058.

\bibitem{feng} J.L.~Feng and T.~Moroi, 
               {\em Phys. Rev.} {\bf D 61} (2000) 095004, 
               hep-ph/9907319.

\bibitem{winotev} J.L.~Feng, T.~Moroi, L.~Randall, M.~Strassler and S.~Su, 
                  {\em Phys. Rev. Lett.} {\bf 83} (1999) 1731, 
                  hep-ph/9904250.

\bibitem{winolsp} C.H.~Chen, M.~Drees and J.~Gunion, 
                  {\em Phys. Rev. Lett.} {\bf 76} (1996) 2002,
                  hep-ph/9512230;
                  E: {\em Phys. Rev.} {\bf D 55} (1997) 330, 
                  hep-ph/9902309;\\
                  J.~Gunion and S.~Mrenna, 
                  {\em Phys. Rev.} {\bf D 62} (2000) 015002,
                  hep-ph/9906270.

\end{thebibliography}
\end{document}